%% file: paper.tex
\pdfoutput=1

\documentclass[british]{article}
\usepackage[T1]{fontenc}
\usepackage[latin9]{inputenc}
\usepackage{float}
\usepackage{textcomp}
\usepackage{amsthm}
\usepackage{amsmath}
\usepackage{amssymb}
\usepackage{graphicx}
\usepackage{esint}
\usepackage[authoryear]{natbib}
\PassOptionsToPackage{normalem}{ulem}
\usepackage{ulem}

\makeatletter

\newcommand{\noun}[1]{\textsc{#1}}
\providecommand{\tabularnewline}{\\}
\newcommand{\lyxdot}{.}

\floatstyle{ruled}
\newfloat{algorithm}{tbp}{loa}
\providecommand{\algorithmname}{Algorithm}
\floatname{algorithm}{\protect\algorithmname}

\numberwithin{equation}{section}
\numberwithin{figure}{section}
\newenvironment{lyxlist}[1]
{\begin{list}{}
{\settowidth{\labelwidth}{#1}
 \setlength{\leftmargin}{\labelwidth}
 \addtolength{\leftmargin}{\labelsep}
 }}
{\end{list}}
\theoremstyle{plain}
\newtheorem{thm}{\protect\theoremname}[section]
\theoremstyle{plain}
\newtheorem{lem}[thm]{\protect\lemmaname}
\ifx\proof\undefined
\newenvironment{proof}[1][\protect\proofname]{\par
\normalfont\topsep6\p@\@plus6\p@\relax
\trivlist
\itemindent\parindent
\item[\hskip\labelsep\scshape #1]\ignorespaces
}{%
\endtrivlist\@endpefalse
}
\providecommand{\proofname}{Proof}
\fi
\theoremstyle{remark}
\newtheorem{rem}[thm]{\protect\remarkname}

\usepackage{fullpage}
\usepackage{caption}
\usepackage{datetime} 
\usepackage{lastpage} 	

\usepackage{graphicx}
\usepackage{sidecap}

\usepackage{bbm} \newcommand{\indfun}{\mathbbm{1}} 
\usepackage{mathabx}

\newcommand{\drv}{ \textup{d} } 
 
\newcommand{\Var}{\textup{Var} }

\newcommand{\transpose}{\intercal} 
\newcommand{\idmat}{\mathit{I} } 

\newcommand{\utwi}[1]{\mbox{\boldmath $#1$}}

\providecommand{\cal}{\mathcal}

\newcommand{\bP}{\utwi{P}}

\newcommand{\mN}{{\cal N}}

\newcommand{\bbR}{\mathbb{R}}
\newcommand{\mR}{\mathbb{R}}

 \newcommand{\BX}{\mathbb{X}}
\newcommand{\mT}{{\cal T}}

\newcommand{\mF}{{\cal F}}

\newcommand{\mK}{{\cal K}}
\newcommand{\mX}{{\cal X}}

\newcommand{\mL}{{\cal L}}

\newcommand{\mB}{{\cal B}}
 
\newcommand{\mP}{{\cal P}}

\def\spacingset#1{\renewcommand{\baselinestretch}{#1}\small\normalsize} \spacingset{1}

\usepackage{lipsum}
\usepackage{eso-pic}

\AddToShipoutPictureBG{
  \AtPageLowerLeft{%
    \raisebox{3\baselineskip}{\makebox[\paperwidth]{
     \begin{minipage}{21cm}
       \centering
      \tiny 
      Parallel and Interacting Stochastic Approximation Annealing algorithms for global optimisation \\
      Georgios Karagiannis, Bledar A. Konomi, Guang Lin, and Faming Liang 
    \end{minipage}}}
  }
}

\@ifundefined{showcaptionsetup}{}{%
 \PassOptionsToPackage{caption=false}{subfig}}
\usepackage{subfig}
\makeatother

\usepackage{babel}
\providecommand{\lemmaname}{Lemma}
\providecommand{\remarkname}{Remark}
\providecommand{\theoremname}{Theorem}

\begin{document}

\title{Parallel and Interacting Stochastic Approximation Annealing algorithms
for global optimisation}

\author{%
\begin{tabular}{c||c}
\multicolumn{2}{c}{%
\begin{tabular}{c}
Georgios Karagiannis\tabularnewline
Department of Mathematics \tabularnewline
Purdue University\tabularnewline
West Lafayette, IN 47907-2067, USA\tabularnewline
gkaragia@purdue.edu ; georgios.stats@gmail.com\tabularnewline
~\tabularnewline
\end{tabular}}\tabularnewline
\multicolumn{2}{c}{%
\begin{tabular}{c}
Bledar A. Konomi \tabularnewline
Department of Mathematical Sciences \tabularnewline
University of Cincinnati\tabularnewline
Cincinnati OH 45221, USA \tabularnewline
alex.konomi@uc.edu\tabularnewline
~\tabularnewline
\end{tabular}%
\begin{tabular}{c}
Guang Lin\tabularnewline
Department of Mathematics and \tabularnewline
School of Mechanical Engineering\tabularnewline
Purdue University\tabularnewline
West Lafayette, IN 47907-2067, USA\tabularnewline
lin491@purdue.edu\tabularnewline
\end{tabular}}\tabularnewline
\multicolumn{2}{c}{%
\begin{tabular}{c}
Faming Liang \tabularnewline
Department of Biostatistics\tabularnewline
University of Florida\tabularnewline
Gainesville, FL 32611-7450, USA\tabularnewline
faliang@ufl.edu\tabularnewline
~\tabularnewline
\end{tabular}}\tabularnewline
\end{tabular}}

\date{4th February 2015}
\maketitle
\begin{abstract}
We present the parallel and interacting stochastic approximation annealing
(PISAA) algorithm, a stochastic simulation procedure for global optimisation,
that extends and improves the stochastic approximation annealing (SAA)
by using population Monte Carlo ideas. The standard SAA algorithm
guarantees convergence to the global minimum when a square-root cooling
schedule is used; however the efficiency of its performance depends
crucially on its self-adjusting mechanism. Because its mechanism is
based on information obtained from only a single chain, SAA may present
slow convergence in complex optimisation problems. The proposed algorithm
involves simulating a population of SAA chains that interact each
other in a manner that ensures significant improvement of the self-adjusting
mechanism and better exploration of the sampling space. Central to
the proposed algorithm are the ideas of (i) recycling information
from the whole population of Markov chains to design a more accurate/stable
self-adjusting mechanism and (ii) incorporating more advanced proposals,
such as crossover operations, for the exploration of the sampling
space. PISAA presents a significantly improved performance in terms
of convergence. PISAA can be implemented in parallel computing environments
if available. We demonstrate the good performance of the proposed
algorithm on challenging applications including Bayesian network learning
and protein folding. Our numerical comparisons suggest that PISAA
outperforms the simulated annealing, stochastic approximation annealing,
and annealing evolutionary stochastic approximation Monte Carlo especially
in high dimensional or rugged scenarios.
\end{abstract}
\newpage 
\spacingset{1.45} 

\section{Introduction \label{sec:Introduction}}

There is a continuous need for development of efficient algorithms
to tackle mathematical optimisation problems often met in several
fields of science. For instance, in computational chemistry, predicting
the native conformation of a protein can be performed by minimising
its potential energy. In classical or Bayesian statistics, inference
can be performed by maximising the likelihood function (a statistical
model assumed to have generated an observed data set) \citep{CasellaBerger1990}
or the associated posterior distribution density (a distribution that
reflects the researcher's belief in the unknown quantities of interest)
\citep{Robert2007}, correspondingly. 

We assume that there is interest in minimising a function $U(x)$,
called cost function, defined on a space $\mathcal{X}\subset\mathbb{R}^{d}$;
i.e. we seek $(x_{*},U(x_{*}))$ such that $x_{*}=\arg\min_{\forall x\in\mathcal{X}}U(x)$.
Hereafter, we will discuss in terms of minimisation because maximisation
of $U(x)$ can be performed equivalently by minimising the function
$\tilde{U}(x):=-U(x)$. Several stochastic optimisation algorithms
have been proposed in the literature, e.g. simulated annealing (SA)
\citep{KirkpatrickVecchi1983,MetropolisRosenbluthRosenbluthTellerTeller1953},
genetic algorithm \citep{Goldberg1989,Holland1975}, annealing stochastic
approximation Monte Carlo (ASAMC ) \citep{Liang2007}, annealing evolutionary
stochastic approximation Monte Carlo (AESAMC) \citep{Liang2011},
stochastic approximation annealing (SAA) \citep{LiangChengLin2013}.
Albeit their success, they encounter various difficulties in converging
to the global minimum, an issue that becomes more severe when $U(\cdot)$
is highly rugged or high dimensional.

Simulated annealing (SA) \citep{KirkpatrickVecchi1983,CernyVladimir1985}
aims at finding the global minimum based on the fact that minimisation
of $U(x)$ can be addressed in statistical terms by simulating the
Boltzmann distribution $f_{\tau_{*}}(x)$, with density $f_{\tau_{*}}(x)\propto\exp(-\frac{1}{\tau_{*}}U(x))$,
at a small value of temperature parameter $\tau_{*}>0$ close to $0$.
SA considers a temperature ladder $\{\tau_{t}\}$ that is a monotonically
decreasing sequence of temperatures with $\tau_{1}$ reasonably large.
A standard version of SA involves simulating consecutively from a
sequence of Boltzmann distributions $\{f_{\tau_{t}}(x);t=0,1,...\}$,
parametrised by the temperature ladder, via Metropolis-Hastings MCMC
updates \citep{Hastings1970,MetropolisRosenbluthRosenbluthTellerTeller1953}.
A standard version of SA is presented in Algorithm \ref{Alg:SA} as
a pseudo-code. At early iterations, the algorithm aims at escaping
from the attraction of local minima by flattening $f_{\tau_{t}}(x)$
through $\tau_{t}$. During the subsequent iterations, $\tau_{t}$
decreases progressively towards $0$, and hence the values simulated
from $f_{\tau_{t}}(x)$ concentrate in a narrower and narrower neighbourhood
of the global mode of $f_{\tau_{t}}(x)$ (or equiv. the global minimum
of $U(x)$). In theory, convergence of SA to the global minimum can
be ensured with probability $1$ if a logarithmic cooling schedule
$O(1/\log(t))$ is adopted \citep{GemanGeman1984,HaarioSaksman1991},
however this rate is too slow to be implemented in practice because
it requires an extremely long CPU time. In practice, linear or geometric
cooling schedules are used, however they do not guarantee convergence
to the global minimum, and hence the algorithm tends to become trapped
to local minima in complex scenarios. 
\begin{algorithm}
\begin{description}
\item [{Requires :}] Seed $x_{0}\in\mathcal{X}$, temperature ladder $\{\tau_{t}\}$,
density $f_{\tau_{t}}(x)\propto\exp(-\frac{1}{\tau_{t}}U(x))$.
\item [{Initialise :}] At $t=0$, set $x_{0}\in\mathcal{X}$, and $\tau_{0}>0$.
\item [{Iterate :}] For $t=1,...,T$,

For $n_{t}$ iterations repeat simulating $f_{\tau_{t}}(\cdot)$ by
using a Metropolis-Hastings algorithm:
\begin{enumerate}
\item Propose $x'\sim Q(\drv\cdot|x)$, where $Q(\drv\cdot|\cdot)$ is a
proposal distribution that can be sampled directly.
\item Accept $x'$ as $x_{t}$ with prob. $a_{\text{MH}}=\min(1,\frac{f_{\tau_{t}}(x')}{f_{\tau_{t}}(x_{t-1})}\frac{Q(x_{t-1}|x')}{Q(x'|x_{t-1})})$.
\end{enumerate}
\end{description}
\protect\caption{Simulated annealing algorithm used to detect the minimum of a cost
function $U(x)$, $x\in\mathcal{X}$\label{Alg:SA} }
\end{algorithm}

The stochastic approximation annealing (SAA) \citep{LiangChengLin2013}
is a stochastic optimisation algorithm that builds upon the SA and
SAMC%
\footnote{SAMC \citep{LiangLiuCarroll2007,LiangLiuCarroll2010,WuLiang2011,BornnJacobMoralDoucet2013,SongWuLiang2014}
is an adaptive MCMC sampler that aims at addressing the local mode
trapping problem that standard MCMC samplers encounter. It is a generalisation
of the Wang-Landau algorithm \citep{WangLandau2001} but equipped
with a stochastic approximation scheme \citep{RobbinsMonro1951} that
adjusts the target distribution. It involves generating a time-inhomogeneous
Markov chain that targets a biased distribution, adjusted as the iterations
evolve, instead of the distribution of interest itself. The biased
distribution is parametrised by a partition scheme and designed such
that the generated chain equally visits each subregion of the partition
with a predetermined frequency as the iterations evolve. For an overview
see \citep{Liang2014}.%
} ideas. It involves simulating a time-inhomogeneous Markov chain via
MCMC transitions targeting a sequence of modified Boltzmann distributions
whose densities adaptively adjust via a stochastic approximation mechanism
(inherited by SAMC). Each distribution of the sequence is biased according
to a partitioning scheme (inherited by SAMC) and parametrised by a
temperature ladder (inherited by SA). SAA aims at gradually forcing
sampling toward the local minima of each subregion of the partition
through lowering the temperature with iterations, while it ensures
that each subregion is visited by the chain according to a predetermined
frequency. This strategy shrinks the sampling space in a soft manner
and enables SAA to escape from local traps. The global minimum is
guaranteed to be reached as the temperature tends to $0$ if the temperature
ladder uses a square root cooling schedule $O(1/\sqrt{t})$ \citep{LiangChengLin2013}.
We emphasise that, compared to SA, SAA ensures convergence to global
minimum at a much faster cooling schedule (square-root). In spite
of these appealing features, the performance of SAA crucially depends
on the efficiency of the self-adjusting mechanism and the exploration
of the sampling space involved. In scenarios that the cost function
is rugged or high-dimensional, the exploration of the sampling space
can be slow because it is performed by a single Markov chain. Moreover,
the information obtained to support the self-adjusting process is
limited which makes the adjustment of the target density quite unstable
and too slow to convergence. When the target distribution is poorly
adjusted, the convergence of the whole algorithm to the global minimum
decays severely, and the chain may be trapped in local minima. This
problematic behaviour can downgrade severely the overall performance
of SAA, or even cause local trapping, in complex optimisation problems.

In this article, we develop the parallel and interacting stochastic
approximation annealing (PISAA), a general purpose stochastic optimisation
algorithm, that extends SAA \citep{LiangChengLin2013} by using population
Monte Carlo ideas \citep{SongWuLiang2014,BornnJacobMoralDoucet2013,LiangWong2000,LiangWong2001,WuLiang2011}.
Essentially, PISAA works on a population of SAA chains that interact
each other in a manner that eliminates the aforementioned problematic
behaviour of SAA, and accelerates the overall convergence. This allows
the proposed algorithm to demonstrate great performance, and address
challenging optimisation problems with high-dimensional and very rugged
cost functions. PISAA is enabled to use advanced MCMC transitions
that incorporate crossover operations. These operations allow the
distributed information across chains of the population to be used
in guiding further simulations, and therefore lead to a more efficient
exploration of the sampling space. Furthermore, PISAA is equipped
with a more accurate and stable self-adjusting mechanism for the target
density, that uses information gained from the whole population, and
therefore accelerates the overall convergence of the algorithm to
the global minimum. The use of multiple chains allows PISAA to initialise
from various locations and search for the global minimum at different
regions of the sampling space simultaneously. PISAA can be implemented
in parallel, if parallel computing environment is available, and hence
the computational overhead due to the generation of multiple chains
can be reduced dramatically. It is worth emphasising that PISAA is
not just an implementation of the SAA running in parallel; its key
feature is the way the parallel chains interact in order to overcome
the aforesaid problematic behaviour and improve performance. Our numerical
examples suggest that the performance of PISAA improves with the size
of the population. Also, in problems where the cost function is rugged
or high-dimensional, PISAA significantly outperforms other competitors,
SA, ASAMC, and SAA, and their population analogues, VFSA, AESAMC,
as it was able to discover the global minimum much quicker.

The layout of the article is as follows. In Section \ref{sec:Stochastic-approximation-annealing},
we give a brief review of SAA and discuss problems concerning the
efficiency of the algorithm; in Section \ref{sec:Parallel-interacting-stochastic-approximation-annealing},
we present the proposed algorithm PISAA; in Section \ref{sec:Applications},
we examine the performance of the proposed algorithm and compare it
with those of other stochastic optimisation algorithms (such as SA,
ASAMC, AESAMC, and SAA) against challenging optimisation problems;
and in Section \ref{sec:Conclusion}, we conclude.

\section{Stochastic approximation annealing: A review \label{sec:Stochastic-approximation-annealing}}

Stochastic approximation annealing (SAA) algorithm \citep{LiangChengLin2013}
casts the optimisation problem in a combined framework of SAMC and
SA, in the sense that the variant distribution is self-adjusted and
parametrised by a sampling space partition and temperature ladder.

Let $\mathcal{E}=\{E_{j};\ j=1,...,m\}$ be a partition of the sampling
space $\mathcal{X}$ with subregions $E_{1}=(x\in\mathcal{X}:\ -\infty<U(x)\le u_{1})$,
..., $E_{j}=(x\in\mathcal{X}:\ u_{j-1}<U(x)\le u_{j})$, ..., $E_{m}=(x\in\mathcal{X}:\ u_{m-1}<U(x)<\infty)$,
and grid $\{u_{j};\ u_{j}\in\mathbb{R},\ j=1:m-1\}$, for $m>1$.
SAA aims at drawing samples from each subregion with a pre-specified
frequency. Let $\pi:=(\pi_{j};\ j=1,...,m)$, such that $\pi_{j}=\Pr(x\in E_{j})$,
$\pi_{j}>0$ and $\sum_{j=1}^{m}\pi_{j}=1$, denote the vector of
desired sampling frequencies of the $m$ subregions $\{E_{j}\}$.
We refer to $\{\pi_{j}\}$ as the desired probability. How to choose
the partition scheme $\mathcal{E}$ for the sampling space or the
desired probability $\{\pi_{j}\}$ are problem dependent. SAA seeks
to draw samples from the modified Boltzmann distribution with density
\begin{align}
f_{\theta_{*},\tau_{*}}(x;\mathcal{E}) & =\sum_{j=1}^{m}\pi_{j}\frac{1}{w_{*}^{(j)}}\exp(-\frac{1}{\tau_{*}}U(x))\indfun(x\in E_{j});\label{eq:dist_SAA}\\
 & \propto\sum_{j=1}^{m}\exp(-\frac{1}{\tau_{*}}U(x)-\theta_{*}^{(j)})\indfun(x\in E_{j}),\nonumber 
\end{align}
at a low temperature value $\tau_{*}$, where $w_{*}:=(w_{*}^{(j)};j=1:m)$,
$w_{*}^{(j)}=\int_{E_{j}}\exp(-\frac{1}{\tau_{*}}U(x))\drv x<\infty$
are called bias weights, and $\theta_{*}^{(j)}$ is such that $\exp(\theta_{*}^{(j)})\propto w_{*}^{(j)}/\pi_{j}$,
for $j=1,...,m$. 

The rational behind SAA is that, if $\{\theta_{*}\}$ were known,
sampling from (\ref{eq:dist_SAA}) could lead to a random walk in
the space of subregions (by regarding each subregion as a point) with
each subregion being sampled with frequency proportional to $\{\pi_{j}\}$.
Ideally, this can ensure that the lowest energy subregion can be reached
by SAA in a long enough run and thus samples can be drawn from the
neighbourhood of the global minimum when $\tau_{*}$ is close to $0$. 

Since $\{w_{*}^{(j)}\}$ are generally unknown, in order to simultaneously
approximate these values and perform sampling, SAA is equipped with
an adaptive MCMC scheme that combines SAMC and SA algorithms. Let
$\{\gamma_{t};\ t=1,...\}$ denote the gain factor, in terms of SAMC
algorithm, that is a deterministic, positive, and non-increasing sequence
such as $\gamma_{t}=t_{0}/t^{\beta}$ with $\beta\in(0.5,1].$ Let
$\{\tau_{t}\}$ denote a temperature ladder, in terms of SA algorithm,
that is a deterministic, positive and non-increasing sequence such
as $\tau_{t}=t_{1}/\sqrt{t}+\tau_{*}$ with $t_{1}>0$, and $\tau_{*}>0$
very small. We consider a sequence $\theta_{t}:=(\theta_{t}^{(j)},\ j=1:m)$,
as a working estimator of $\{\theta_{*}\}$, where $\theta_{t}\in\Theta$
and $\Theta\subseteq\mathbb{R}^{m}$ is a compact set, e.g. $\Theta=[10^{-10},10^{10}]^{m}$.
A truncation mechanism is also considered in order to ensure that
$\{\theta_{t}\}$ remains in compact set $\Theta$. We define $\{M_{c};\ c=1,...\}$
as a positive, increasing sequence of truncation bounds for $\{\theta_{t}\}$,
and $\{c_{t}\}$ as the total number of truncations until iteration
$t$. 

SAA algorithm proceeds as a recursion which consists of three steps,
at iteration $t$: The sampling update, where a sample $x_{t}$ is
simulated from a Markov chain transition probabilities $P_{\theta_{t-1},\tau_{t}}(x_{t},\drv\cdot;\mathcal{E})$
(e.g. a Metropolis-Hastings kernel) with invariant distribution $f_{\theta_{t-1},\tau_{t}}(\drv\cdot;\mathcal{E})$;
the weight update, where the unknown bias weights of the target density
are approximated through a self-adjusting mechanism; and the truncation
step, where $\{\theta_{t}\}$ is ensured to be in a compact set of
$\Theta$. Given the notation above, SAA is presented as a pseudo-code
in Algorithm \ref{Alg:SAA}. 
\begin{algorithm}[H]
\begin{description}
\item [{Requires :}] Insert $\{\tau_{t}\}$, $\{\gamma_{t}\}$, $\{E_{j}\}$,
$\{\pi_{j}\}$, $\{M_{k}\}$, $\tilde{\theta}_{0}$
\item [{Initialise :}] At $t=0$, set $x_{0}\in\mathcal{X}$, $\tilde{\theta}_{0}\in\Theta$,
such that $\|\tilde{\theta}_{0}\|_{2}<M_{0}$, and $c_{0}=0$.
\item [{Iterate :}] For $t=1,...,n$,

\begin{enumerate}
\item Sampling update:

Simulate $x_{t}$ from the Metropolis-Hastings transition probability
$P_{\theta_{t-1},\tau_{t}}(x_{t-1},\drv\cdot;\mathcal{E})$ that targets
$f_{\theta_{t-1},\tau_{t}}(\drv\cdot;\mathcal{E})$

\item Weight update:

Compute $\theta'=\theta_{t-1}+\gamma_{t}H_{\tau_{t}}(\theta_{t-1},x_{t})$,
where $H_{\tau_{t}}(\theta_{t},x_{t})=[p_{t}-\pi]$, $p_{t}:=(p_{t}^{(j)},j=1:m)$,
and $p_{t}^{(j)}=\indfun(x_{t}\in E_{j})$ for $j=1,...,m$.

\item Truncation step:

Set $\theta_{t}=\theta'$, and $c_{t}=c_{t-1}$ if $\|\theta'^{(j)}\|_{2}\le M_{c_{t}}$,
or set $\theta_{t}=\tilde{\theta}_{0}$, and $c_{t}=c_{t-1}+1$ if
otherwise.

\end{enumerate}
\end{description}
\protect\caption{Stochastic approximation annealing algorithm\label{Alg:SAA} }
\end{algorithm}

\noindent Note that additive transformations of $\{\theta_{t}\}$
leave $f_{\theta_{t-1},\tau_{t}}(\cdot;\mathcal{E})$ invariant. Therefore,
it is possible to apply a $\theta$-normilisation step at the end
of the run, such that $\tilde{\theta}_{n}\leftarrow\theta_{n}+z$,
where $\sum_{j=1}^{m}\exp(\theta_{n}^{(j)}+z)=Z$, and $Z$ is a pre-specified
constant, e.g. $z=(-\log(\sum_{j=1}^{m}\exp(\theta_{n}^{(j)}));\ j=1:m)$
for $Z=1$. Appropriate conditions under which SAA is a valid adaptive
MCMC algorithm that converges to the global minimum are reported in
detail in \citep[Conditions A1-A3 in ][]{LiangChengLin2013}.

SAA presents a number of appealing features when employed to minimise
complex systems with rugged cost functions. SAA can work with an affordable
square-root cooling schedule $O(1/\sqrt{t})$ for $\{\tau_{t}\}$,
which guarantees the global minimum to be reached as the temperature
tends to $\tau_{*}\approx0$, $\lim_{t\rightarrow\infty}\tau_{t}=\tau_{*}$.
It is able to locate the minima of each subregion simultaneously (including
the global minimum), after a long run, if $\tau_{*}$ is close to
0 \citep[Corollary 3.1 in ][]{LiangChengLin2013}. It is worth mentioning
that the square-root rate is much faster than the logarithmic rate
that guarantees convergence in the SA algorithm. SAA gradually forces
sampling toward the local minima of each subregion of the partition
through lowering the temperature with iterations while it ensures
that each subregion is visited by the chain according to the predetermined
frequency $\{\pi_{j}\}$; this reduces the risk of getting trapped
into local minima. 

The superiority of SAA is subject to its self-adjusting mechanism
that operates based on the past samples in order to estimate the unknown
$\{\theta_{*}$\}. This remarkable mechanism, which distinguishes
SAA from SA, proceeds as follows: Given that the current state of
the Markov chain is at the subregion $E_{j}$ and that a proposal
has been made to jump to subregion $E_{j'}$, if the proposal is rejected
during the sampling update, the working value $\theta_{t}^{(j')}$
will be adjusted to increase during the weight update and make it
easier to be accepted in the next iteration; if otherwise, $\theta_{t}^{(j')}$
will be adjusted to decrease during the weight update step and make
it harder to be accepted in the next iteration. Essentially, it penalises
the over-visited subregions and rewards the under-visited subregions,
and hence makes easier for the system to escape from local traps.
This striking mechanism makes the algorithm appealing to address optimisation
problems with rugged cost functions. 

Although SAA can be quite effective, its success depends crucially
on whether the unknown bias weights $\{\theta_{t}\}$ can be estimated
accurately enough through the adjustment process, and whether the
Markov chain, generated through the sampling step, can explore the
sampling space adequately. In complex problems where the ruggedness
or the dimensionality of the cost function are high, the convergence
of $\{\theta_{t}\}$ is usually slow; an issue that significantly
downgrades the overall performance of SAA. The reason is that, at
each iteration, the self-adjusting process relies on limited information
obtained based on a single draw from the sampling step. Essentially,
the function $H_{\tau_{t}}(\theta_{t-1},x_{t})$ is computed by only
one single observation: at iteration $t$, $p_{t}$ in Algorithm \ref{Alg:SAA}
is an $m$-dimensional vector of $0$ \& $1$ (occurrence \& absence)
indicating to which subregion the sample $x_{t}$ belongs. Even after
a long run, this can cause a large variation on the estimate of $\{\theta_{t}\}$
and slow down severely the convergence of $\{\theta_{t}\}$, especially
if the number of subregions $m$ is large. Consequently, the adjustment
of the target density becomes quite unstable and the self-adjusting
mechanism becomes less effective. That can slow down the convergence
of SAA, or even cause the chain to be trapped in local minima. This
problematic behaviour can downgrade severely the ability of SAA to
discover the global minumun in challenging optimisation problems.

Because SAA presents appealing properties, it is of great importance
to design an improved algorithm that inherits the aforementioned desired
features and eliminates the aforementioned problematic behaviour of
SAA.

\section{Parallel and interacting stochastic approximation annealing \label{sec:Parallel-interacting-stochastic-approximation-annealing}}

The parallel and  interacting stochastic approximation annealing (PISAA)
builds on the main principles of SAA \citep{LiangChengLin2013} and
the ideas of population MC \citep{SongWuLiang2014,BornnJacobMoralDoucet2013}.
It works on a population of parallel SAA chains that interact each
other appropriately in order to facilitate the the search for the
global minimum by improving the self-adjusting mechanism and the exploration
of the sampling space. In what follows, we use the notation introduced
in Section \ref{sec:Stochastic-approximation-annealing}.

\subsection{The procedure\label{sub:Description}}

PISAA works with a population of samples at each iteration. At iteration
$t$, let $x_{t}^{(1:\kappa)}:=(x_{t}^{(i)};\ i=1:\kappa)$ denote
the population of samples (abbr. population) which is defined on the
population sample space $\mathcal{X}^{\kappa}:=\mathcal{X}\times\ldots\times\mathcal{X}$.
We refer to $x_{t}^{(i)}$ as population individual and assume that
$x_{t}^{(i)}\in\mathcal{X}$, for $i=1,...,\kappa$, where $\mathcal{X}\in\mathbb{R}^{d}$
is called marginal sample space. The total number of population individuals
$\kappa\ge1$ is called population size. 

We assume that the whole population shares the same common partition
scheme $\mathcal{E}=\{E_{j};\ j=1:m\}$ with subregions $\{E_{j}\}$
defined according to a grid $\{u_{j};\ u_{j}\in\mathbb{R},\ j=1:m-1\}$,
as in Section \ref{sec:Stochastic-approximation-annealing}. For each
individual, PISAA aims at drawing samples from each subregion $\{E_{j}\}$
with a desired probability $\pi:=(\pi_{j};\ j=1,...,m)$ defined as
in Section \ref{sec:Stochastic-approximation-annealing}. Thus, under
these specifications, we define a population modified Boltzmann distribution
with density 
\begin{align}
f_{\theta_{*},\tau_{*}}^{(\kappa)}(x^{(1:\kappa)};\mathcal{E}) & =\prod_{i=1}^{\kappa}f_{\theta_{*},\tau_{*}}(x^{(i)};\mathcal{E});\label{eq:target_pisaa_pop}\\
 & =\prod_{i=1}^{\kappa}\sum_{j=1}^{m}\pi_{j}\frac{1}{w_{*}^{(j)}}\exp(-\frac{1}{\tau_{*}}U(x^{(i)}))\indfun(x^{(i)}\in E_{j});\nonumber \\
 & \propto\prod_{i=1}^{\kappa}\sum_{j=1}^{m}\exp(-\frac{1}{\tau_{*}}U(x^{(i)})-\theta_{*}^{(j)})\indfun(x^{(i)}\in E_{j}),\nonumber 
\end{align}
 where $\{w_{*}^{(j)}\}$, and $\{\theta_{*}^{(j)}\}$ are defined
as in Section \ref{sec:Stochastic-approximation-annealing}. Note
that, the individuals $x^{(i)}$ of the population $x^{(1:\kappa)}$
are independent and identically distributed (i.i.d.) such that each
individual $x^{(i)}$ has marginal distribution $f_{\theta_{*},\tau_{*}}(x^{(i)};\mathcal{E})=\int_{\mathcal{X}^{n-1}}f_{\theta_{*},\tau_{*}}^{(\kappa)}(x^{(1:\kappa)};\mathcal{E})\drv(x^{(1:i-1)},x^{(i+1:\kappa)})$
--the SAA target distribution. Moreover, that the total number of
the unknown weights $\{\theta_{*}^{(j)}\}$ is invariant to the population
size. The reason why we consider the individuals to be i.i.d. (share
common $\mathcal{E}$, $\{\pi_{j}\}$, $\{\theta_{*}^{(j)}\}$) will
become more clear later in the section.

PISAA aims at simulating from the distribution $f_{\theta_{*},\tau_{*}}^{(\kappa)}(\drv\cdot;\ \mathcal{E})$
at a low temperature $\tau_{*}>0$. The reason is similar to that
of SAA: if $\{\theta_{*}^{(j)}\}$ were known, sampling from (\ref{eq:target_pisaa_pop})
could lead to a random walk in the space of subregions with each subregion
being sampled with frequency proportional to $\{\pi_{j}\}$, for each
individual. Ideally, this can ensure that the lowest energy subregion
can be reached, and thus samples can be drawn from the neighbourhood
of the global minimum when $\tau_{*}$ is close to $0$. Because $\{\theta_{*}^{(j)}\}$
are unknown, PISAA employs a population SAMC \citep{SongWuLiang2014,BornnJacobMoralDoucet2013}
embedded with the SA in order to simultaneously approximate their
values and sample the population. Therefore, we consider a sequence
of population modified Boltzmann distributions $\{f_{\theta_{t-1},\tau_{t}}^{(\kappa)}(\drv\cdot;\ \mathcal{E})\}$
with density 
\begin{equation}
f_{\theta_{t-1},\tau_{t}}^{(\kappa)}(\drv\cdot;\ \mathcal{E})\propto\prod_{i=1}^{\kappa}\sum_{j=1}^{m}\exp(-\frac{1}{\tau_{t}}U(x^{(i)})-\theta_{t}^{(j)})\indfun(x^{(i)}\in E_{j}),\label{eq:target_pisaa_pop_working}
\end{equation}
 where the temperature sequence $\{\tau_{t}\}$, gain factor $\{\gamma_{t}\}$,
working estimates $\{\theta_{t}\}$ are defined as in Section \ref{sec:Stochastic-approximation-annealing}.
PISAA is a recursive procedure that iterates three steps: the sampling
update, the weight update, and the truncation step. Although the structure
of PISAA is similar to that of SAA, the sapling update and weight
update are different and in fact significantly more efficient.

The sampling update, at iteration $t$, involves simulating a population
of $\kappa$ chains from a Markov transition probability $P_{\theta_{t-1},\tau_{t}}^{(\kappa)}(\cdot,\drv\cdot;\mathcal{E})$
that admits $f_{\theta_{t-1},\tau_{t}}^{(\kappa)}(\drv\cdot;\ \mathcal{E})$
as the invariant distribution. The Markov transition probabilities
$\{P_{\theta_{t-1},\tau_{t}}^{(\kappa)}(\cdot,\drv\cdot;\mathcal{E})\}$
can be designed as a mixture of different MCMC kernels. Because it
uses a population of chains, PISAA allows the use of advanced updates
for the design of these MCMC kernels which facilitate the exploration
of the sampling space and the search for the global minimum. Two types
of such operation updates are the mutation, and the crossover operations. 
\begin{itemize}
\item Mutation operations update the population individual-by-individual
through Metropolis-Hastings within Gibbs algorithm \citep{Muller1991,RobertCasella2004}
by viewing the population as a long vector. Because the population
individuals in (\ref{eq:target_pisaa_pop_working}) are independent
and identically distributed, in practice the whole population can
be updated simultaneously (in parallel) by using the same operation
with the same bias weights for each individual. This eliminates the
computational overhead due to the generation of multiple chains. Parallel
chains allow breaking the sampling into parallel simulations, possibly
initialised from different locations, which allows searching for global
minimum at different subregions of the sampling space simultaneously.
Moreover, it avoids the need to move a single chain across a potentially
large and high modal sampling space. Therefore, it facilitates the
search for the global minimum and the exploration of both the sample
space and partition space, while it discourages local trapping. They
include the random walk Metropolis \citep{MetropolisRosenbluthRosenbluthTellerTeller1953},
hit-and-run \citep{Smith1984,ChenSchmeiser1993}, $k$-point \citep{Liang2011,LiangWong2001,LiangWong2000},
Gibbs \citep{Muller1991,GemanGeman1984} updates etc.
\item Crossover operations, originated in genetic algorithms \citep{Holland1975},
update the population through a Metropolis-Hastings algorithm that
operates on the population space and constructs the proposals by using
information from different population chains. Essentially, the distributed
information across the population is used to guide further simulations.
This allows information among different chains of the population to
be exchanged in order to improve mixing. As a result, crossover operations
can facilitate the exploration of the sample space. Crossover operations
include the $k$-point \citep{Liang2011,LiangWong2001,LiangWong2000},
snooker \citep{Liang2011,LiangWong2001,LiangWong2000,GilksRobertsGeorge1994},
linear \citep{Liang2011,LiangWong2001,LiangWong2000,GilksRobertsGeorge1994}
crossover operations etc.
\end{itemize}

The weight update aims at estimating $\{\theta_{*}^{(j)}\}$ by using
a mean field approximation at each iteration with the step size controlled
by the gain factor. It is performed by using all the population of
chains: At iteration $t$, the update of $\{\theta^{(j)}\}$ is performed
as $\theta'=\theta_{t-1}+\gamma_{t}H_{\tau_{t}}^{(\kappa)}(\theta_{t-1},x_{t}^{(1:\kappa)})$,
where $H_{\tau_{t}}^{(\kappa)}(\theta_{t-1},x_{t}^{(1:\kappa)})=\frac{1}{\kappa}\sum_{i=1}^{\kappa}H_{\tau_{t}}(\theta_{t-1},x_{t}^{(i)})=[p_{t}^{(\kappa)}-\pi]$,
$p_{t}^{(\kappa)}:=(p_{t}^{(\kappa,j)},j=1:m)$, and $p_{t}^{(\kappa,j)}=\frac{1}{\kappa}\sum_{i=1}^{\kappa}\indfun(x_{t}^{(i)}\in E_{j})$,
for $j=1,...,m$. Intuitively, because all the population chains share
the same partition $\mathcal{E}$ and bias weights $\{\theta_{*}\}$,
and the population individuals are independent and identically distributed,
the indicator functions of $p_{t}$ (used in Algorithm \ref{Alg:SAA})
can be replaced here by the proportion $p_{t}^{(\kappa)}$ of the
population in the associated subregions at each iteration. Namely,
the indicator functions of $p_{t}$ (in Algorithm \ref{Alg:SAA})
is replaced by the law of the MCMC chain associated with the current
parameter. A theoretical analysis in Appendix \ref{sec:Theoretical-justification}
shows that the multiple-chain weight update (in Algorithm \ref{Alg:PISAA})
is asymptotically more efficient that the single-chain one (in Algorithm
\ref{Alg:SAA}).

The truncation step applies a truncation on $\theta_{t}$ to ensure
that $\theta_{t}$ lies in a compact set $\Theta$ as in SAA; hence
we consider quantities $\tilde{\theta}_{0}$, $\{M_{c}\}$, and $\{c_{t}\}$
as in Section \ref{sec:Stochastic-approximation-annealing}.

The proposed algorithm works as follows: At iteration $t$, we assume
that the Markov chain is at state $x_{t-1}^{(1:\kappa)}$ with a working
estimate $\theta_{t-1}$. Firstly, simulate a population sample $x_{t}^{(1:\kappa)}$
from the Markov transition probability $P_{\theta_{t-1},\tau_{t}}^{(\kappa)}(x_{t-1}^{(1:\kappa)},\drv\cdot;\mathcal{E})$
. Secondly, update the working estimate $\theta_{t}$ according to
$\theta'=\theta_{t-1}+\gamma_{t}H_{\tau_{t}}^{(\kappa)}(\theta_{t-1},x_{t}^{(1:\kappa)})$,
where $H_{\tau_{t}}^{(\kappa)}(\theta_{t-1},x_{t}^{(1:\kappa)})=[p_{t}^{(\kappa)}-\pi]$,
$p_{t}^{(\kappa)}:=(p_{t}^{(\kappa,j)},j=1:m)$, and $p_{t}^{(\kappa,j)}=\frac{1}{\kappa}\sum_{i=1}^{\kappa}\indfun(x_{t}^{(i)}\in E_{j})$,
for $j=1,...,m$, by using the whole population $\{x_{t}^{(1:\kappa)}\}$.
Thirdly, if $\|\theta'^{(j)}\|_{2}\le M_{c_{t}}$, truncate such that
$\theta_{t}=\tilde{\theta}_{0}$, and $c_{t}=c_{t-1}+1$. At the end
of the run, $t=n$, it is possible to apply a $\theta$-normalisation
step (see Section \ref{sec:Stochastic-approximation-annealing}) --an
alternative $\theta$-normalisation step can be $\tilde{\theta}_{n}^{(j)}\leftarrow\theta_{n}^{(j)}+z$,
where $z=-\log(\sum_{j=1}^{m}\pi_{j}\exp(\theta_{n}^{(j)}))$. PISAA
is summarised as a pseudo-code in Algorithm \ref{Alg:PISAA}. A more
rigorous analysis about the convergence and the stability of PISAA
is given in Appendix \ref{sec:Theoretical-justification} and summarised
in Section \ref{sub:Justification}.

\begin{algorithm}
\begin{description}
\item [{Requires :}] Insert $\{\tau_{t}\}$, $\{\gamma_{t}\}$, $\{E_{j}\}$,
$\{\pi_{j}\}$, $\{M_{c}\}$, $\kappa$, $\tilde{\theta}_{0}$
\item [{Initialise :}] At $t=0$, set $x_{0}^{(1:\kappa)}\in\mathcal{X}^{\kappa}$,
$\tilde{\theta}_{0}\in\Theta$, such that $\|\tilde{\theta}_{0}\|_{2}<M_{0}$,
and $c_{0}=0$.
\item [{Iterate :}] For $t=1,...,n$,

\begin{enumerate}
\item Sampling update:

Simulate $x_{t}^{(1:\kappa)}$ from the Metropolis-Hastings transition
probability $P_{\theta_{t-1},\tau_{t}}^{(\kappa)}(x_{t-1},\drv\cdot;\mathcal{E})$
that targets $f_{\theta_{t-1},\tau_{t}}^{(\kappa)}(\drv\cdot;\mathcal{E})$ 

\item Weight update:

Compute $\theta'=\theta_{t-1}+\gamma_{t}H_{\tau_{t}}^{(\kappa)}(\theta_{t-1},x_{t}^{(1:\kappa)})$,
where $H_{\tau_{t}}^{(\kappa)}(\theta_{t-1},x_{t}^{(1:\kappa)})=[p_{t}^{(\kappa)}-\pi]$,
$p_{t}^{(\kappa)}:=(p_{t}^{(\kappa,j)},j=1:m)$, and $p_{t}^{(\kappa,j)}=\frac{1}{\kappa}\sum_{i=1}^{\kappa}\indfun(x_{t}^{(i)}\in E_{j})$,
for $j=1,...,m$.

\item Truncation step:

Set $\theta_{t}=\theta'$, and $c_{t}=c_{t-1}$ if $\|\theta'^{(j)}\|_{2}\le M_{c_{t}}$,
or set $\theta_{t}=\tilde{\theta}_{0}$, and $c_{t}=c_{t-1}+1$ if
otherwise.

\end{enumerate}
\end{description}
\protect\caption{Parallel and  interacting stochastic approximation annealing algorithm
\label{Alg:PISAA}}
\end{algorithm}

\subsection{Theoretical analysis: a synopsis\label{sub:Justification}}

Regarding the convergence of the proposed algorithm, PISAA inherits
a number of desirable theoretical results from SAA \citep{LiangChengLin2013}
and pop-SAMC \citep{SongWuLiang2014}. A brief theoretical analysis
related to the convergence of PISAA is included in Appendix \ref{sec:Theoretical-justification},
where we show that theoretical results of \citealt{SongWuLiang2014}
for pop-SAMC hold in the PISAA framework as well, and we present theoretical
results in \citet{LiangChengLin2013} for SAA that hold for PISAA
as well. The Theorems \ref{mainth1}, \ref{SAAth2}, \ref{normalitytheorem},
and \ref{efftheorem}, as well as related conditions on PISAA, are
included in the Appendix \ref{sec:Theoretical-justification}. We
recall, the temperature ladder: $\tau_{t}=t_{1}/\sqrt{t}+\tau_{*}$,
$t_{1}>0$, the gain function: $\gamma_{t}=t_{0}/t^{\beta}$, $t_{0}>0$,
$\beta\in(0.5,1)$, and consider that $X_{t}^{(1:\kappa)}:=(X_{t}^{(i)};i=1,...,\kappa)$
denotes a draw from PISAA at the $t$-th iteration.

PISAA can achieve for any individual the following convergence result:
For any $\epsilon>0$, as $t\rightarrow\infty$, and $\tau_{*}\rightarrow0$
\[
\text{P}(U(X_{t}^{(i)})\le u_{j}^{*}+\epsilon|J(X_{t}^{(i)})=j)\rightarrow1,\quad a.s.,
\]
 where $J(x)=j$ if $x\in E_{j}$, and $u_{j}^{*}=\min_{x\in E_{j}}U(x)$,
for $j=1,...,m$. Namely, as the number of iterations $t$ becomes
large, PISAA is able to locate the minima of each subregion in a single
run if $\tau_{*}$ is small. This comes as a consequence of \citet[ Corollary 3.1]{LiangChengLin2013}
and the Theorems \ref{mainth1}, and \ref{SAAth2} in Appendix \ref{sec:Theoretical-justification}.
Theorem \ref{mainth1} in Appendix \ref{sec:Theoretical-justification}
(a restatement of Theorems 3.1 and 3.2 of \citet{LiangChengLin2013})
indicates that the weights $\{\theta_{t}\}$ remain in a compact subset
of $\Theta$ and hence $\theta_{*}=(\theta_{*}^{(j)};j=1,...,m)$
can be expressed in the form $\theta_{*}^{(j)}=c+\log(\int_{E_{j}}\exp(-U(x^{(i)})/\tau_{*})\drv x^{(i)})-\log(\pi_{j})$,
for $j=1,...m$, and any $i=1,...,\kappa$, where $c\in\mathbb{R}$
is an arbitrary constant. Namely, as $t\rightarrow\infty$, $f_{\theta_{t},\tau_{t+1}}^{(\kappa)}(x^{(1:\kappa)}|\mathcal{E})\rightarrow f_{\theta_{*},\tau_{*}}^{(\kappa)}(x^{(1:\kappa)}|\mathcal{E})$,
$a.s.$; since $f_{\theta,\tau}^{(\kappa)}(x^{(1:\kappa)}|\mathcal{E})$
is invariant to transformations $\theta\leftarrow\theta+c$. Furthermore,
Theorem \ref{SAAth2} in Appendix \ref{sec:Theoretical-justification}
(a restatement of Theorem 3.3 of \citet{LiangChengLin2013}) implies
that $X_{t+1}^{(1:\kappa)}\sim f_{\theta_{t},\tau_{t+1}}^{(\kappa)}(x^{(1:\kappa)}|\mathcal{E})$,
in a SLLN fashion; where $X_{t+1}^{(1:\kappa)}$ a draw from PISAA
at the $(t+1)$-th iteration. 

It is not trivial to show that the results of \citep{SongWuLiang2014}
for pop-SAMC hold in the PISAA framework as well. The reason is that,
unlike in pop-SAMC, in the PISAA framework the target distribution
is parametrised by an additional control parameter the temperature
ladder $\{\tau_{t}\}$, and hence the density of the target distribution
changes at each iteration. I.e.  $f_{\theta_{t},\tau_{t}}(\cdot|\mathcal{E})\ne f_{\theta_{t'},\tau_{t'}}(\cdot|\mathcal{E})$
if $t\ne t'$ in the PISAA framework. In Appendix \ref{sec:Theoretical-justification},
Lemma \ref{lem51} considers the decomposition of the noise in the
PISAA framework, and allows us to be able to extent the main theoretical
results of \citep{SongWuLiang2014} to the PISAA framework as stated
in Theorems \ref{normalitytheorem} and \ref{efftheorem} in the Appendix
\ref{sec:Theoretical-justification}. Theorem \ref{normalitytheorem}
implies that the weights $\{\theta_{t}\}$ generated by PISAA are
asymptotically distributed according to the Gaussian distribution,
and constitutes an extension of \citep[Theorem 2,][]{SongWuLiang2014}
in the PISAA framework. Theorem \ref{efftheorem} considers the relative
efficiency of the bias weight estimate $\{\theta_{t}^{p}\}$ generated
by the self-adjusting mechanism of the multiple-chain PISAA (with
population size $\kappa$) at iteration $t$, against estimate $\{\theta_{\kappa t}^{s}\}$
generated by the self-adjusting mechanism of the single-chain SAA
at iteration $\kappa\cdot t$. Theorem \ref{efftheorem} implies that
$(\theta_{t}^{p}-\theta_{*})/\sqrt{\gamma_{t}}$ and $(\theta_{\kappa t}^{s}-\theta_{*})/\sqrt{\kappa\gamma_{t}}$
follow the same distribution asymptotically with convergence rate
ratio $\kappa^{\beta-1}$, where $\beta\in(0.5,1]$, and hence is
the extension of \citep[Theorem 4,][]{SongWuLiang2014} in the PISAA
framework.

In other words, when $\beta<1$, the multiple-chain PISAA estimator
of the bias weights is asymptotically more efficient than that of
the single-chain SAA; while when $\beta=1$, the two estimators present
similar efficiency. In practice, PISAA estimator is expected to outperform
the single-chain SAA estimator even when $\beta=1$ because of the
so called population effect; the use of multiple-chains to explore
the sampling space and approximate the unknown $\{\theta_{t}\}$.
Theorem \ref{efftheorem} implies rigorously that the adjustment process
in PISAA is more stable than that in SAA.

\subsection{Practical implementation and remarks}

\citet{LiangChengLin2013} discussed several practical issues on the
implementation of SAA (including the algorithmic settings $\{\pi_{j}\}$,
$\{\gamma_{t}\}$, $\{\tau_{t}\}$ $\mathcal{E}$, $\{M_{c}\}$, $n$
) that are still applicable to PISAA. Here, we adopt these algorithmic
settings, i.e.: $\pi_{j}\propto\exp(-\lambda(j\text{\textminus}1))$
with $\zeta\ge0$; $\gamma_{t}=(\frac{n^{(\gamma)}}{\max(t,n^{(\gamma)})})^{\beta}$
with $\beta\in(0.5,1]$; $\tau_{t}=\tau_{h}\sqrt{\frac{n{}^{(\tau)}}{\max(t,n{}^{(\tau)})}}+\tau_{*}$
where $\tau_{*}>0$, $\tau_{h}>0$, and $n{}^{(\tau)}>0$; and $M_{c}=10^{10}M_{c-1}$
with $M_{0}=10^{100}$. We briefly discuss additional practical details
of PISAA:
\begin{itemize}
\item The population seed $x_{0}^{(1:\kappa)}$ controls the initialisation
of the population. It is preferable, but not necessary, for the population
of chains to initiate from various locations, possibly around different
local minima. This could benefit the exploration of the space and
the search for the global minimum. This can be achieved, for example,
by sampling from a flat distribution e.g. $f_{\tau_{0}}(x)\propto\exp(-U(x)/\tau_{0})$,
with $\tau_{0}>0$ large enough, via a random walk Metropolis algorithm.
\item The MCMC operations must result in reasonable expected acceptance
probabilities because they can affect the sampling update. It is possible
to calibrate the scale parameter of the proposals adaptively (on-the-fly)
by using an adaptation scheme \citep{AndrieuThoms2008}, during the
first few iterations. 
\item The rates of the operations in the MCMC sweep at each iteration are
problem dependent. One may favour specific operations by increasing
the corresponding rates if it is believed that they are more effective
or cheaper to run for the particular application. 
\end{itemize}
PISAA can be modified to deal with empty subregions similar to SAA.
Let $S_{t}$ denote the set of non-empty subregions until iteration
$t$, $\theta_{t}^{S_{t}}$ denote the sub-vector of $\theta_{t}$
corresponding to elements of $S_{t}$, and $\Theta^{S_{t}}$ denote
the sub-space of $\Theta$ corresponding to elements of $S_{t}$.
Yet, let $y^{(1:n)}$ denote the proposed population value generated
during the sampling update, and $J(x)=j$ if $x\in E_{j}$. Then Algorithm
\ref{Alg:PISAA} can be modified as follows: 
\begin{itemize}
\item (Sampling~update): Simulate $x_{t}^{(1:\kappa)}\sim P_{\theta_{t-1},\tau_{t}}^{(\kappa)}(x_{t-1}^{(1:\kappa)},\drv\cdot;\mathcal{E})$
(as in Algorithm \ref{Alg:PISAA}), and set $S_{t}\leftarrow S_{t-1}\cup\{J(y^{(i)});\, i=1,...,\kappa\}$.
\item (Weight~update): Compute $\theta'^{(j)}=\theta_{t-1}^{(j)}+\gamma_{t}H_{\tau_{t}}^{(\kappa)}(\theta_{t-1}^{(j)},x_{t}^{(1:\kappa)})$,
for $j\in S^{(t)}$.
\item (Truncation~step): Set $\theta_{t}=\theta'$, and $c_{t}=c_{t-1}$
if $\|\theta'^{S_{t}}\|_{2}\le M_{c_{t}}$, or set $\theta_{t}=\tilde{\theta}_{0}$,
and $c_{t}=c_{t-1}+1$ if otherwise.
\end{itemize}
This modification ensures $\{\theta_{t}\}$ to remain in a compact
set. Note that the desired sampling distribution becomes actually
$\{\pi_{j}+\pi_{\text{e}};\text{for }j=1:m\}$, and 
\[
\theta_{*}^{(i)}=\begin{cases}
C+\log(\int_{E_{i}}\exp(-U(x)/\tau_{*})\drv x)-\log(\pi_{i}+\pi_{\text{e}}) & \text{, if }E_{i}\ne\emptyset\\
\tilde{\theta}_{0}^{(i)} & \text{, if }E_{i}=\emptyset
\end{cases},
\]
 where $\pi_{\text{e}}=\sum_{j\notin S_{\infty}}\pi_{j}/\left\Vert S_{\infty}\right\Vert $,
and $S_{\infty}$ is the limiting set of $S_{t}$.

For population size $\kappa=1$, PISAA is identical to the single-chain
SAA.

PISAA can be used, in the same spirit as the tempered transitions
\citep{Neal1996}, for sampling from a multi-modal distribution $f(\drv\cdot)$.
One can run PISAA with $U(x):=-\log(f(x))$, $\tau_{t}=\tau_{h}\sqrt{\frac{n{}^{(\tau)}}{\max(t,n{}^{(\tau)})}}+\tau_{*}$,
$\tau_{h}>1$, $\tau_{*}=1$, and collect the sample $x_{n}^{(1:\kappa)}$.
Then, inference can be performed by importance sampling methods due
to Theorems \ref{mainth1} and \ref{SAAth2} in Appendix \ref{sec:Theoretical-justification}.

\section{Applications \label{sec:Applications}}

We compare the performance of PISAA with those of other stochastic
optimisation procedures such as the simulated annealing (SA) \citep{KirkpatrickVecchi1983},
very fast simulated re-annealing (VFSA) \citep{Ingber1989,SenStoffa1996,JacksonSenStoffa2004},
annealing stochastic approximation Monte Carlo (ASAMC) \citep{Liang2007},
annealing evolutionary stochastic approximation Monte Carlo (AESAMC)
\citep{Liang2011}, and stochastic approximation annealing (SAA) \citep{LiangChengLin2013}.

As a performance measure, we consider the average best function value
discovered by the algorithm. We perform $48$ independent realisations
for each simulation, and average out the values of the performance
measures, in order to eliminate nuisance variation in the output of
the algorithms (caused by their stochastic nature or random seeds).
To monitor the convergence of PISAA and the stability of its self-adjusting
mechanism, we consider the MSE of the bias weights as in \citep{SongWuLiang2014}
$\text{MSE}:=\left\Vert \theta_{t}^{(\kappa)}-w_{t}\right\Vert $,
where $w_{t}:=(w_{t}^{(j)};j=1:m)$, $w_{t}^{(j)}:=\int_{E_{j}}\frac{1}{\tau_{t}}U(x)\drv x$
are the real values of the bias weights, and $\theta_{t}^{(\kappa)}$
are the estimates of $w_{t}$ approximated by the self-adjusting mechanism
of PISAA with population size $\kappa$. 

The mutation operations and crossover operations, used in the examples,
are presented in Appendix \ref{sec:Appendix_MCMC_operations} as pseudo-codes.

\subsection{Gaussian mixture model \label{sub:Gaussian-mixture-model}}

We consider the Gaussian mixture with density 
\begin{equation}
f_{1}(x)=\sum_{i=1}^{20}\varpi_{i}\text{N}_{2}(x|\mu_{i},\sigma^{2})\indfun(x\in\mathcal{X}),\label{eq:Nmix_pdf}
\end{equation}
 where $x\in\mathbb{R}^{2}$, $\mathcal{X}=[-10^{10},10^{10}]^{2}$,
$\sigma^{2}=0.001$, $\{\varpi_{i}=1/20;\ i=1:20\}$, and $\{\mu_{i}\}$
are given in \citep[Table 1 in ][]{LiangWong2001}. Sampling from
(\ref{eq:Nmix_pdf}) is challenging because this distribution is multi-modal
and has several isolated modes. Here, our purpose is to check the
validity of PISAA instead of optimisation.

We consider default algorithmic settings for PISAA: (i) energy function
$U_{1}(x)=-\log(f_{1}(x))$, (ii) uniformly spaced grid $\{u_{j}\}$
with $m=19$, $u_{1}=0$, and $u_{19}=9.0$, (iii) gain factor $\{\gamma_{t}\}$
with $n^{(\gamma)}=100,$ $\beta=0.55$, (iv) temperature sequence
$\{\tau_{t}\}$ with $n^{(\tau)}=1$, $\tau_{h}=5$, and $\tau_{*}=1-\tau_{h}\sqrt{1/n}$,
and (v) MCMC transition probability that uses mutation operations
(Metropolis, hit-and-run, $k$-point) and crossover operations ($k$-point,
snooker, linear), with equal operation rates, and proposal scales
calibrated so that the expected acceptance probabilities to be around
$0.234$. At the end of the simulation, at iteration $n=10^{6}$,
the temperature will be $\tau_{n}=1$; and hence one may see this
example as tempered transition sampling from multi-modal distribution
$f_{1}(\drv\cdot)$ via PISAA.

We run PISAA with different combinations of population size $\kappa\in\{1,...,30\}$
and gain factor power $\beta\in\{0.55,0.65,0.75,0.85,0.95,1.0\}$.
Each of these runs was repeated for $100$ realisations in order to
compute the estimates, error bars, mean square error $\text{MSE}:=\left\Vert \theta_{t}^{(\kappa)}-w_{t}\right\Vert $,
and relative efficiency $\text{RE}(\kappa;\beta):=\left\Vert \theta_{\left\lfloor n/\kappa\right\rfloor }^{(\kappa)}-w_{*}\right\Vert /\left\Vert \theta_{n}^{(1)}-w_{*}\right\Vert $
of the bias weights. Note, that the bias weights are estimated by
the self-adjusting mechanism of PISAA using the $\theta$-normalisation
step $\sum_{j=1}^{m}\exp(\theta_{n}^{(\kappa,j)}+z)=1$.

\begin{figure}
\center

\subfloat[Estimate and $95\%$ error bars of $\{w_{n}^{(\kappa,j)};j=1:6\}$
(at $n=10^{6}$). \label{fig:West_pisaa_Nmix}]{\includegraphics[scale=0.2]{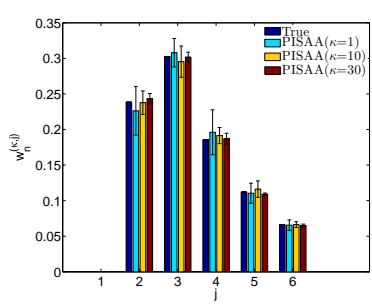}

}\subfloat[Progression curves of the MSE estimate of $w^{(\kappa,j)}$ for different
population sizes. ($\beta=0.55$). \label{fig:MSEs_t_n_pisaa_Nmix}]{\includegraphics[scale=0.2]{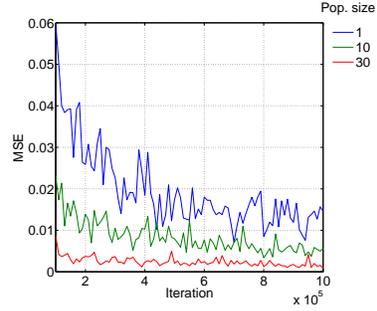}

}

\subfloat[MSE against the population size, for different power $\beta$ of gain
factor (at $n=10^{6}$). \label{fig:MSEs_n_s_pisaa_Nmix}]{\includegraphics[scale=0.2]{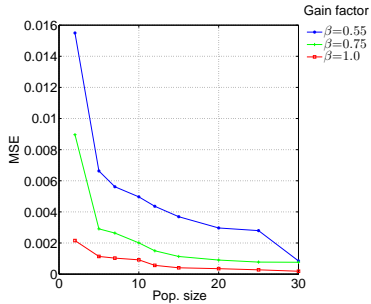}

}\subfloat[Relative efficiency of the bias weight estimate as a function of the
population size, for different power of gain factor. \label{fig:RelativeEfficiency_Mixt2D20C}]{\includegraphics[scale=0.2]{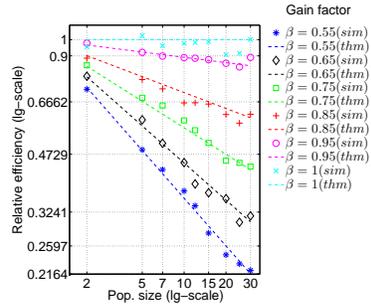}

}

\protect\caption{(Section \ref{sub:Gaussian-mixture-model}) Estimates, MSEs, and relative
efficiency of the bias weights $\{w_{n}^{(j)}\}$ produced by PISAA
at different iteration, population sizes, and power of gain factor
$\beta$.}

\end{figure}

Figure \ref{fig:West_pisaa_Nmix} presents the estimates of the bias
weights $\theta_{\left\lfloor n/\kappa\right\rfloor }^{(\kappa)}$,
for $j=1,...,6$, and $n=10^{6}$, as produced by the self-adjusting
mechanism of PISAA with different population sizes $n\in\{1,10,30\}$.
We observe that the $\{\theta_{n}^{(\kappa,j)}\}$ of PISAA have converged
to the true values at any of the population sizes considered, and
that the associated error bars are narrower for larger population
sizes. Figure \ref{fig:MSEs_t_n_pisaa_Nmix} presents the MSEs produced
by PISAA at different iteration steps, and for different population
sizes. We observe that PISAA with larger population sizes has produced
smaller MSEs throughout the whole simulation time. Yet, MSE decays
as the iterations evolve; this behaviour, although not surprising,
may be non-trivial due to the heterogeneous nature of the sequence
$\{w_{t}\}$ (that is $w_{t}\ne w_{t'}$, for $t\ne t'$). Figure
\ref{fig:MSEs_n_s_pisaa_Nmix} presents the progression of the MSEs
produced by PISAA for different gain factor powers. We observe that
MSE decreases when the population size increases. Furthermore, we
observe that this behaviour is more significant for slower decaying
gain factors --namely when the power of the gain factor is smaller
and close to $0.5$. 

Figure \ref{fig:RelativeEfficiency_Mixt2D20C} presents the relative
efficiency $\text{RE}(\kappa;\beta)$ of the self-adjusting process
estimator for the biased weights $w_{*}$ as a function of the population
size $\kappa\in\{2,...,30\}$, and for different powers of gain factors
$\beta\in\{0.55,0.65,0.75,0.85,0.95,1\}$. In serial computing environments,
the computational cost can be defined as the iterations times the
population size. For the computation of relative efficiency $\text{RE}(\kappa;\beta)$,
we considered constant computational cost, and hence PISAA with population
size $\kappa$ ran for $\left\lfloor n/\kappa\right\rfloor $ iterations,
where $n=10^{6}$. In Figure \ref{fig:RelativeEfficiency_Mixt2D20C},
the marks refer to the estimated relative efficiency, the dashed lines
are lines with slop $\beta-1$ and refer to the theoretical behaviour
of the relative efficiency (i.e. $\lg(\text{RE}(\kappa;\beta))\approx(\beta-1)\lg(\kappa)$)
from Theorem \ref{efftheorem}, while the different colours correspond
to different values of $\beta$. We observe that the empirical results
are consistent with Theorem \ref{efftheorem} since the marks lie
close to their corresponding lines. The efficiency of the estimates
of the bias weights produced by the self-adjusting mechanism of PISAA
improves as $\kappa$ increases. Thus, increasing the population size
improves the stability of PISAA even in the case that a serial computing
environment is used and a fixed computational budget is given. We
observe that this behaviour is even more significant for slower decaying
gain factors.

The results support that, PISAA produces the `real' estimates for
$w_{*}$ as $\tau_{t}\rightarrow\tau_{*}$, the MSE of those estimates
reduces as $\kappa$ increases, and the efficiency of the self-adjusting
mechanism improves as $\kappa$ increases. 

\subsection{Rastrigin's function\label{sub:Rastrigin's-function}}

We test the proposed algorithm on a benchmark optimisation problem
where the goal is to minimise the rotated Rastrigin's function $U_{2}(\cdot)$
\begin{align}
U_{2}(x) & =\text{Ra}(y(x));\label{eq:Rotated Rastrigin's function}\\
\text{Ra}(y) & =10d+\sum_{k=1}^{d}(y_{k}^{2}-10\cos(2\pi y_{k});\label{eq:Rastrigin's function}\\
y(x) & =Rx,\label{eq:Rastrigin Salomom rotation}
\end{align}
$x\in\mathcal{X}$, on space $\mathcal{X}=[-5.12,5.12]^{d}$, $d\in\mathbb{N}-\{0\}$,
where $\text{Ra}:\mathcal{X}\rightarrow\mathbb{R}$ is the Rastrigin's
function \citep{TornZilinskas1989,MuhlenbeinSchomischBorn1991,Liang2011},
and $R$ is a rotation matrix generated according to the Salomon's
method, see details in \citep[Appendix B in ][]{Salomon1996}. The
global minimum of (\ref{eq:Rotated Rastrigin's function}) is $\text{Ra}(x_{*})=0$
at $x_{*}=(0,...,0)$, for $d\in\mathbb{N}-\{0\}$ \citep{MuhlenbeinSchomischBorn1991}. 

Rastrigin's function has been used by several researchers as a hard
benchmark function to test experimental optimisation algorithms \citep{DieterichHartke2012,TornZilinskas1989,MuhlenbeinSchomischBorn1991,Liang2011,LiangQinSuganthanBaskar2006,AliKhompatrapornZabinsky2005}.
It presents features that can complicate the search for the global
minimum: it is non-convex, non-linear, relatively flat and presents
several local minima that increase with dimension; e.g. about $50$
local minima for $d=2$ \citep{AliKhompatrapornZabinsky2005}. The
rotation transformation (\ref{eq:Rastrigin Salomom rotation}) is
a well established technique that transforms originally separable
test functions, such as the Rastrigin's one, into non-separable. Non-separability
makes the optimisation task even harder by preventing the optimisation
of a multidimensional function to be reduced into many separate lower-dimensional
optimisation tastks. For instance, in (\ref{eq:Rastrigin Salomom rotation}),
all the dimensions in vector $y$ are affected when one dimension
in vector $x$ changes in value.

Here, if not stated otherwise, we consider default settings for PISAA:
(i) $n=10^{6}$ iterations, (ii) uniformly spaced grid $\{u_{j}\}$
with $m=400$, $u_{1}=-0.01$, $u_{400}=40$, (iii) desirable probability
with parameter $\lambda=0.1$, (iv) temperature ladder $\{\tau_{t}\}$
with $\tau_{h}=1$, $n^{(\tau)}=1,$ $\tau_{*}=10^{-2}$, (iv) gain
factor $\{\gamma_{t}\}$ with $n^{(\gamma)}=10^{5},$ $\beta=0.55$.
One MCMC sweep is considered to be a random scan of mutation operations
(Metropolis, hit-and-run, $k$-point) and crossover operations ($k$-point,
snooker, linear), with equal operation rates, and scale parameters
calibrated so that the expected acceptance ratio to be around $0.234$.

\begin{figure}
\subfloat[Average progression curves of the best function values, for different
population sizes. \label{fig:esaa_FvsTcondP}]{\includegraphics[scale=0.19]{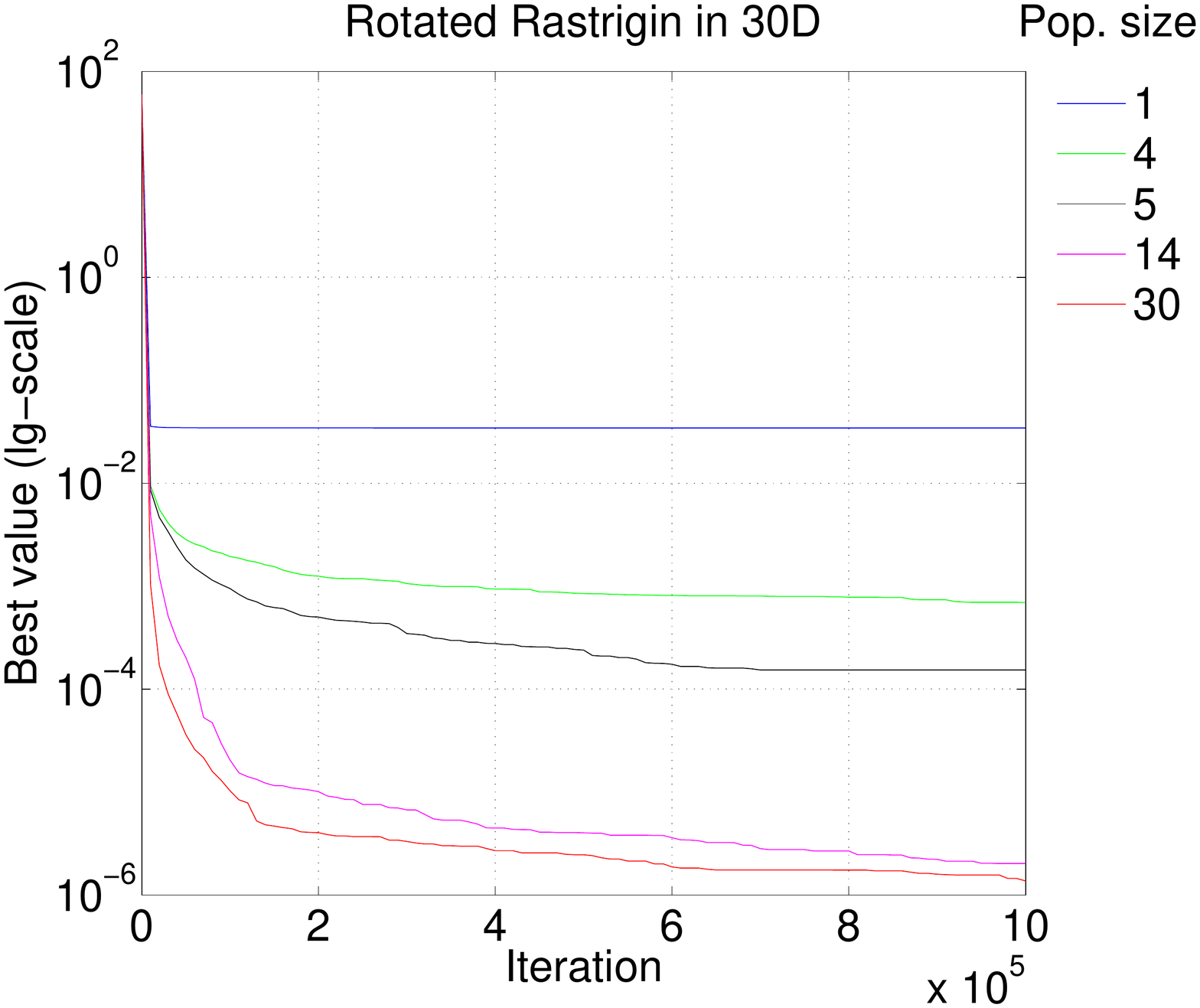}

}\subfloat[Average best function values against the population size, for different
dimensionality. \label{fig:esaa_FvsDcondP}]{\includegraphics[scale=0.19]{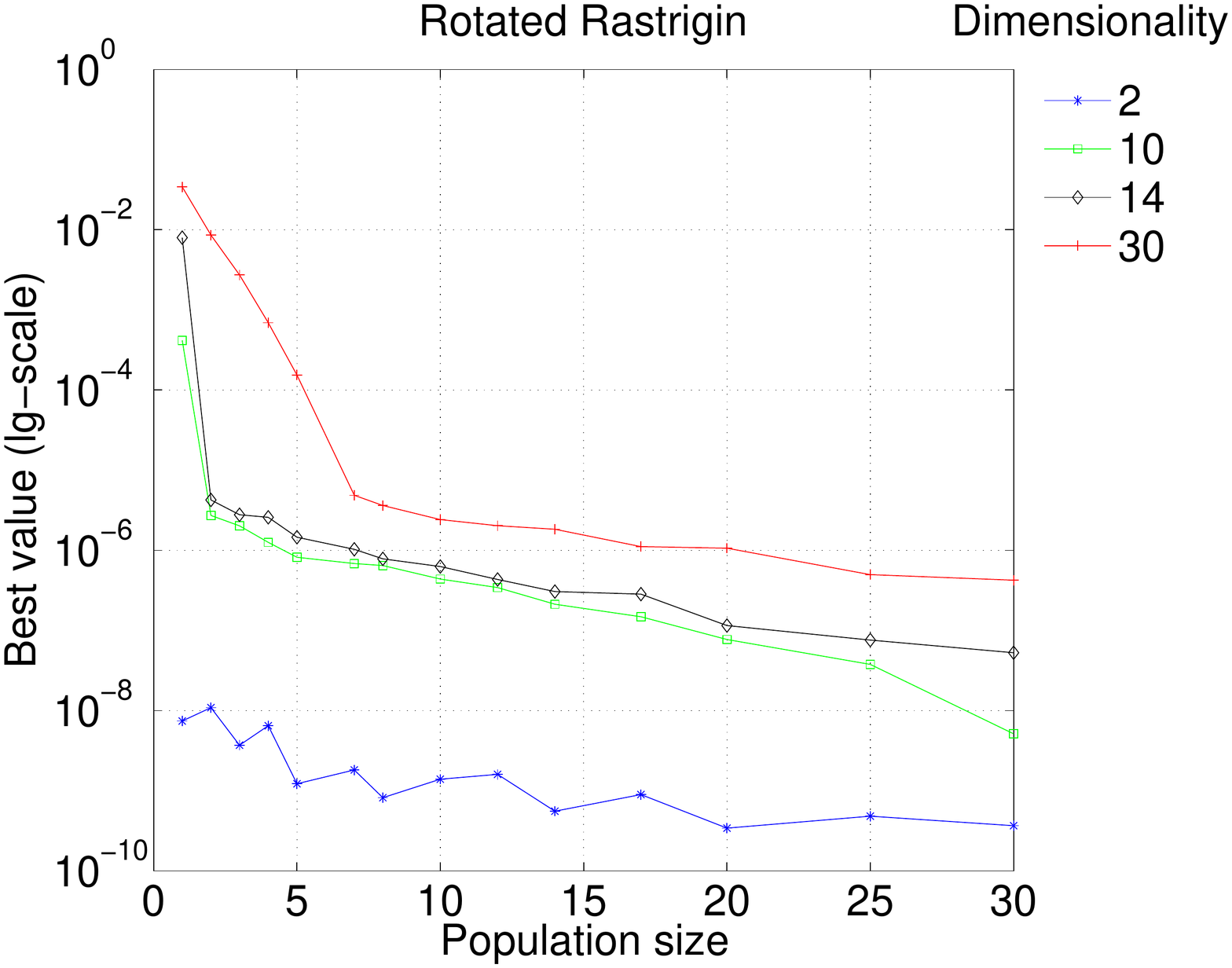}

}\subfloat[Progression curves of the MSE estimate of $w_{t}$, in $\lg$-scale,
for different population sizes. ($\beta=0.55$). \label{fig:Progression-MSE-theta-ras}]{\includegraphics[scale=0.19]{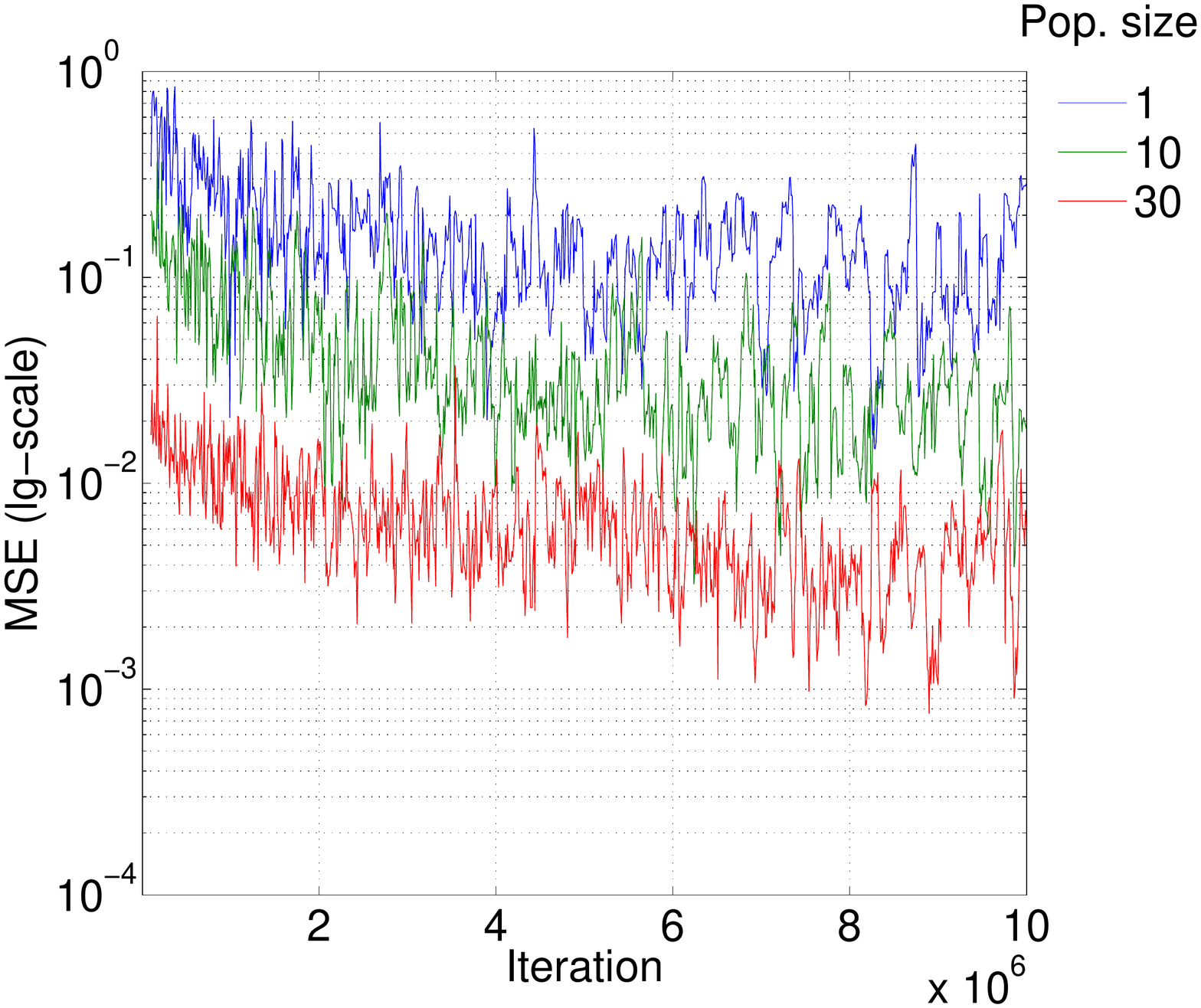}

}

\subfloat[Average best function values as functions of the grid size, for different
values $\lambda$ of the desired distribution.\label{fig:FvsMcondZ_Ra}]{\includegraphics[scale=0.19]{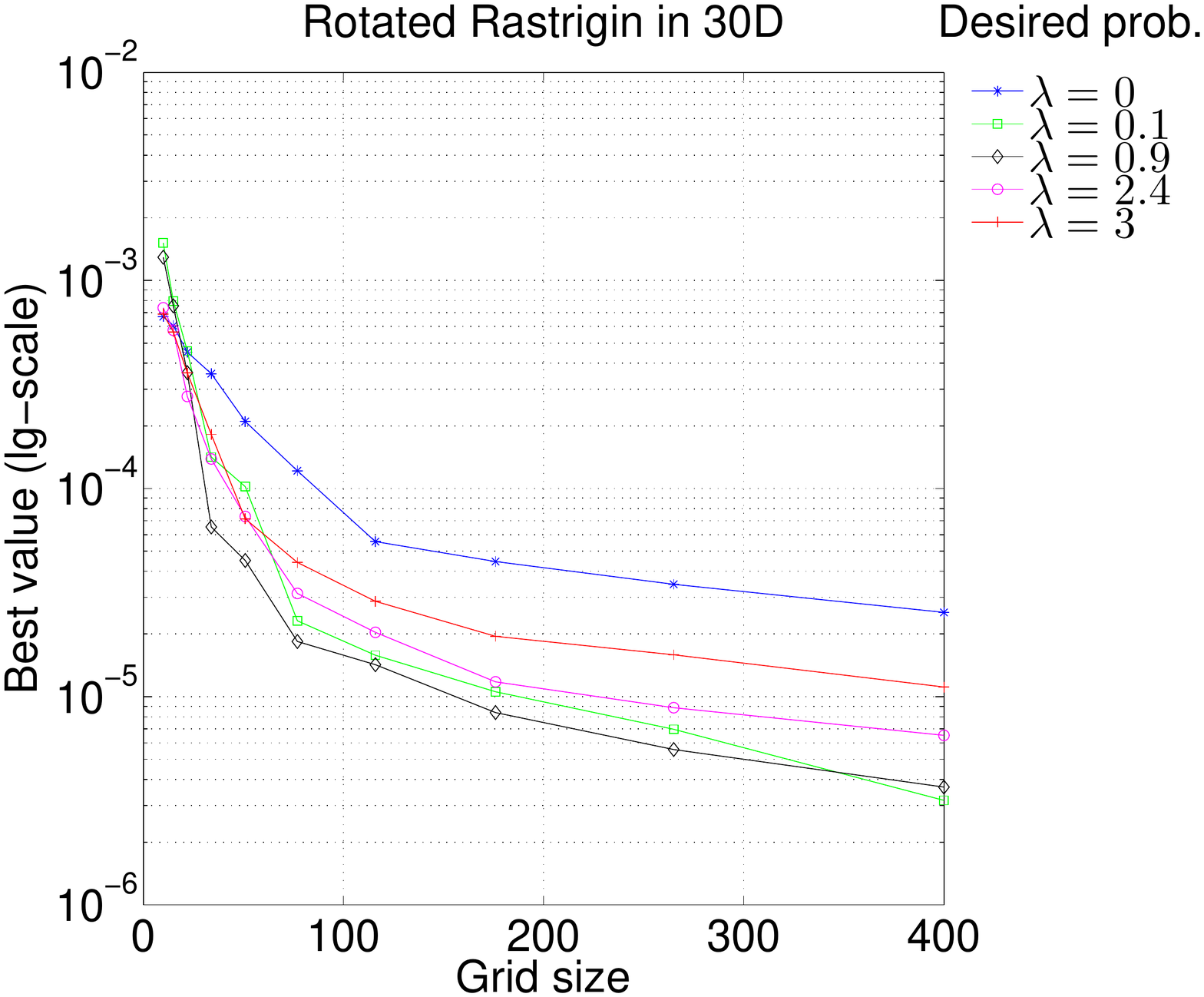}

}\subfloat[Average best function values as functions of the population size,
for different values $\lambda$ of the desired distribution.\label{fig:FvsNcondZ_Ra}]{\includegraphics[scale=0.19]{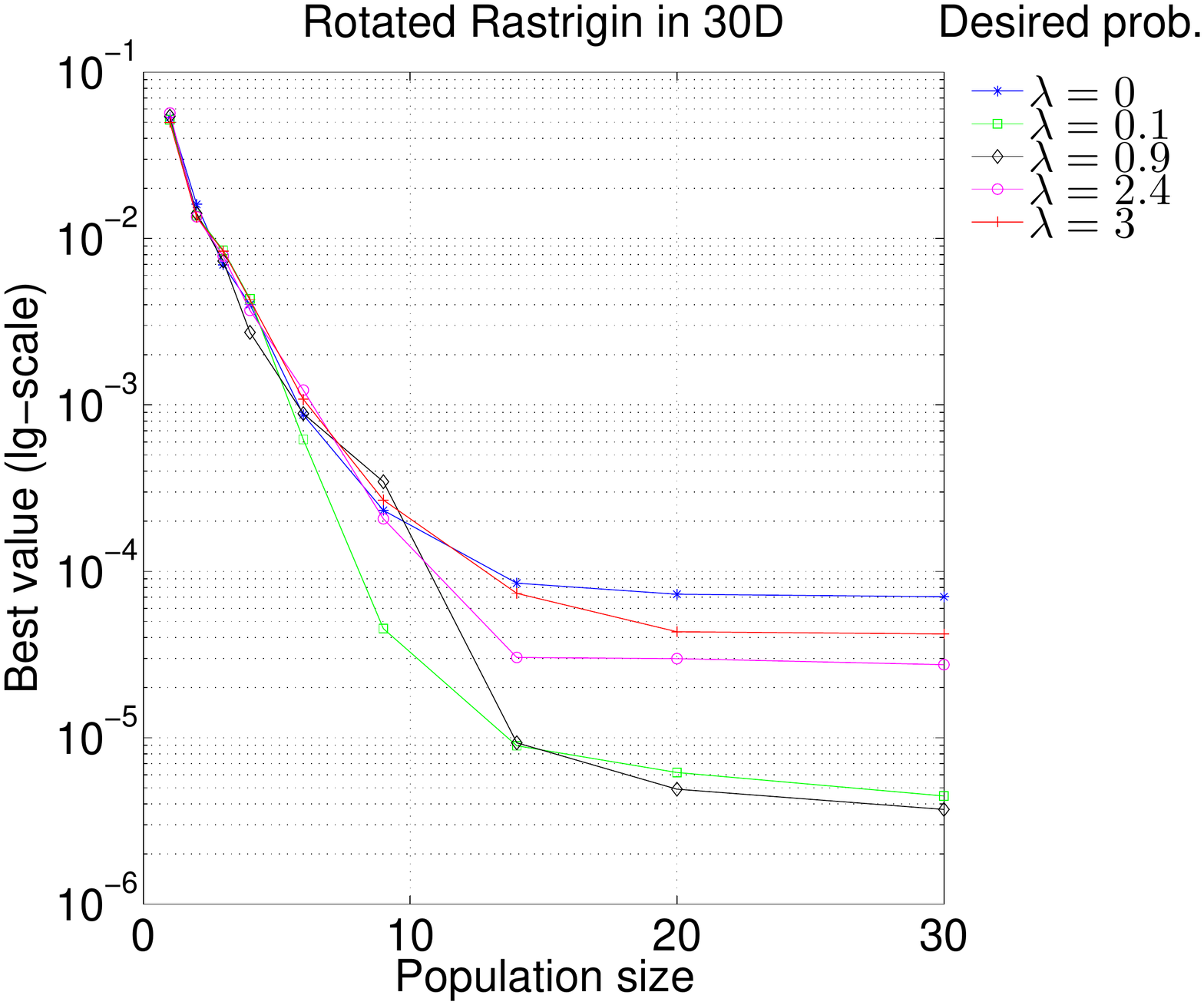}

}\subfloat[Average best function values as functions of the population size,
for different grid sizes. \label{fig:FvsPvsM_Ra}]{\includegraphics[scale=0.19]{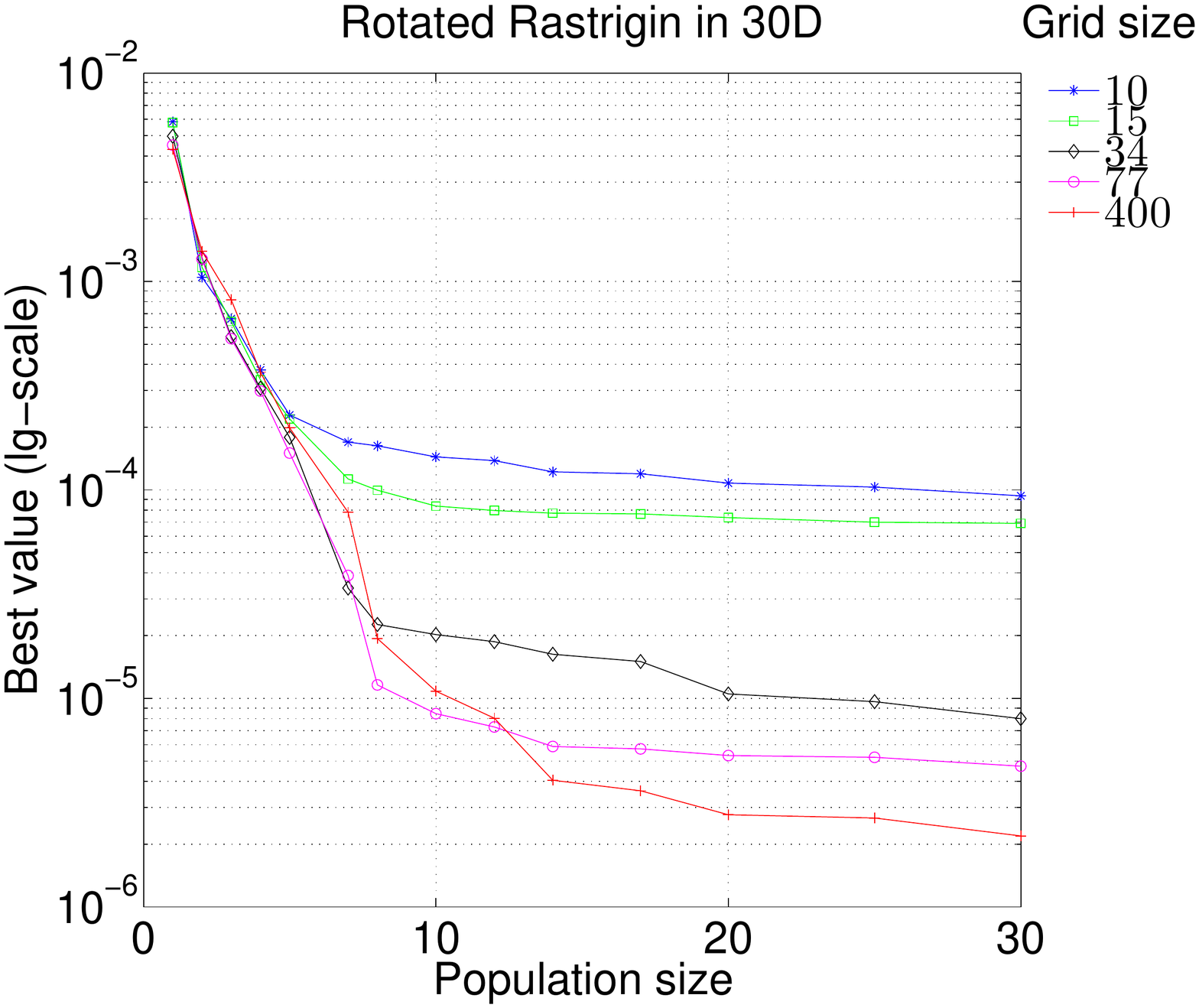}

}

\protect\caption{(Section \ref{sub:Rastrigin's-function}) Performance plots of PISAA.
The results reported consider averaged values over $48$ independent
runs. \label{fig:PISAA_Rastrigin}}
\end{figure}

In Figure \ref{fig:esaa_FvsTcondP}, we present the average progression
curves of the best function value (best value), discovered by PISAA
for different population sizes $\kappa\in\{1,4,5,14,30\}$. We observe
that by using larger population sizes, the algorithm quicker discovers
smaller best values, and quicker converges towards the global minimum.
The difference in performance between SAA (aka PISAA with $\kappa=1$)
and PISAA using a moderate population size, such as $\kappa=5$, is
significant. In Figure \ref{fig:esaa_FvsDcondP}, we plot the best
value against the population size for different dimensionality of
the Rastrigin's function. We observe that PISAA discovers smaller
best values as the population size increases for the same number of
iterations. Increasing the population size improves the performance
of the algorithm significantly at any dimensionality considered, while
it is particularly effective in large or moderate dimensionalities.
We highlight that the most striking performance improvement is observed
in the range of small population sizes. In Figure \ref{fig:Progression-MSE-theta-ras},
we observe that the MSE of the bias weights approximated by the self-adjusting
mechanism of PISAA becomes smaller when larger population sizes are
used. This indicates that increasing the population size makes the
self-adjusting mechanism of PISAA more stable.

The performance of PISAA with respect to the grid size, for different
desired probabilities, is presented in Figure \ref{fig:FvsMcondZ_Ra}.
In particular, we ran PISAA with a large enough population size ($\kappa=30$)
to ensure that all the subregions are visited. We observe that larger
grid sizes lead to a better performance for PISAA, given that the
population size is large enough. We observe that the choice of the
desired probability has bigger impact for large grid sizes ($m>50$)
than for smaller grid sizes ($m<50$). However, for any grid size,
we observe that a moderately biased desired distribution ($\lambda\approx0.1,...,0.9$)
is preferable. The performance of PISAA against the population size
for different desired probabilities is presented in Figure \ref{fig:FvsNcondZ_Ra}.
We observe that increasing the population size is more effective for
desired probabilities with ($\lambda\approx0.1,...,0.9$). Hence,
although biasing towards low energy subregions is preferable for optimisation
problems, over-biasing can slow down the convergence towards the global
minimum. Figure \ref{fig:FvsPvsM_Ra} presents the performance of
PISAA against the population size for different grid sizes. The performance
improvement of PISAA due to the population size increase becomes more
significant when finer grids (larger grid sizes) are used. As mentioned,
finer grids improve the exploration of the sampling space, however
they require a more efficient self-adjusting mechanism to fight against
possible larger variance in the approximation of $\{\theta_{t}\}$
due to the increased number of subregions. Here, we observed that
increasing the population size allows the use of finer grids, while
it reduces the aforesaid consequence.

\begin{figure}
\subfloat[Average progression curves of the best function values generated by
PISAA, AESAMC, and VFSA with population size $5$.\label{fig:esa_aesamc_esaa_FvsTcondA_2}]{\includegraphics[scale=0.19]{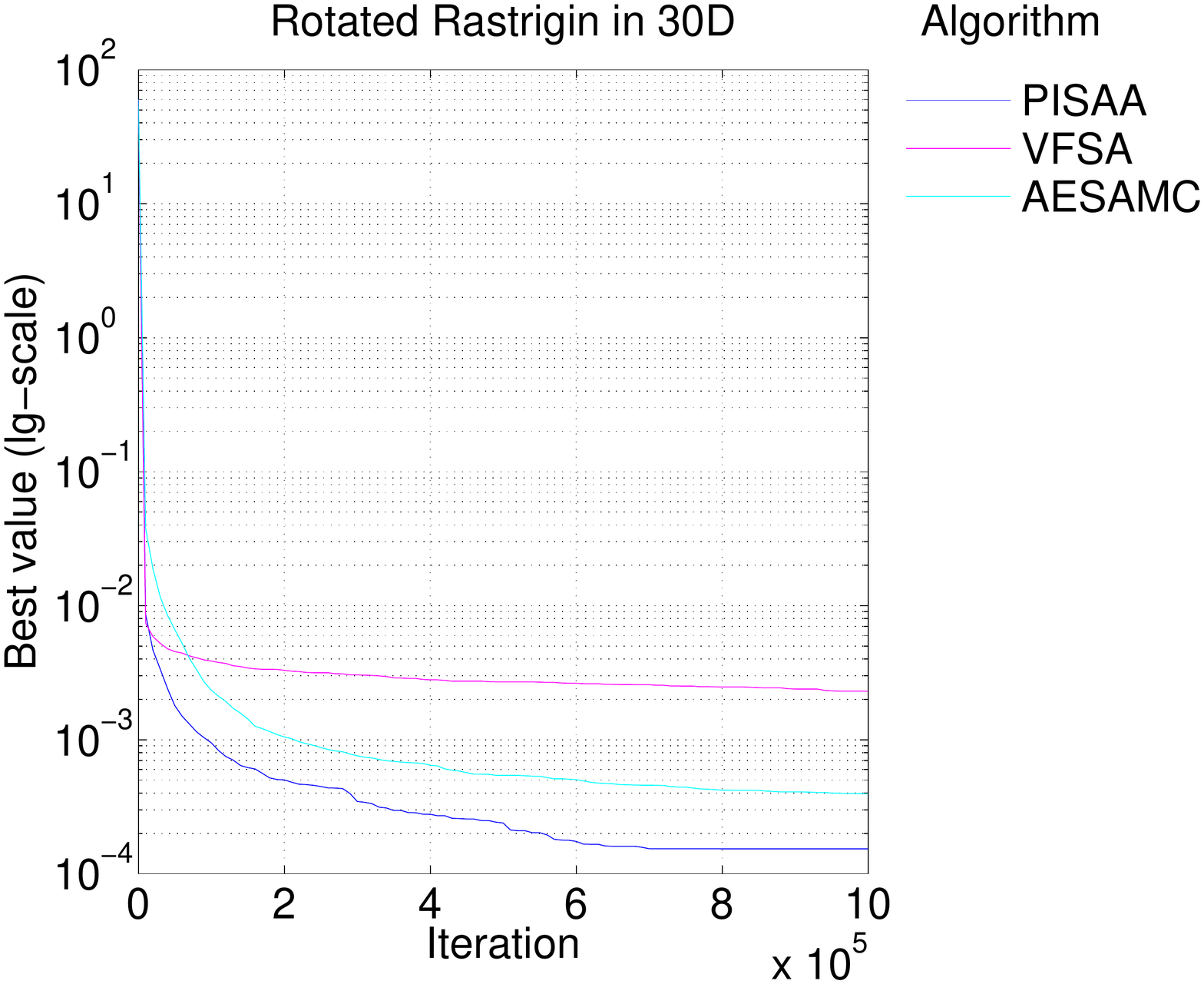}

}\subfloat[Average progression curves of the best function values generated by
PISAA, AESAMC, and VFSA with population size $14$.\label{fig:esa_aesamc_esaa_FvsTcondA_12}]{\includegraphics[scale=0.19]{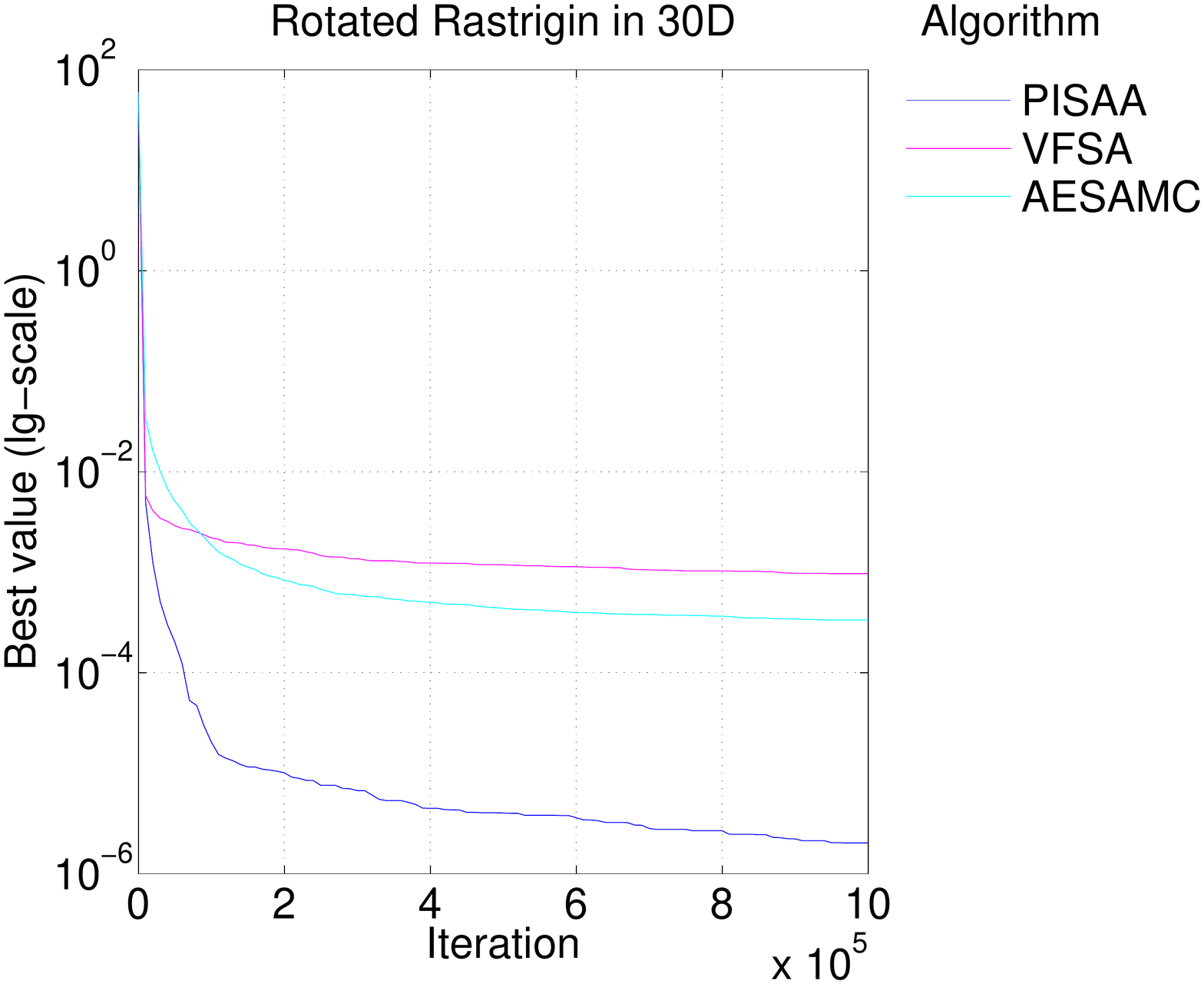}

}\subfloat[Average best function values generated by PISAA, AESAMC, and VFSA
against the population size.\label{fig:PISAA_AESAMC_VFSA_Rastrigin}]{\includegraphics[scale=0.19]{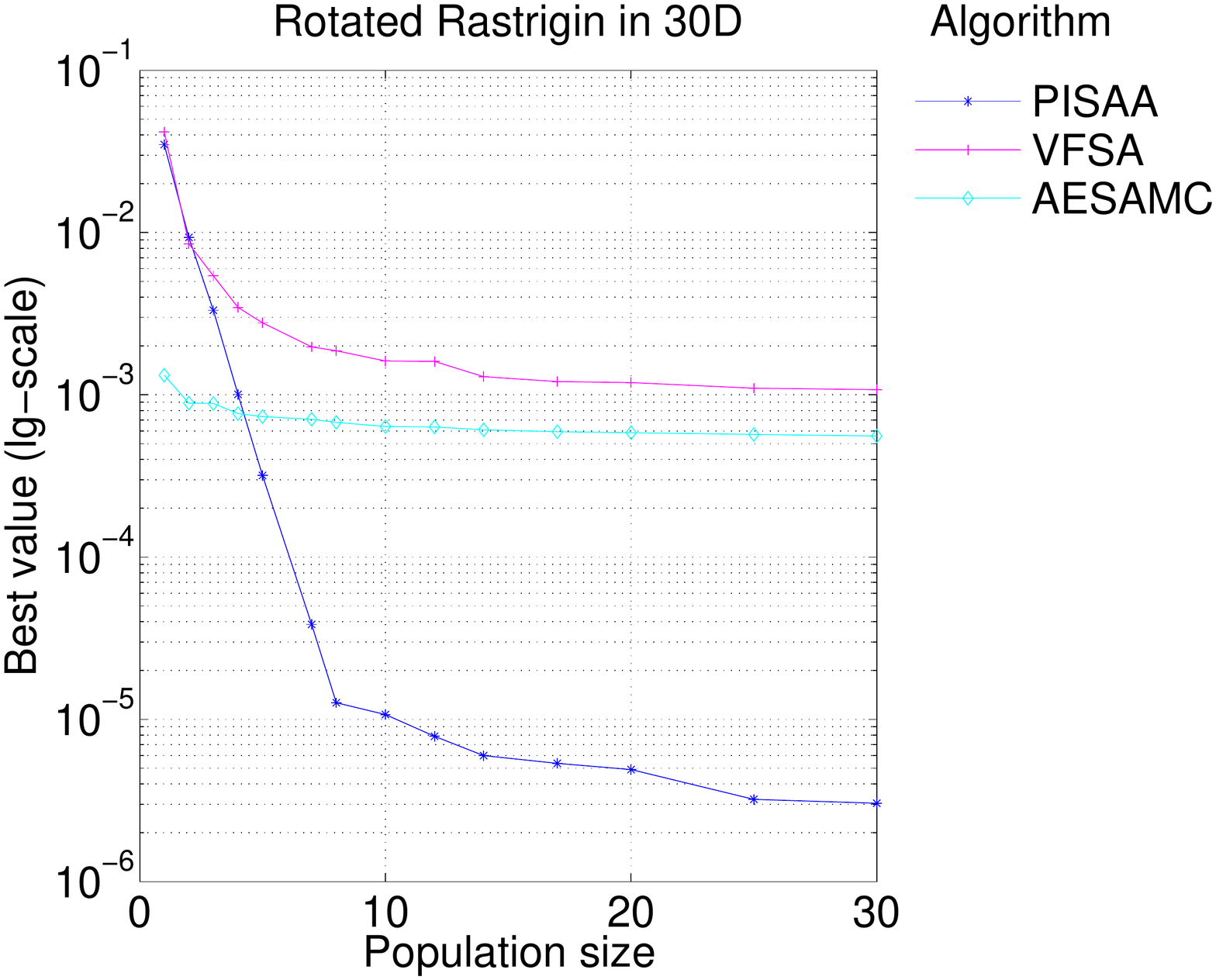}

}

\protect\caption{(Section \ref{sub:Rastrigin's-function}) Average best values (averaged
over $48$ independent runs) discovered by PISAA, AESAMC, and VFSA. }
\end{figure}

We compare PISAA with VFSA using the same operations and temperature
ladder as PISAA, and with AESAMC using the settings used by \citep{Liang2011},
against the $30$D Rastrigin's function. In Figures \ref{fig:esa_aesamc_esaa_FvsTcondA_2}
and \ref{fig:esa_aesamc_esaa_FvsTcondA_12}, we plot the average progression
curves of the best values discovered by each algorithm for population
sizes $\kappa=5$, and $14$ respectively. We observe that PISAA converges
quicker to global minimum than VFSA and AESAMC in both cases. Figure
\ref{fig:PISAA_AESAMC_VFSA_Rastrigin} presents the performance of
the algorithms against the population size. We observe that increasing
the population size improves the performance of PISAA, in terms of
average best values discovered, significantly faster than the performance
of VFSA and AESAMC. It is observed that, although the population size
increases, VFSA and AESAMC stop improving after $\kappa=10$, while
PISAA continues to improve even after $\kappa>10$ but at a slower
rate. This is because the underline adjustment process of $\{w_{t}\}$
keeps on improving, in terms of variance, and converges faster as
$\kappa$ increases. Therefore, PISAA outperforms significantly VFSA,
and AESAMC.

\subsection{Protein folding\label{sub:Protein-folding}}

Proteins are essential to the living organisms as they can carry out
a multitude of biological processes, e.g. production of enzymes, antibodies
etc. In biophysics, understanding the protein folding mechanism is
important because the native conformation of a protein strongly determines
its biological function. Predicting the native conformation of a protein
from its sequence can be treated as an optimisation problem that involves
finding the coordinates of atoms so that the potential energy of the
protein is minimised. This is a challenging optimisation problem \citep{Liang2004},
because (i) the dimensionality of the system is usually high, and
(ii) the landscape of the potential energy is rugged and characterised
by a multitude of local energy minima separated by high energy barriers.

To understand the relevant mechanics of protein folding, simplified,
but still non-trivial, theoretical protein models exist; among them
is the off-lattice AB protein model \citep{StillingerHeadGordonHirshfeld1993}.
The off-lattice AB protein model incorporates only two types of monomers
A and B, in place of the $20$ that occur naturally, which have hydrophobic
and hydrophilic behaviours respectively. The atom sequence $S_{i}$,
$i\in\{2,3,...\}$, of a $N_{i}$-mer, can be determined by a Fibonacci
sequence \citep{StillingerHeadGordonHirshfeld1993,StillingerHeadGordon1995,HsuMehraGrassberger2003}
which is defined recursively as $S_{0}=A$, $S_{1}=B$, $S_{i}=S_{i-2}S_{i-1}$
and has length given by the Fibonacci number $N_{i}=N_{i-2}+N_{i-1}$,
$i\ge2$. The atoms are assumed to be linked consecutively by rigid
bonds of unit length to form a linear chain which can bend continuously
between any pair of successive links. The chain can reside in the
$2$--, or $3$-- dimensional physical space which defines the $2$D,
or $3$D off-lattice AB model, correspondingly.

For the $2$D AB model \citep{StillingerHeadGordonHirshfeld1993,StillingerHeadGordon1995,Liang2004},
the potential energy is 
\begin{gather}
U_{3,1}(\theta_{2:N-1})=\sum_{i=1}^{N-2}V_{\theta}(i)+\sum_{i=1}^{N-2}\sum_{j=i+2}^{N}V_{LJ}(i,j);\label{eq:Cost function Protein 2D}\\
V_{\theta}(i):=0.25(1-u_{i}^{\transpose}\cdot u_{i+1}),V_{LJ}(i,j):=4(r_{i,j}^{-12}-C_{2\text{D}}(i,j)r_{i,j}^{-6}),\nonumber 
\end{gather}
 where $C_{2\text{D}}(i,j)$ is $1$, $1/2$, and $-1/2$, for AA,
BB, and AB pairs respectively, $u_{i}:=(\cos(\theta_{i}),\sin(\theta_{i}))^{\transpose}$
is the unit vector joining monomer $i$ to monomer $i+1$, $r_{i,j}:=r_{i,j}(\theta_{2:N-1})$
denotes the distance between monomers $i$ and $j$, and $\theta_{1}=0$,
$\theta_{i}\in[0,2\pi)$, for $i=2,...,N-1$, are polar coordinates.
The dimensionality of the problem is $d=N-2$. For the $3$D AB model
\citep{IrbackPetersonPotthastSommelius1997,HsuMehraGrassberger2003,BachmannArkinJanke2005,KimLeeLee2005,Liang2004},
the potential energy is 
\begin{gather}
U_{3,2}(\theta_{2:N-1},\phi_{3:N-1})=\sum_{i=1}^{N-2}V_{\theta}(i)+\sum_{i=1}^{N-3}V_{\tau}(i)+\sum_{i=1}^{N-2}\sum_{j=i+2}^{N}V_{LJ}(i,j);\label{eq:Cost function Protein 3D}\\
V_{\theta}(i):=u_{i}\cdot u_{i+1},\ V_{\tau}(i):=-0.5(u_{i}\cdot u_{i+2}),\ V_{LJ}(i,j):=4(r_{i,j}^{-12}-C_{3\text{D}}(i,j)r_{i,j}^{-6}),\nonumber 
\end{gather}
 where $C_{3\text{D}}(i,j)$ is $1$, for AA, and $1/2$, for BB,
and AB pairs, $u_{i}:=(\cos(\theta_{i})\sin(\phi_{i}),\sin(\theta_{i})\sin(\phi_{i}),\cos(\phi_{i}))^{\transpose}$,
$\theta_{i}$ is the azimuthal angle, and $\phi_{i}$ is the polar
angle of $u_{i}$ such that $\theta_{1}=\phi_{1}=\phi_{2}=0$, $\theta_{i}\in[0,2\pi)$,
$\phi_{i}\in[0,\pi]$, for $i=1,...,N-1$. The dimensionality of the
problem is $d=2N-5$. Here, for the purpose of demonstration, we concentrate
on the $13$--, $21$--,$34$--, and $55$-- mers AB.

We consider default settings for PISAA (valid if not stated otherwise):
(i) $n=2\cdot10^{7}$ iterations, (ii) uniformly spaced grid $\{u_{j}\}$
with $m=101$, (iii) desirable probability with parameter $\lambda=0.1$,
(iv) temperature ladder $\{\tau_{t}\}$ with $\tau_{h}=10$, $n^{(\tau)}=10^{6},$
$\tau_{*}=10^{-2}$, (iv) gain factor $\{\gamma_{t}\}$ with $n^{(\gamma)}=10^{3},$
$\beta=0.55$, and (v) MCMC transition probability with mutation operations
(Metropolis, hit-and-run, $k$-point) and crossover operations ($k$-point,
snooker, linear), equal operation rates, and operation scale parameters
$\sigma j/(m+1)$, where $\sigma$ is calibrated so that the expected
acceptance ratio to be around $0.234$, and $j$ is the label of the
subregion the current state belongs to. Each experiment ran $48$
times independently to eliminate possible variation in the output
caused by nuisance factors.

\begin{figure}
\subfloat[Average progression curves of the best function values, for different
population sizes.\label{fig:esaa_FvsTcondP-2D55mer-Ex2}]{\includegraphics[scale=0.19]{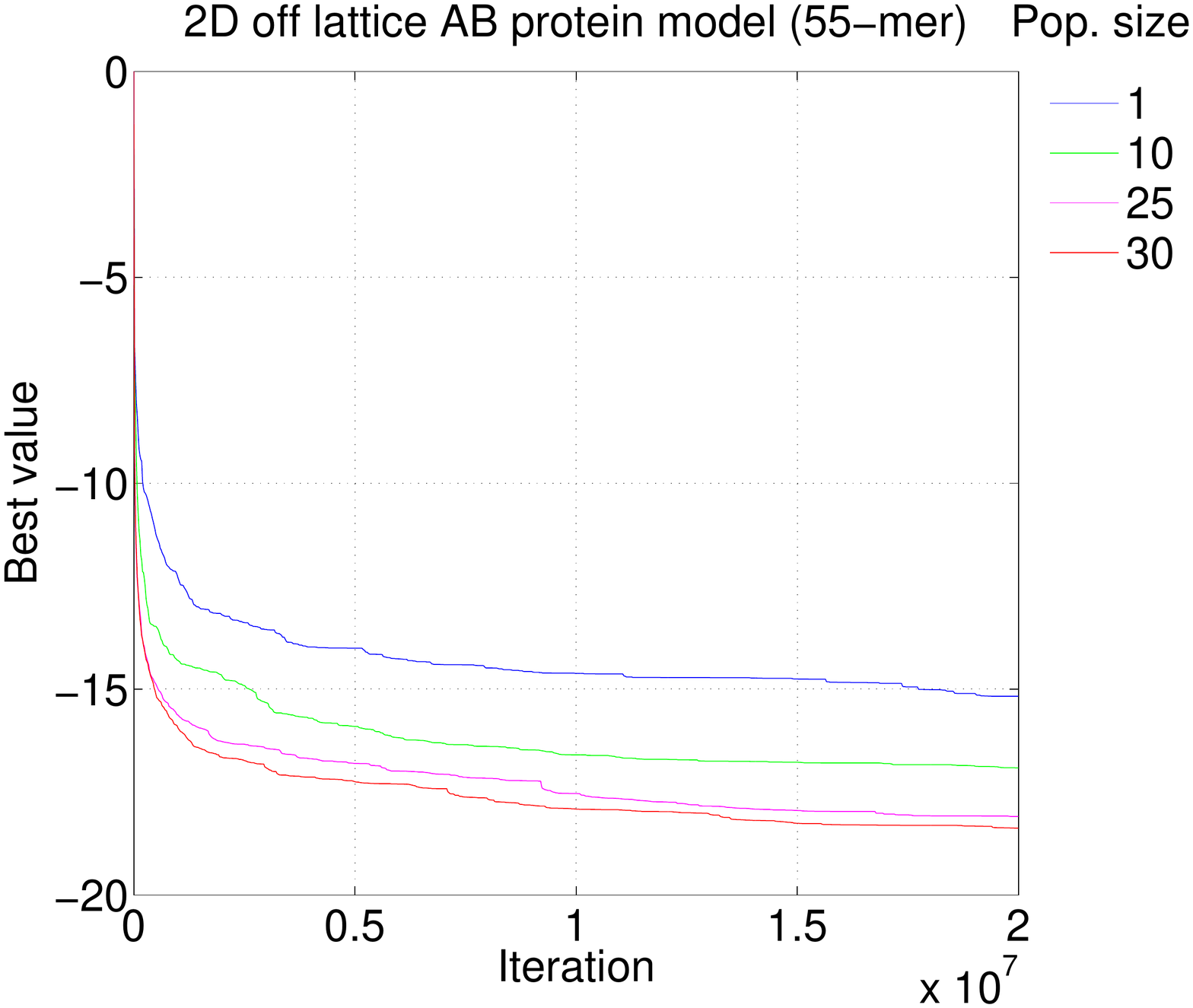}

}\subfloat[Average best function values against the population size, for different
lengths of polymer. \label{fig:esaa_FvsPcondD-2D-Ex2}]{\includegraphics[scale=0.19]{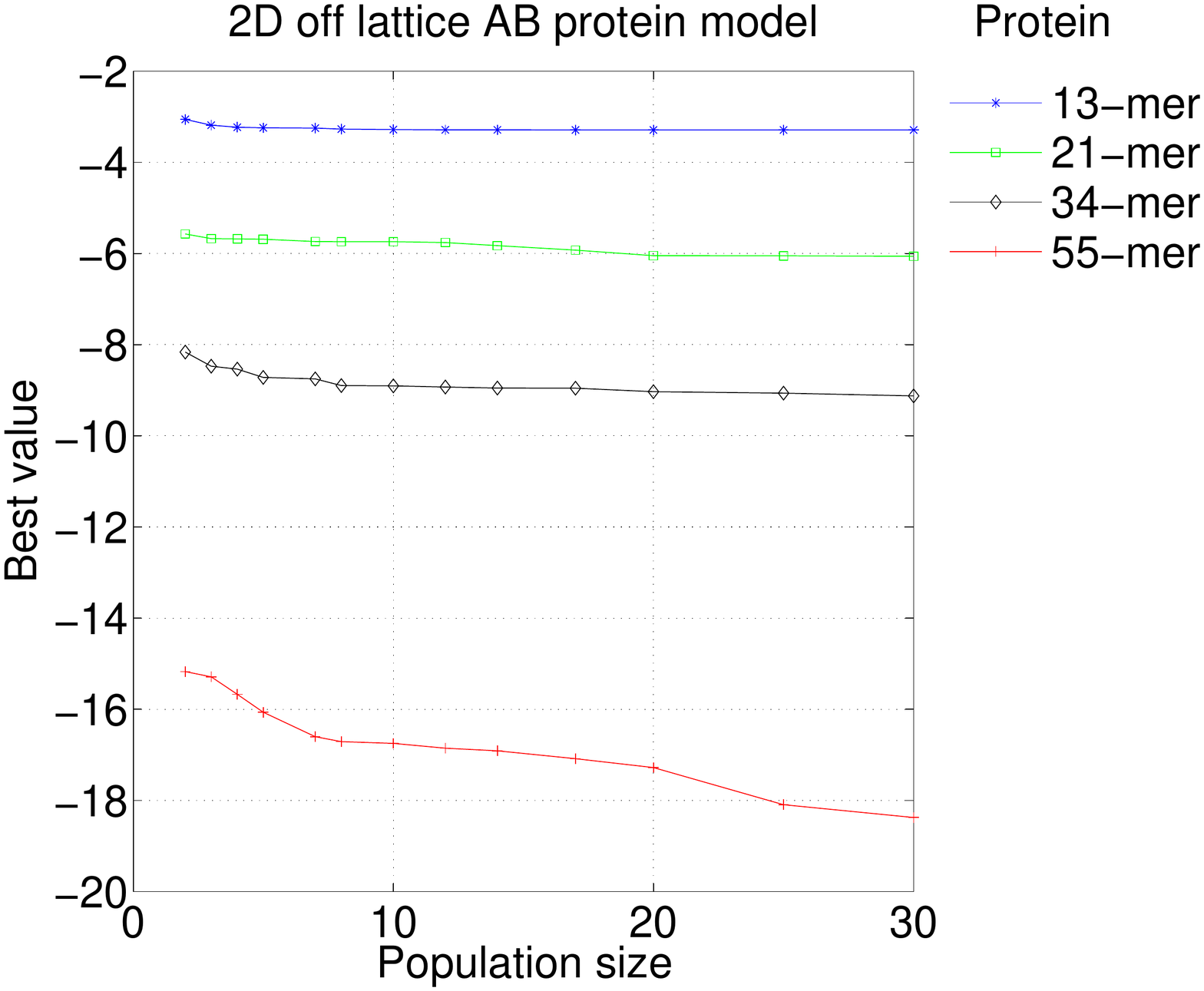}

}\subfloat[Progression curves of the MSE estimate of $w_{t}$ for different population
sizes. ($\beta=0.55$). \label{fig:Progression-MSE-theta-Pr2D}]{\includegraphics[scale=0.19]{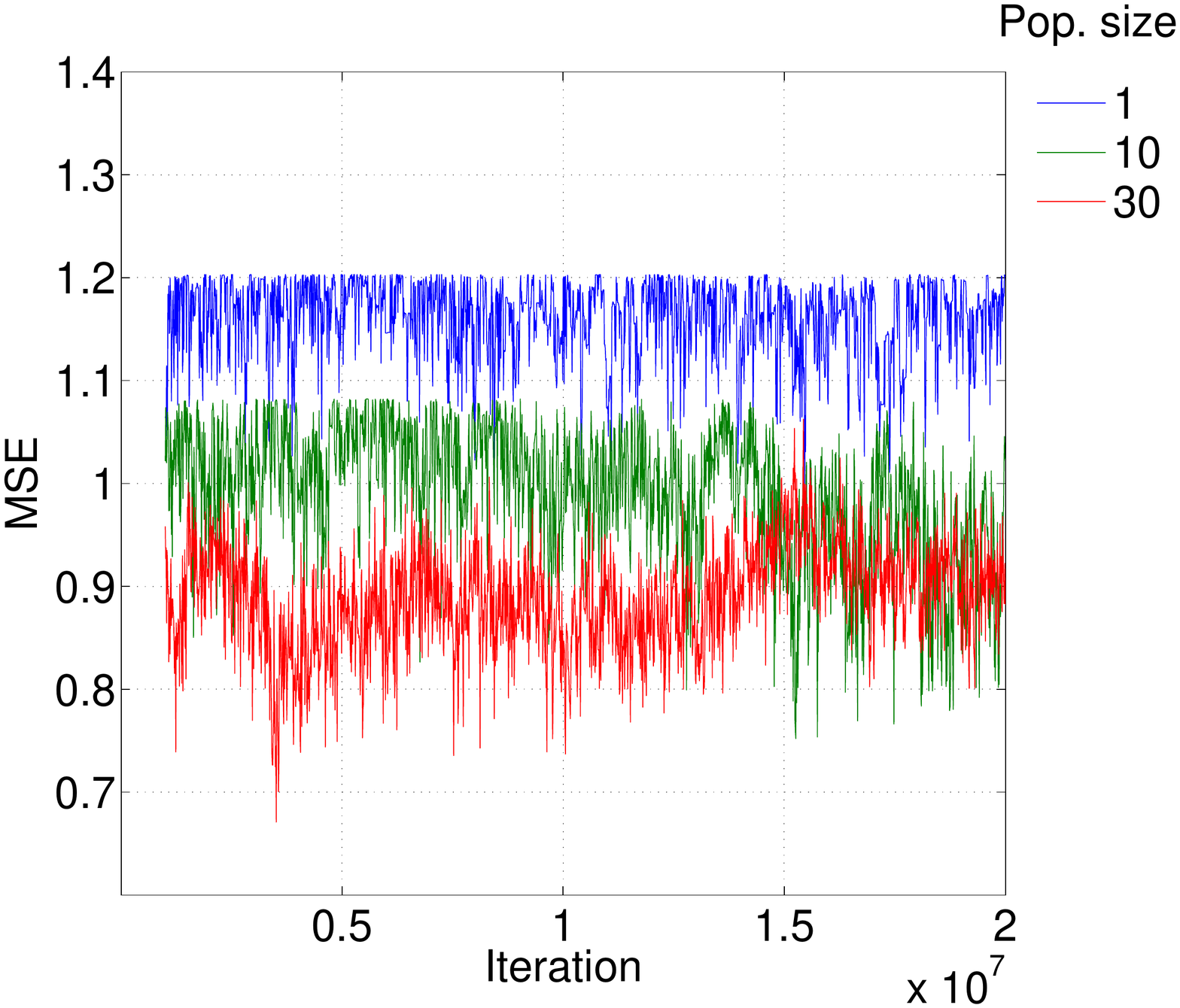}

}

\subfloat[Average progression curves of the best function values, for different
population size. \label{fig:esaa_FvsTcondP-3D55mer-Ex2}]{\includegraphics[scale=0.19]{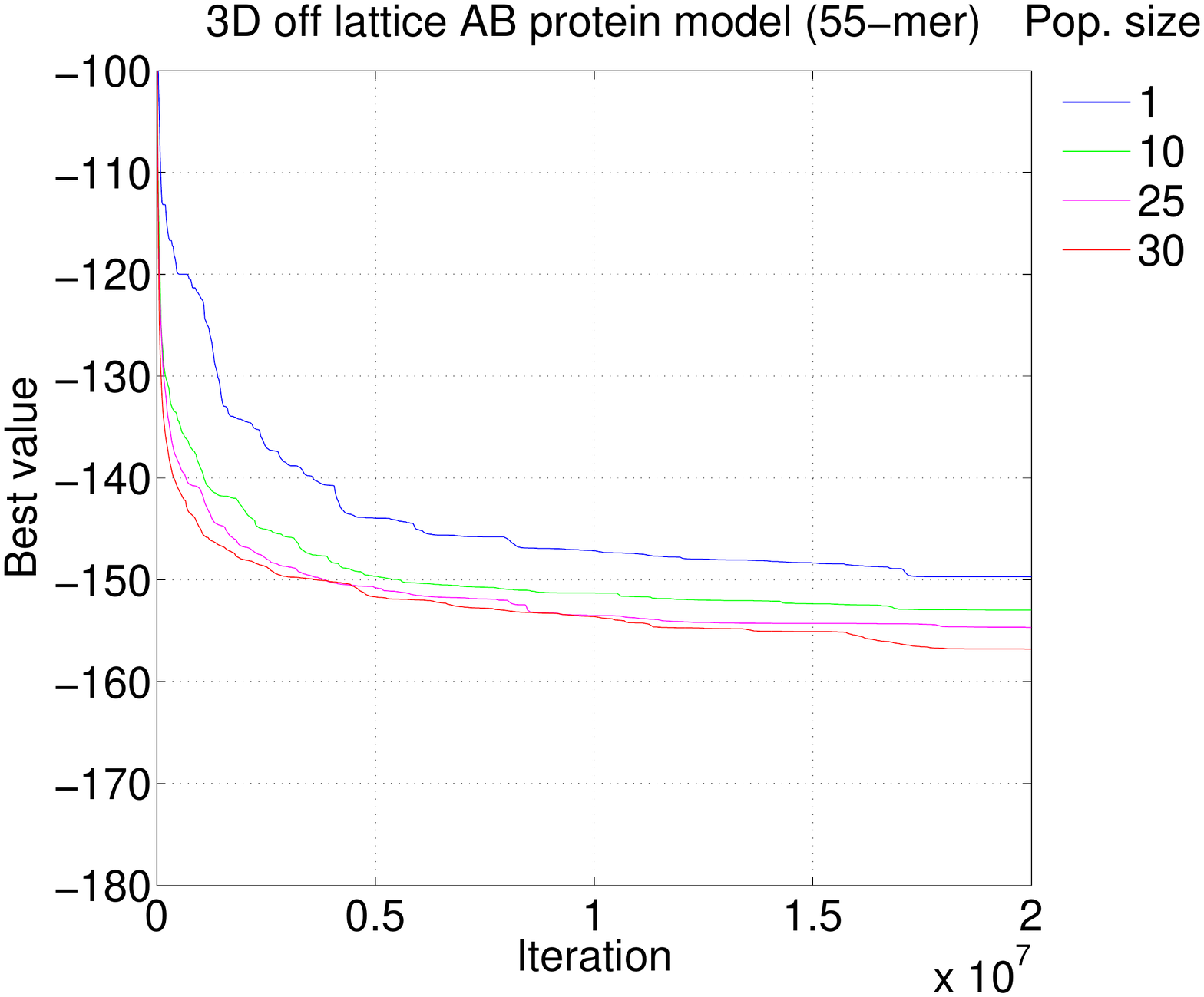}

}\subfloat[Average best function values against the population size, for different
lengths of polymer. \label{fig:esaa_FvsPcondD-3D-Ex2}]{\includegraphics[scale=0.19]{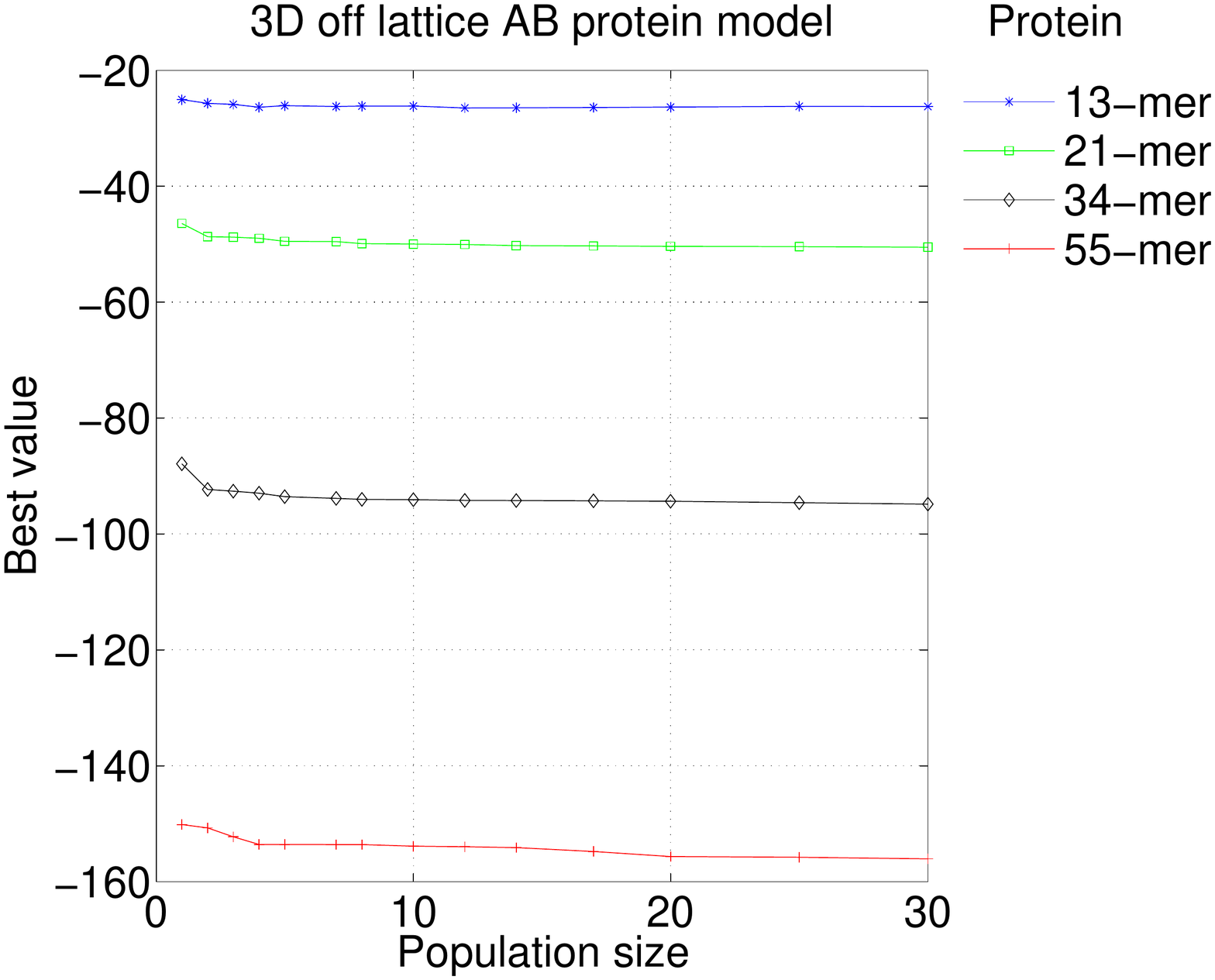}

}\subfloat[Progression curves of the MSE estimate of $w_{t}$ for different population
sizes. ($\beta=0.55$). \label{fig:Progression-MSE-theta-Pr3D}]{\includegraphics[scale=0.19]{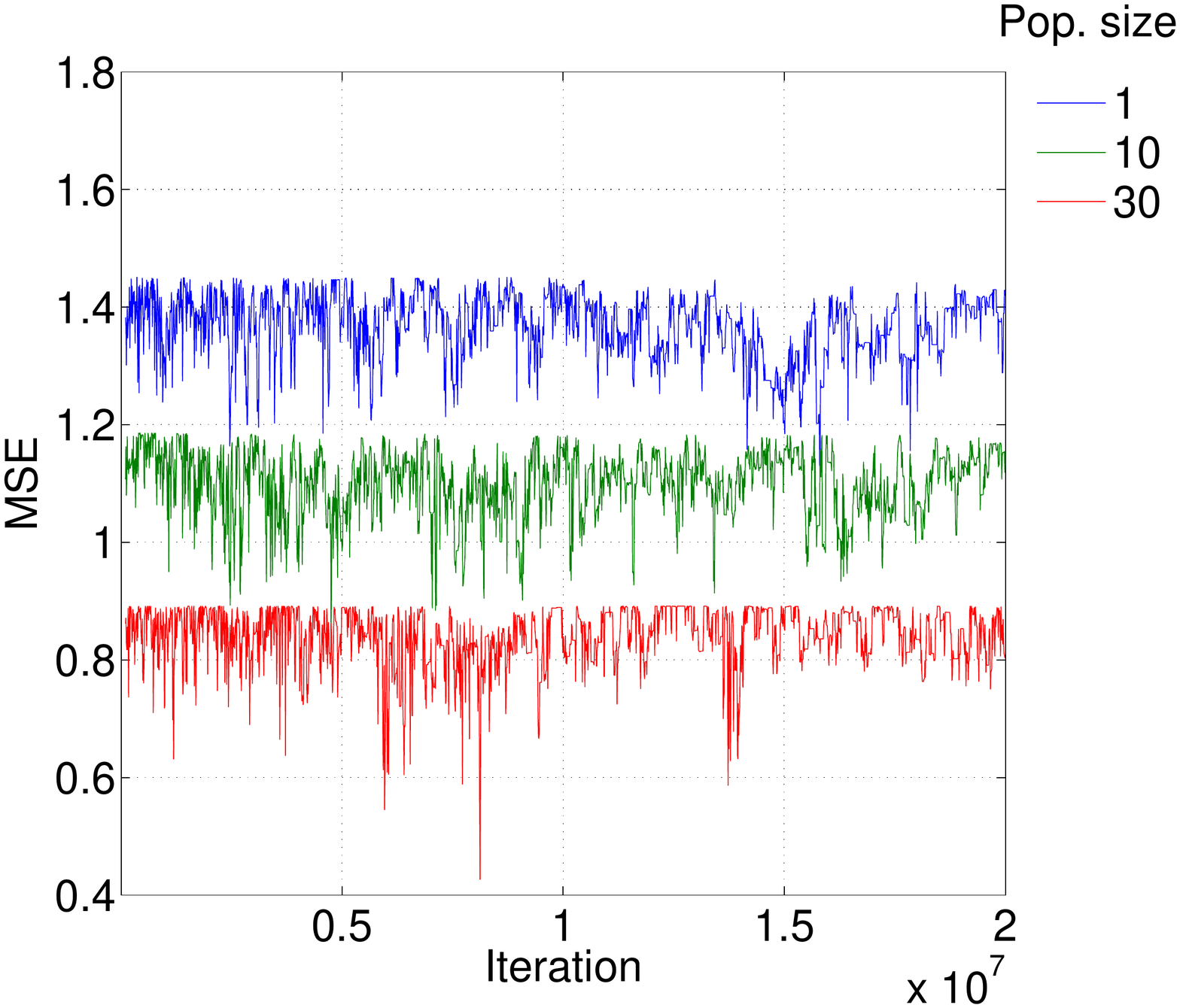}

}

\protect\caption{(Section \ref{sub:Protein-folding}) Performance plots of PISAA. The
results reported consider averaged values over $48$ independent runs.
The $1$st and $2$nd rows refer to the $2$D and $3$D AB models
correspondingly. \label{fig:PISAA_Ex2}}
\end{figure}

We examine the performance of PISAA as a function of the iterations
and the population size. In Figures \ref{fig:esaa_FvsTcondP-2D55mer-Ex2}
and \ref{fig:esaa_FvsTcondP-3D55mer-Ex2}, we illustrate the average
progressive curves of the best values discovered by PISAA using different
population sizes against the $55$-mer $2$D and $3$D AB models.
We observe that PISAA using larger population sizes converges quicker
towards smaller average best values. In Figures \ref{fig:esaa_FvsPcondD-2D-Ex2}
and \ref{fig:esaa_FvsPcondD-3D-Ex2}, we present the performance of
PISAA with respect to the `best values' discovered until the iteration
$n=2\cdot10^{7}$ as a function of the population size against the
$2$D and $3$D AB models, respectively. In our simulations, we have
considered the $13$--, $21$--,$34$--, and $55$--mers AB sequences.
We observe that increasing the population size of PISAA is particularly
effective in longer AB sequences (and so higher in dimension problems),
while moderate population sizes are adequate in shorter AB sequences
(and so moderate in dimension problems). In fact, PISAA improves significantly
as the population size increases in the high dimensional case of $55$--mers,
however it performs acceptably even with a moderate population size
($\kappa\approx10$) in the lower dimensional cases of $13$--, $21$--,$34$--mers.
Compared to the standard SAA (aka PISAA with $\kappa=1$), PISAA (with
$\kappa>1$) presents significantly improved performance, in the $55$-mer
$2$D and $3$D AB models when the same number of iterations is considered.
Note that increasing the population size of PISAA does not necessarily
mean that the CPU time required for the algorithm to run increases
significantly because PISAA can be implemented in parallel computational
environment if available. In Figures \ref{fig:Progression-MSE-theta-Pr2D}
and \ref{fig:Progression-MSE-theta-Pr3D}, we observe that when PISAA
uses larger population sizes, the bias weights generated by the self-adjusting
mechanism of PISAA have smaller MSE, and hence the algorithm tends
to present a more stable self-adjusting process.

We compare the performance of PISAA with those of VFSA and AESAMC,
against the $55$-mer $2$D, and $3$D off-lattice AB models. We run
each simulation $48$ times independently to eliminate possible variation
in the output caused by nuisance factors. About the algorithmic settings:
PISAA uses the aforementioned settings, VFSA shares common settings
with PISAA, and AESAMC uses an equally spaced partition of $10^{4}$
subregions, temperature $\tau=1.0$, and threshold values $\aleph=10$.
VFSA and AESAMC use the same crossover and mutation operations as
PISAA.

\begin{figure}
\subfloat[Average progression curves of the best function values discovered
by PISAA, AESAMC, and VFSA with population size $3$.\label{fig:esa_aesamc_esaa_FvsTcondA_Ex2_2D_3}]{\includegraphics[scale=0.19]{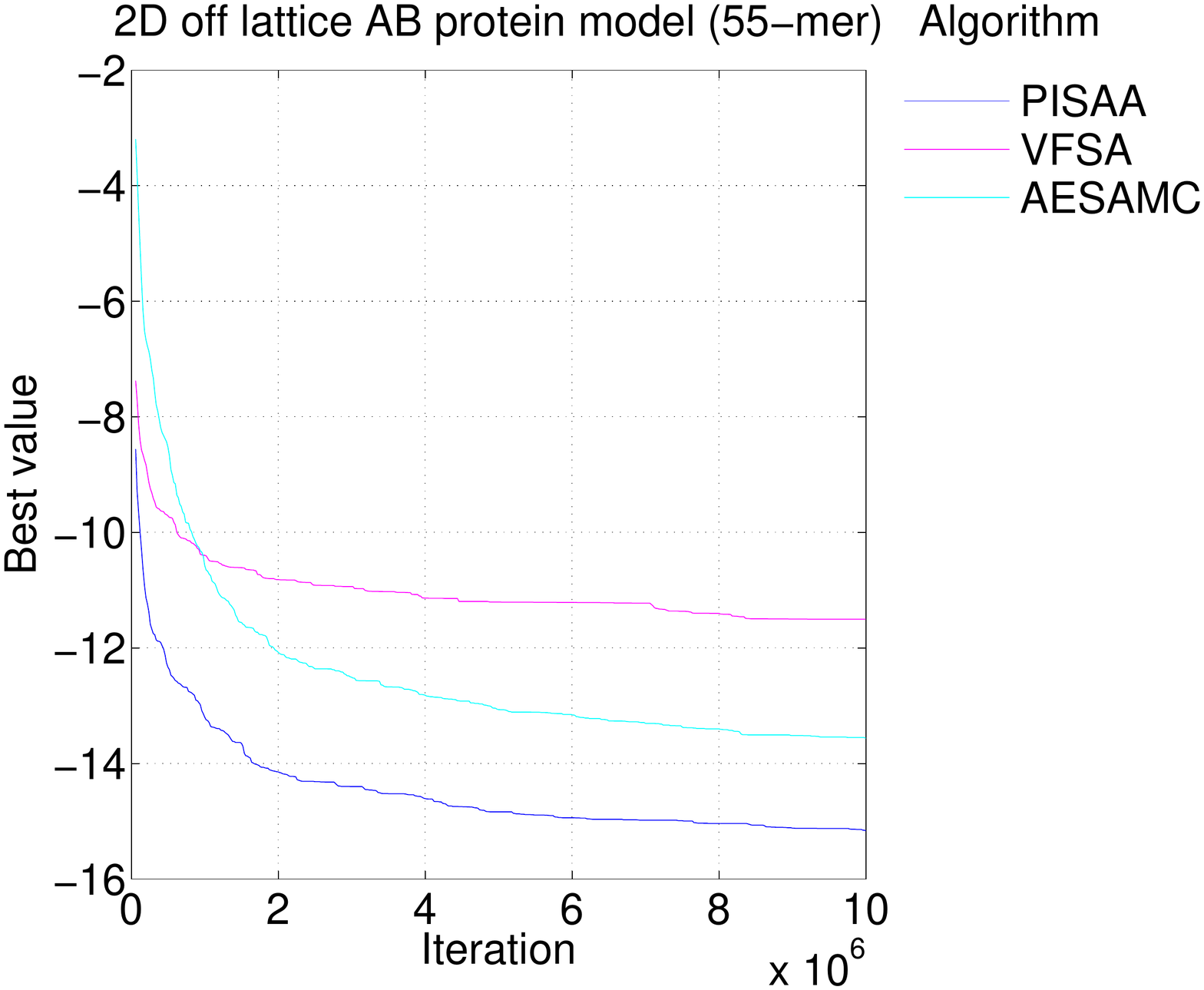}

}\subfloat[Average progression curves of the best function values discovered
by PISAA, AESAMC, and VFSA with population size $30$.\label{fig:esa_aesamc_esaa_FvsTcondA_Ex2_2D_30}]{\includegraphics[scale=0.19]{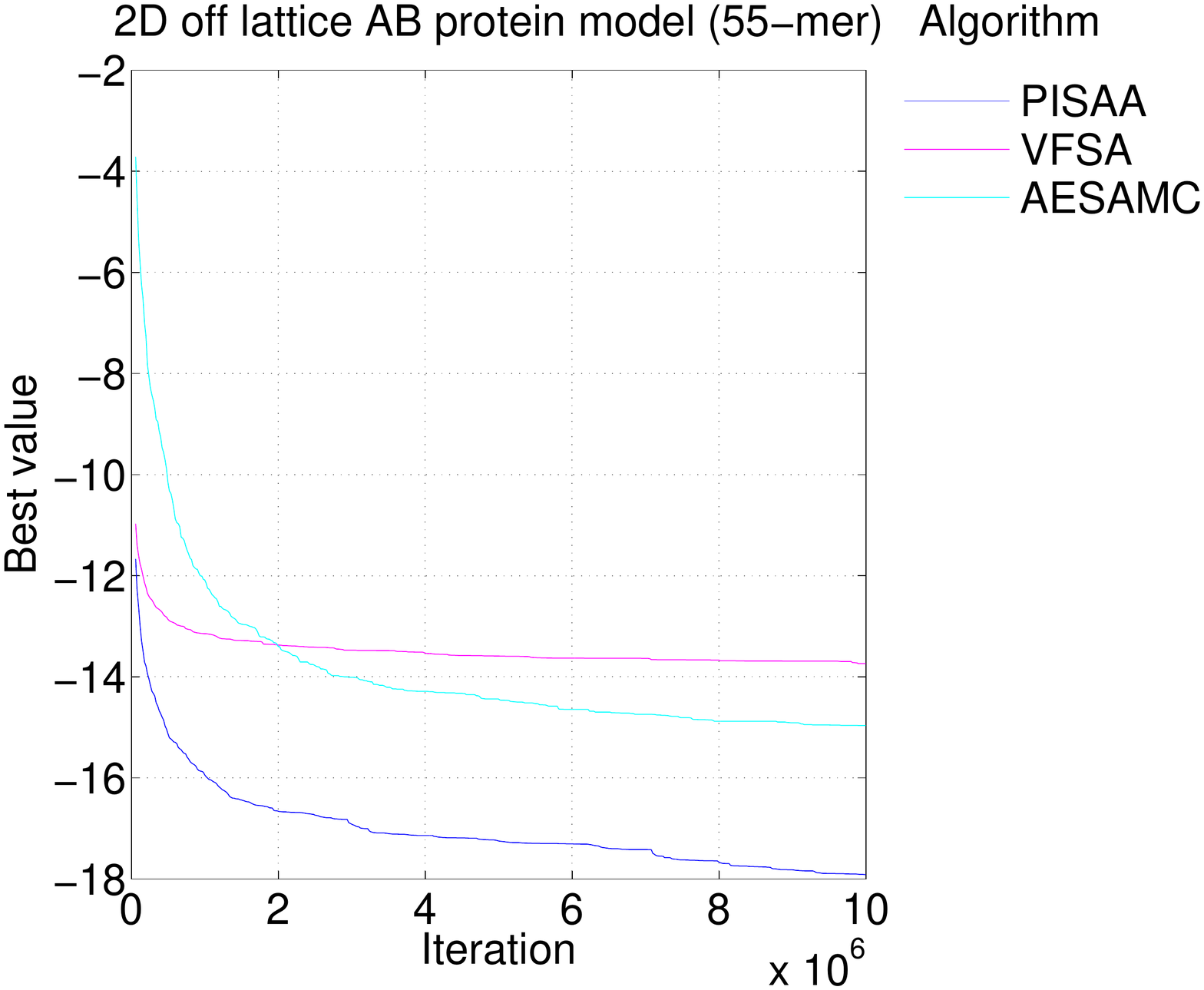}

}\subfloat[Average best function values discovered by PISAA, AESAMC, and VFSA
against the population size ($n=2\cdot10^{7}$).\label{fig:PISAA_AESAMC_VFSA_Ex2_2D}]{\includegraphics[scale=0.19]{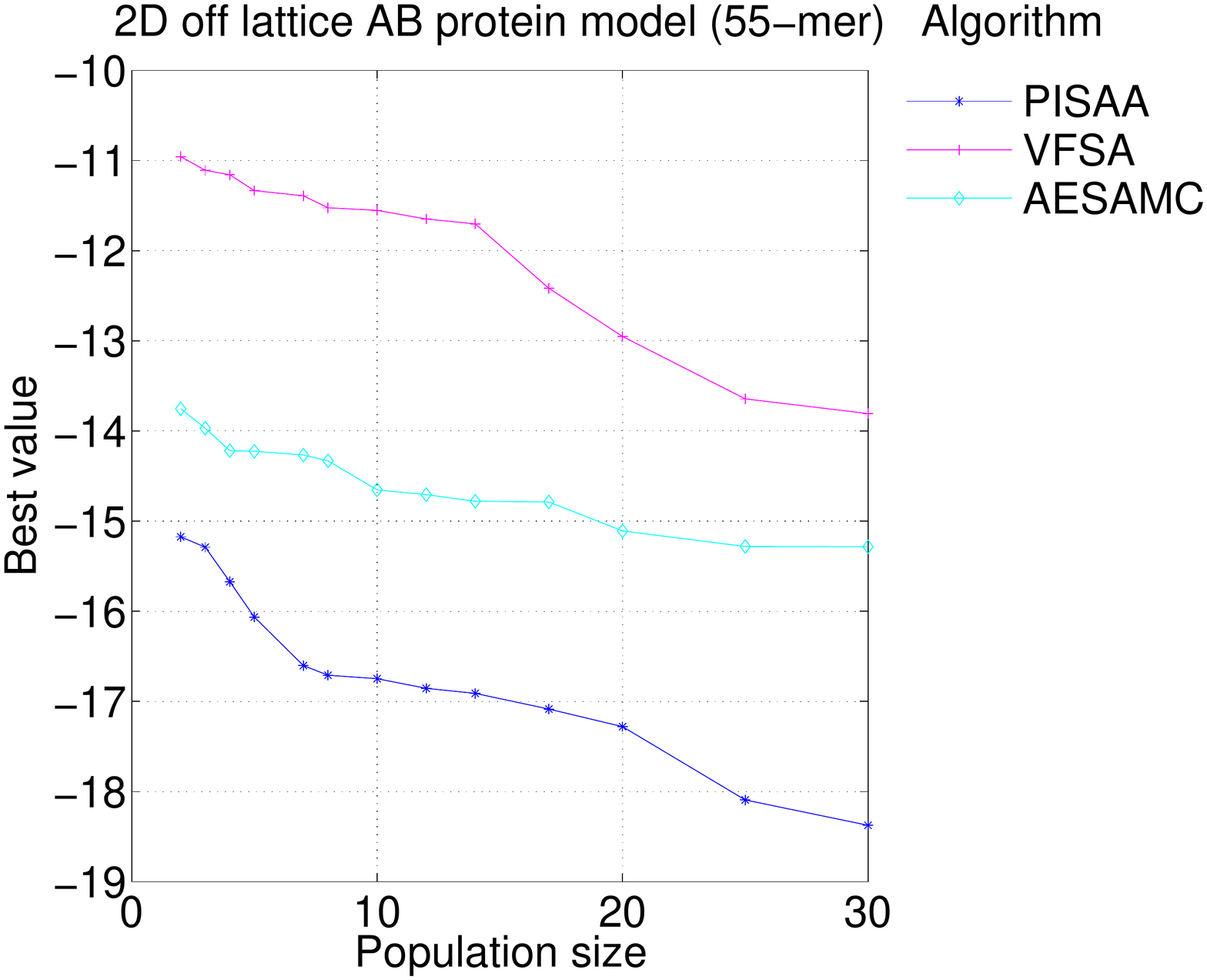}

}

\subfloat[Average progression curves of the best function values discovered
by PISAA, AESAMC, and VFSA with population size $10$.\label{fig:esa_aesamc_esaa_FvsTcondA_Ex2_3D_10}]{\includegraphics[scale=0.19]{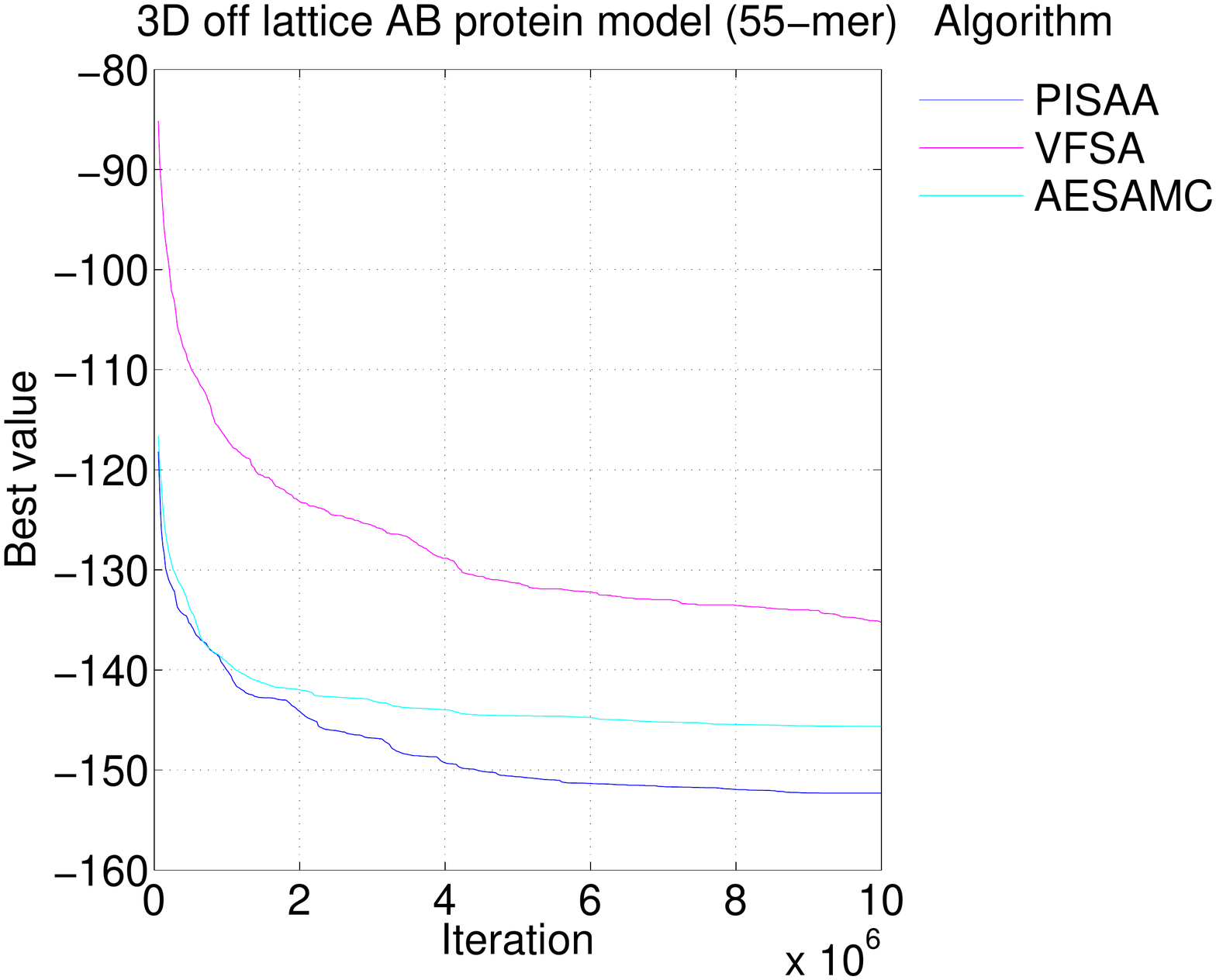}

}\subfloat[Average progression curves of the best function values discovered
by PISAA, AESAMC, and VFSA with population size $30$.\label{fig:esa_aesamc_esaa_FvsTcondA_Ex2_3D_30}]{\includegraphics[scale=0.19]{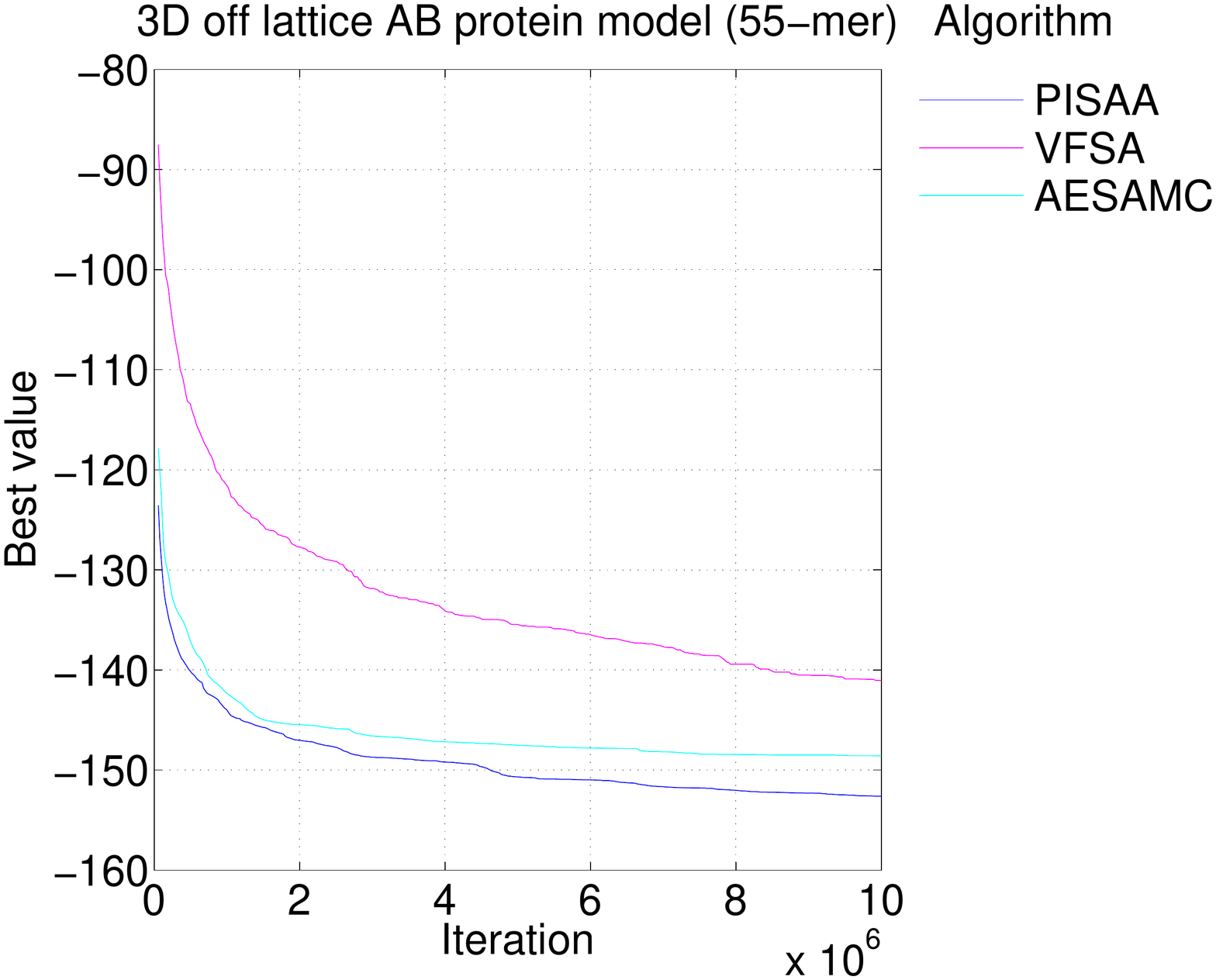}

}\subfloat[Average best function values generated by PISAA, AESAMC, and VFSA
against the population size ($n=2\cdot10^{7}$).\label{fig:PISAA_AESAMC_VFSA_Ex2_3D}]{\includegraphics[scale=0.19]{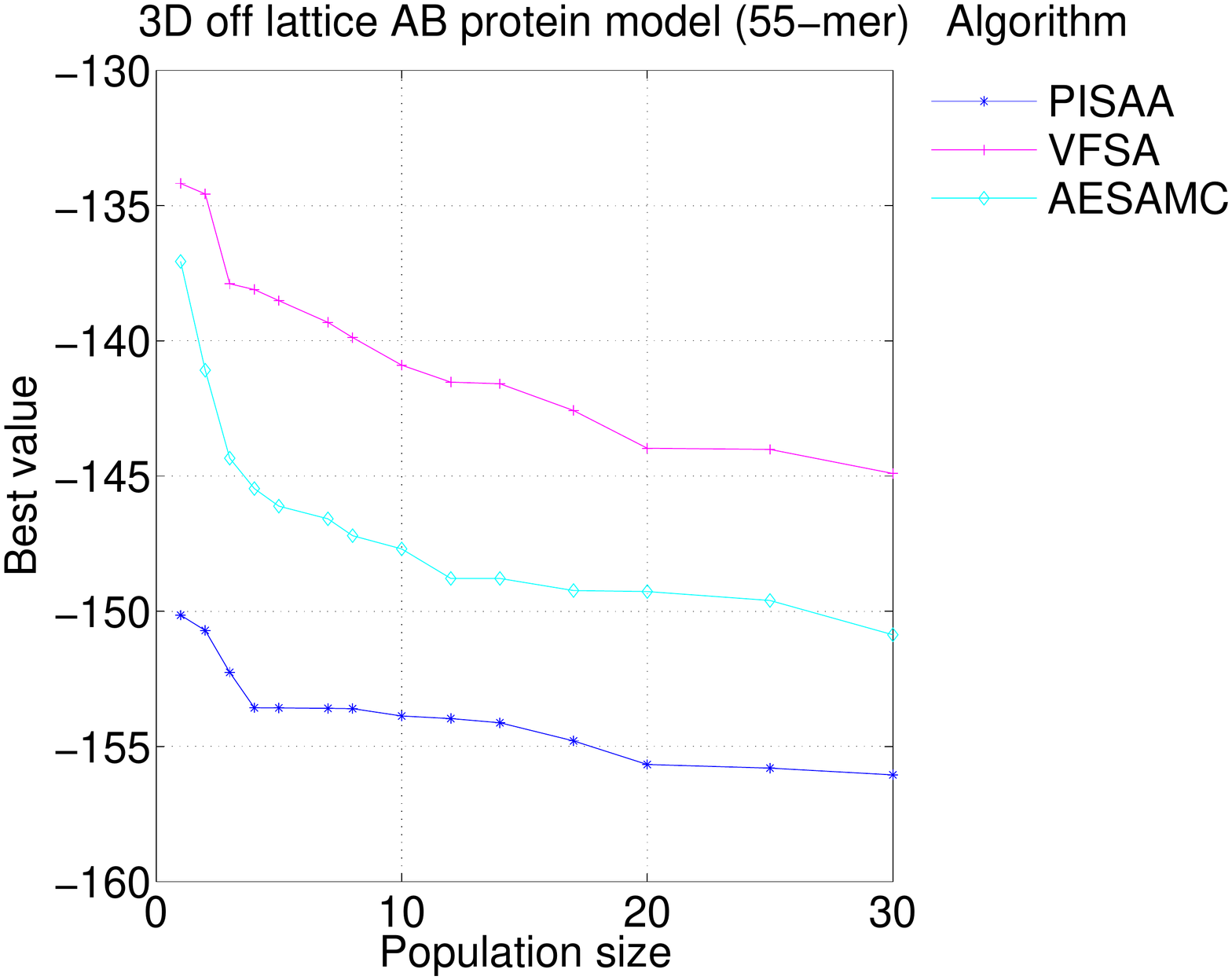}

}

\protect\caption{(Section \ref{sub:Protein-folding}) Average best values (averaged
over $48$ independent runs) discovered by PISAA, AESAMC, and VFSA.
We consider the $55$-mer AB model in $2$D and $3$D, $1$st and
$2$nd rows correspondingly. \label{fig:PISAA_VFSA_AESAMC_Ex3}}
\end{figure}

The results from the empirical comparison of PISAA, AESAMC, and VFSA
associated to the $2$D and $3$D AB models are summarised in the
1st and 2nd rows of Figure \ref{fig:PISAA_VFSA_AESAMC_Ex3}, respectively.
Figures \ref{fig:esa_aesamc_esaa_FvsTcondA_Ex2_2D_3}, \ref{fig:esa_aesamc_esaa_FvsTcondA_Ex2_2D_30},
\ref{fig:esa_aesamc_esaa_FvsTcondA_Ex2_3D_10} and \ref{fig:esa_aesamc_esaa_FvsTcondA_Ex2_3D_30}
show the average progression curves, up to iteration $n=10^{7}$,
generated by the algorithms under comparison. We observe that the
average progression curves generated by PISAA converge quicker towards
smaller `best values' than those generated by AESAMC, and VFSA. This
behaviour is observed in both large population sizes ($\kappa=30$)
and small population sizes ($\kappa=3$ and $\kappa=10$), in $2$D
and $3$D AB models. PISAA does not appear to become trapped into
local minima although, during the first iteration steps, the curves
generated by PISAA reduce at a faster rate than those generated by
AESAMC and VFSA. Possibly, the reason is because compared to AESAMC,
PISAA uses a smoother shrink strategy towards areas of minima, while
compared to VFSA, PISAA uses an enhanced self-adjusting mechanism. 

In Figures \ref{fig:PISAA_AESAMC_VFSA_Ex2_2D} and \ref{fig:PISAA_AESAMC_VFSA_Ex2_3D},
we compare the performance of PISAA, AESAMC, and VFSA with respect
to the averaged best values discovered as a function of the population
size, in the $2$D and $3$D AB models. We observe that PISAA has
discovered smaller `best values' than AESAMC and VFSA for any population
size considered in both $2$D, and $3$D AB models. However, if parallel
environment is available, PISAA is expected to further outperform
AESAMC, for a given budget of execution time, because at each iteration
PISAA can generate the population simultaneously by using several
CPU cores in parallel while AESAMC has to do it serially. Thus, we
observe that PISAA significantly outperforms AESAMC and VFSA. 

\subsection{Spatial imaging\label{sub:Spatial-imaging}}

\begin{figure}
\centering

\subfloat[Gray scale image ($480\times640$ pixels), and the fragment under
consideration ($240\times320$ pixels) \label{fig:FP_Original-image}]{\includegraphics[angle=-90,scale=0.25]{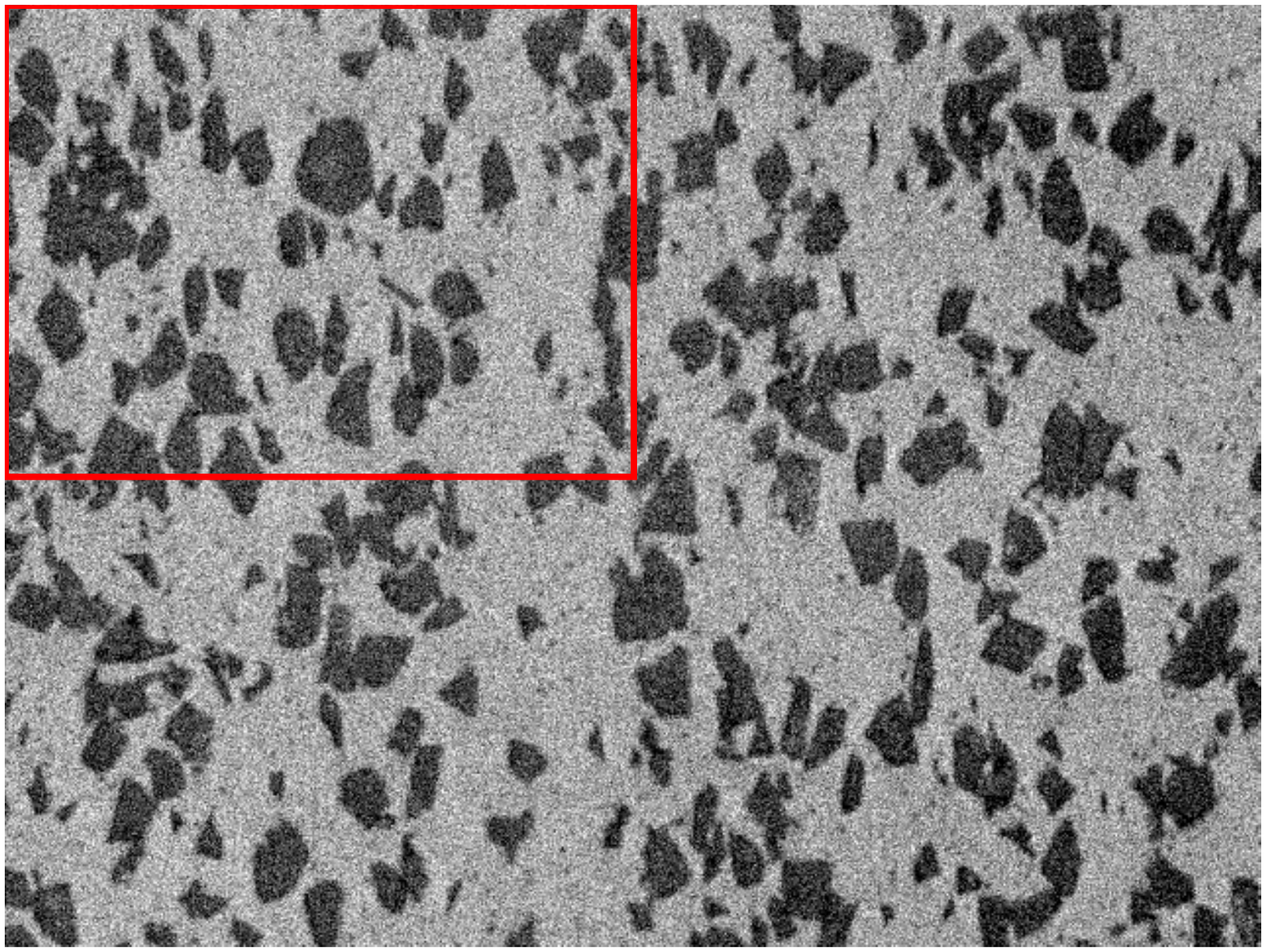}

}\subfloat[MAP estimate of image fragment ($240\times320$ pixels) \label{fig:FP_MAP-estimate}]{\includegraphics[angle=-90,scale=0.25]{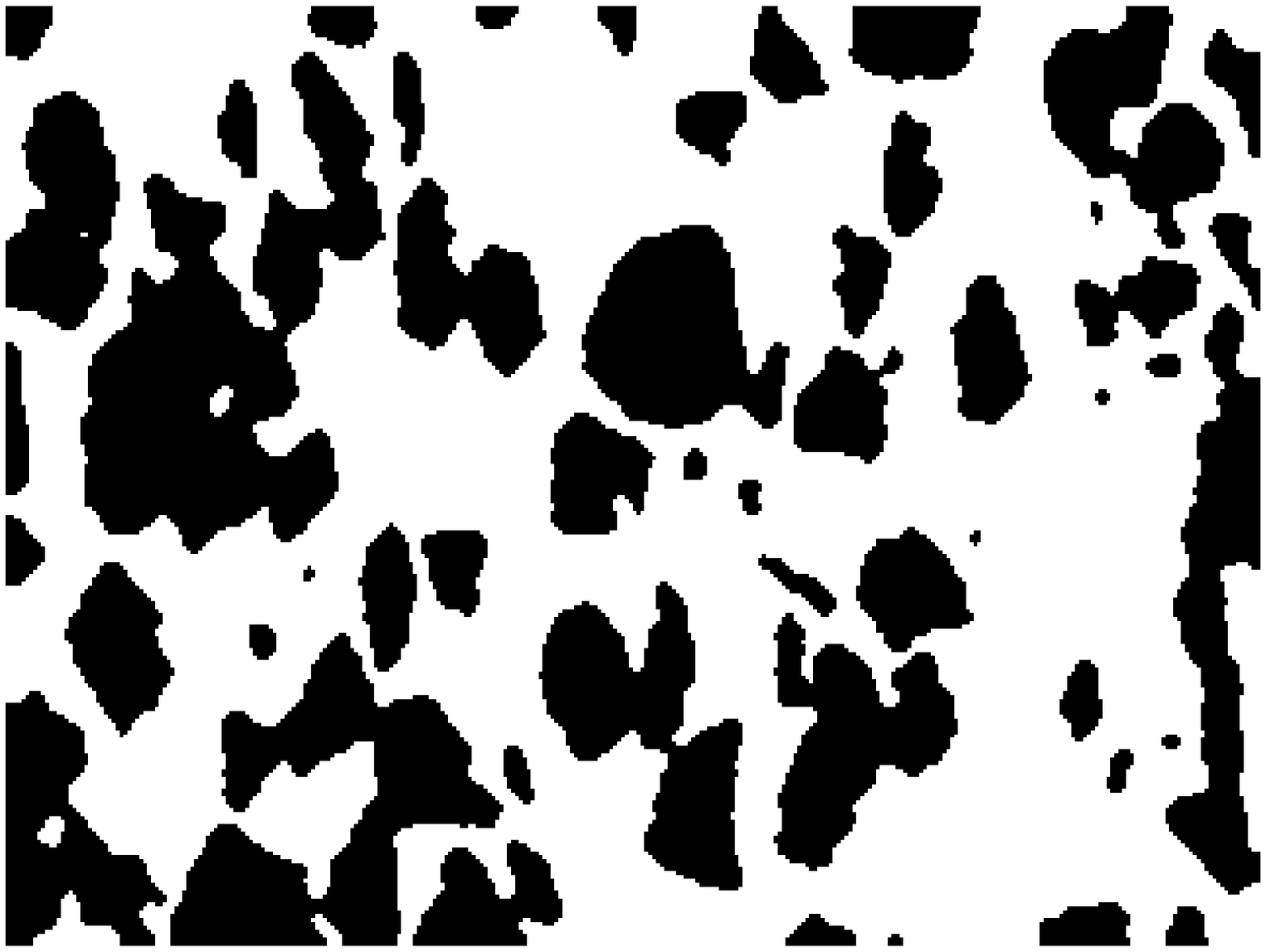}

}

\protect\caption{(Section \ref{sub:Spatial-imaging}) Gray scale digital photo-micrograph
of the micro-structure of the Ferrite-Pearlite steel, and the MAP
estimate of its, red in colour, framed fragment.}
\end{figure}

We consider an image restoration problem where there is need to remove
the noise from a $2$D binary image. The image under consideration
was obtained from PNNL\textquoteright s project supported by the U.S.
Department of Energy's Office of Energy Efficiency and Renewable Energy
to improve advanced transportation technologies. The image is a gray-scale
photo-micrograph of the micro-structure of the Ferrite-Pearlite steel
(Figure \ref{fig:FP_Original-image}), where the lighter part is ferrite
while the darker part is pearlite. It can help us investigate how
the micrograph of the microstructure of the Ferrite-Pearlite steel
(and hence strength level) develops during hot rolling \citep{GladshteinLarionovaBelyaev2012},
and therefore better understand how to control the strength of a strip
steel. We focus our analysis on the first quarter fragment of size
$240\times320$ pixels (red frame in Figure \ref{fig:FP_Original-image}).
Since the image is contaminated by noise, our purpose is to restore
the original image $x$ given the degraded (observed) image $y$. 

We employ the Bayesian image restoration model of \citep{Besag1977,GemanGeman1984,Besag1986}
which is based on the Ising model \citep{Ising1925} and has posterior
distribution with density $\pi(x|y)$ such that 
\begin{equation}
\pi(x|y)\propto\exp(a\sum_{\forall i}\indfun_{\{y_{i}\}}(x_{i})+b\sum_{\forall i\sim j}\indfun_{\{x_{j}\}}(x_{i})),\label{eq:log_posterior_Ising}
\end{equation}
 where $a>0$, and $b>0$ are fixed parameters (here, $a=1.1$, and
$b=0.9$). The symbol `$\sim$' denotes the neighbourhood of the eight
adjacencies (vertical, horizontal, and diagonal) of each interior
pixel. In Eq. \ref{eq:log_posterior_Ising}, the first term is associated
to the likelihood and encourages states $x_{i}$ to be identical to
the observed pixel $y_{i}$, while the second term is associated to
the Ising prior model, encourages neighbouring pixels to be equal
and hence provides smoothing. In this context, image restoration can
be achieved by computing the maximum a posteriori (MAP) estimate of
the original image which can be found by minimising the negative log
posterior density, $U_{4}(x):=-\log(\pi(x|y))$ \citep{GemanGeman1984}. 

Computational difficulties raise when algorithms based on standard
MCMC samplers with component-wise structure of single-pixel updates
are employed \citep{Higdon1998}. Such an update design tends to either
converge slow or get trapped because the prior term in (\ref{eq:log_posterior_Ising})
strongly prefers large blocks of pixels. This issue becomes even more
serious for large values of $b$ which favour strong dependencies.
Against this application, we compare the PISAA, VFSA, and PSAA. PSAA
refers to the parallel SAA, a multiple-chain implementation of SAA
that involves running a number of standard SAA procedures with the
same algorithmic settings in parallel and completely independently.
PISAA and PSAA use the following algorithmic settings: (i) $n=5\cdot10^{5}$
iterations, (ii) uniformly spaced grid $\{u_{j}\}$ with $m=200$,
$u_{1}=-826315.5$, $u_{100}=-971500.5$, (iii) desirable probability
with parameter $\lambda=0.1$, (iv) temperature ladder $\{\tau_{t}\}$
with $\tau_{h}=5$, $n^{(\tau)}=10^{3},$ $\tau_{*}=10^{-2}$, (iv)
gain factor $\{\gamma_{t}\}$ with $n^{(\gamma)}=10^{3},$ $\beta=0.55$.
The MCMC kernel of PISAA is designed to be a random scan of a Gibbs
update (updating one pixel at a time) and $k$-point crossover operations
(where $k=2$). VFSA and SAA use only Gibbs updates.

\begin{figure}
\subfloat[Average progression curves of the best function values discovered
by PISAA with different population sizes. \label{fig:FP_Average-progression}]{\includegraphics[scale=0.19]{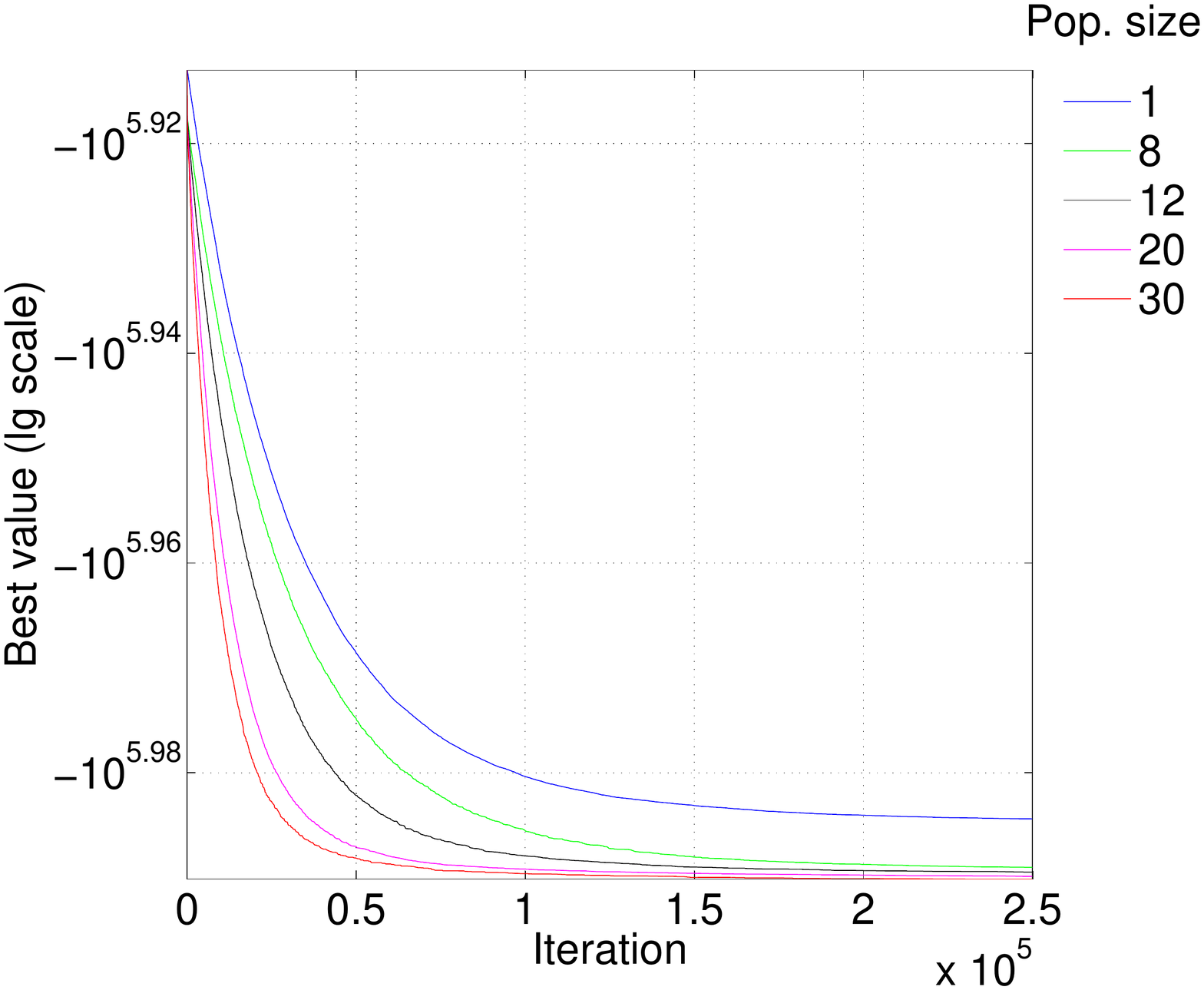}

}\subfloat[Average progression curves of the best function values discovered
by PISAA, PSAA, and VFSA with population size $20$. \label{fig:FP_Average-progression-curves}]{\includegraphics[scale=0.19]{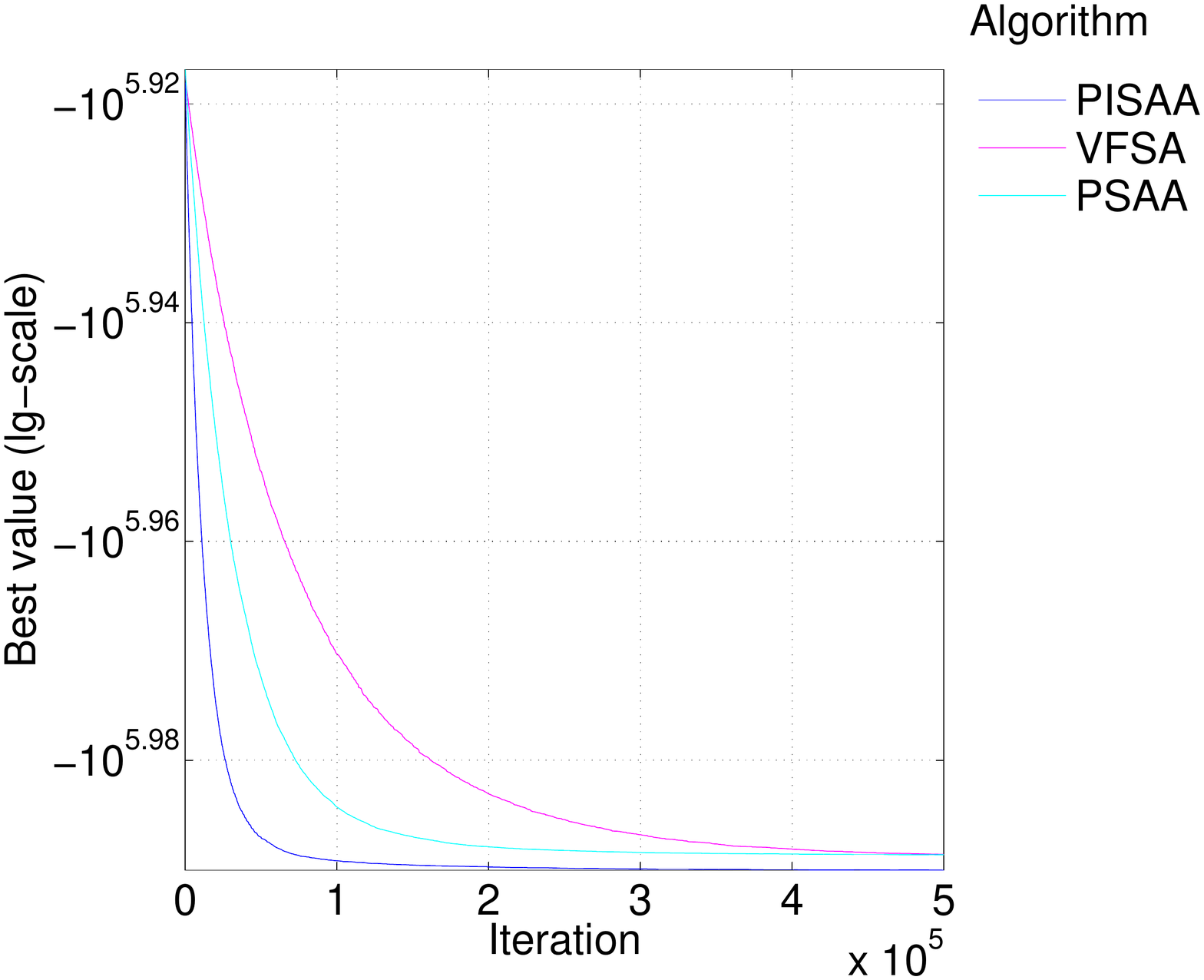}

}\subfloat[Average best function values generated by PISAA, PSAA, and VFSA against
the population size. \label{fig:FP_Best-function-values}]{\includegraphics[scale=0.19]{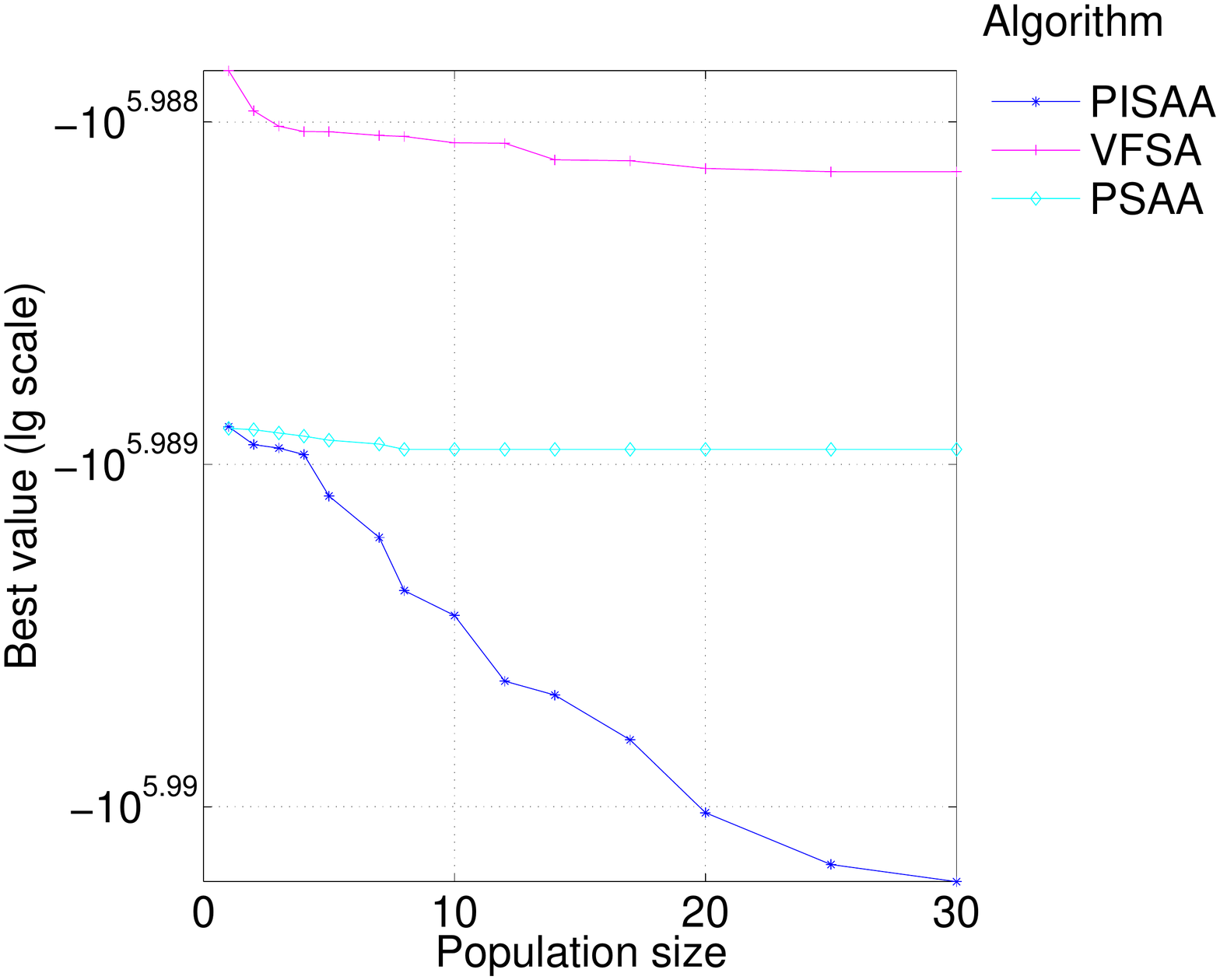}

}

\protect\caption{(Section \ref{sub:Spatial-imaging}) Performance and comparison plots
of PISAA, PSAA, and VFSA. \label{fig:FP_Performance-and-comparison}}
\end{figure}

We observe that PISAA discovers quicker smaller best values when the
population size increases (Figures \ref{fig:FP_Average-progression},
and \ref{fig:FP_Best-function-values}). Figure \ref{fig:FP_Best-function-values}
shows that PISAA converges quicker than PSAA as the number of the
parallel chains involved increases. This implies that it is preferable
to run a PISAA with a population size $\kappa>1$ rather than run
$\kappa$ SAA procedures completely independent from each other. Moreover,
it shows that the interacting character of PISAA is a necessary ingredient
for significantly improving the performance of the algorithm by increasing
the population size. By `interacting character' of PISAA, we refer
to the distinctive way that the crossover operations and self-adjusting
mechanism of PISAA use the distributed information gained from all
the population chains to operate. Moreover, we observe that PISAA
outperforms VFSA (Figures \ref{fig:FP_Average-progression-curves},
and \ref{fig:FP_Best-function-values}). Finally, the MAP estimate
of the original image as computed by PISAA with population size $30$
is shown in Figure \ref{fig:FP_MAP-estimate}.

\subsection{Bayesian network learning\label{sub:Bayesian-network-learning}}

The Bayesian network \citep{EllisWong2008} is a directed acyclic
graph (DAG) whose nodes represent variables in the domain, and edges
correspond to direct probabilistic dependencies between them. It is
a powerful knowledge representation and reasoning tool under conditions
of uncertainty that is typical of real-life applications. Mathematically,
it can be defined as a pair $B=(\mathcal{G},\rho)$, where $\mathcal{G}=(\mathcal{V},\mathcal{E})$
is a DAG representing the structure of the network, $\mathcal{V}$
denotes the set of nodes, $\mathcal{E}$ denotes the set of edges,
and $\rho$ is the vector of the associated conditional probabilities.
In the discrete case we consider here, $V:=\{V_{i};i=1:d\}\in\mathcal{V}$
denotes a node that takes values in a finite set $\{v_{j};j=1:r_{i}\}$,
$r_{i}\in\mathbb{N}-\{0\}$ and hence $V$ is assumed to be a categorical
variable. Therefore, there are $q_{i}=\prod_{V_{j}\in\text{pa}(V_{i})}r_{j}$
possible values for the joint state of the parents of $V$, where
$\text{pa}(V_{i})$ denotes the set of parents of $V_{i}$ node. In
this example, we consider the prior model of \citet{EllisWong2008,LiangZhang2009},
and hence we focus our interest in the marginal posterior probability
$\Pr(\mathcal{G}|\mathcal{D})$ such that 
\begin{equation}
\Pr(\mathcal{G}|\mathcal{D})\propto\prod_{i=1}^{d}(\frac{b}{1-a})^{\left|\text{pa}(V_{i})\right|}\prod_{k=1}^{q_{i}}\frac{\Gamma(a_{i,j,k})}{\Gamma(\sum_{j=1}^{r_{i}}a_{i,j,k}+n_{i,j,k})}\prod_{j=1}^{r_{i}}\frac{\Gamma(a_{i,j,k}+n_{i,j,k})}{\Gamma(a_{i,j,k})},\label{eq:DBN_margpostpdf}
\end{equation}
 where $\mathcal{D}=\{V_{i};i=1:N\}$ denotes the data set, considered
to be IID samples, $n_{i,j,k}$ denotes the number of samples for
which $V_{i}$ is in state $j$ and $\text{pa}(V_{i})$ is in state
$k$, $a_{i,j,k}=(r_{i}q_{i})^{-1}$ \citep{EllisWong2008}, and $b\in(0,1)$
(here, $b=0.1$ \citep{LiangZhang2009}). The negative log-posterior
distribution function, or else energy function, of the Bayesian network
is $U_{5}(\mathcal{G}):=-\log(\Pr(\mathcal{G}|\mathcal{D}))$. 

Existing methods for learning Bayesian networks include conditional
independence tests \citep{WermuthLauritzen1982}, optimisation \citep{HeckermanGeigerChickering1995},
and MCMC simulation \citep{MadiganRaftery1994,LiangZhang2009} approaches.
Often interest lies in finding the maximum a posteriori (MAP) putative
network that can be performed by minimising the negative log-posterior
distribution density $U_{5}(\cdot)$. Deterministic optimisation procedures
often stop at local optima structures. Standard MCMC based approaches,
although seemingly more attractive \citep{LiangZhang2009}, are still
prone to get trapped in local energy minima indefinitely. This is
because the energy landscape of the Bayesian network can be quite
rugged, with a multitude of local energy minima being separated by
high energy barriers, especially when the network size is large. Here,
we examine the performance of PISAA against this challenging optimisation
problem.

We consider the Single Proton Emission Computed Tomography (SPECT)
data set \citep{CiosWeddingLiu1997,KurganCiosTadeusiewiczOgielaGoodenday2001},
available at UC Irvine Machine Learning Repository %
\footnote{\texttt{http://archive.ics.uci.edu/ml}, unless changed%
} that describes diagnosing of cardiac SPECT images. It includes $267$
SPECT image sets (patients) processed to obtain $22$ binary feature
patterns that summarise the original SPECT images. Each patient is
classified into two categories: normal, and abnormal. 

We examine the performance of PISAA as a function of the iterations
and the population size, and compare it with those of PSAA, and VFSA.
PISAA uses algorithmic settings: (i) $n=2\cdot10^{8}$ iterations,
(ii) uniformly spaced grid $\{u_{j}\}$ with $m=2001$, $u_{1}=2000$,
$u_{2001}=3999$, (iii) desirable probability with parameter $\lambda=0.05$,
(iv) temperature ladder $\{\tau_{t}\}$ with $\tau_{h}=50$, $n^{(\tau)}=1,$
$\tau_{*}=10^{-1}$, (iv) gain factor $\{\gamma_{t}\}$ with $n^{(\gamma)}=10^{6},$
$\beta=0.55$. The MCMC kernel is designed to be a random scan of
mutation operations only (temporal order, skeletal and double skeletal
suggested by \citep{LiangZhang2009,WallaceKorb1999}) with equal operation
rates. PSAA and VFSA share common settings with PISAA. Each simulation
runs for $48$ times to eliminate output variations caused by nuisance
factors.

\begin{figure}
\subfloat[Average progression curves of the best function values discovered
by PISAA with different population sizes. \label{fig:BNLDD_Average-progression_PISAA}]{\includegraphics[scale=0.19]{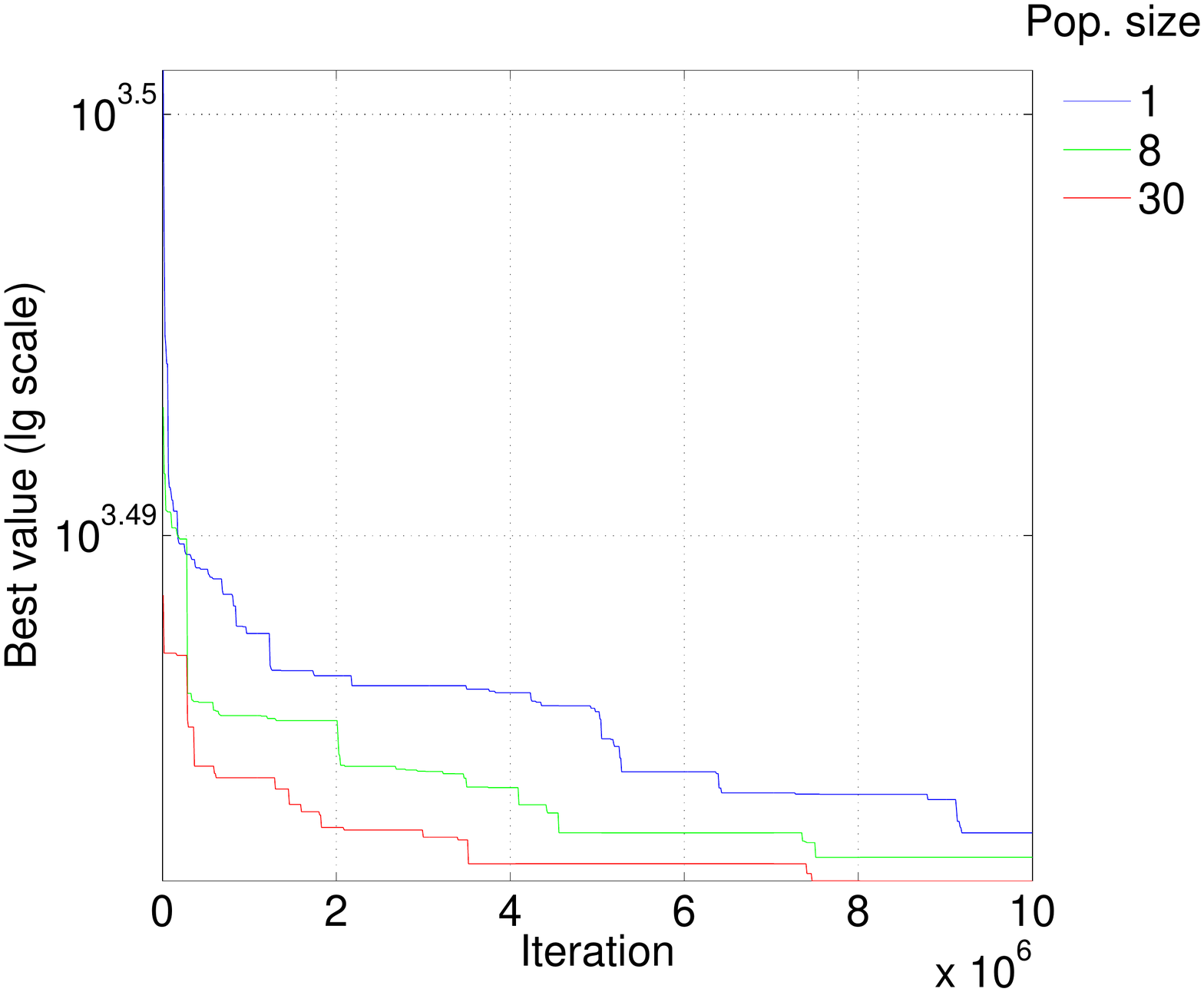}

}\subfloat[Average progression curves of the best function values discovered
by PISAA, PSAA, and VFSA with population size $30$. \label{fig:BNLDD_Average-progression_ALGS}]{\includegraphics[scale=0.19]{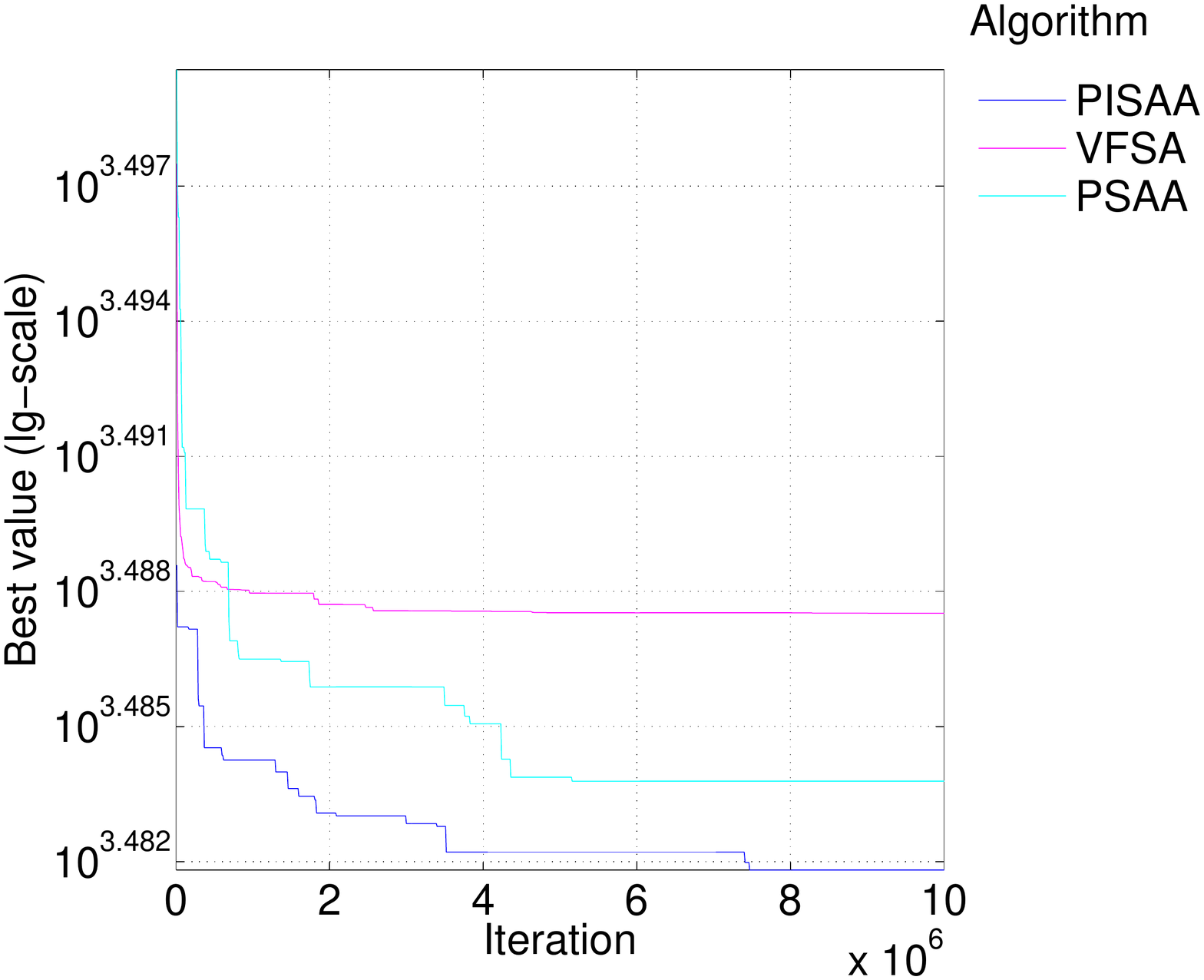}

}\subfloat[Average best function values discovered by PISAA, PSAA, and VFSA against
the population size. \label{fig:BNLDD_Best-function-values_POP}]{\includegraphics[scale=0.19]{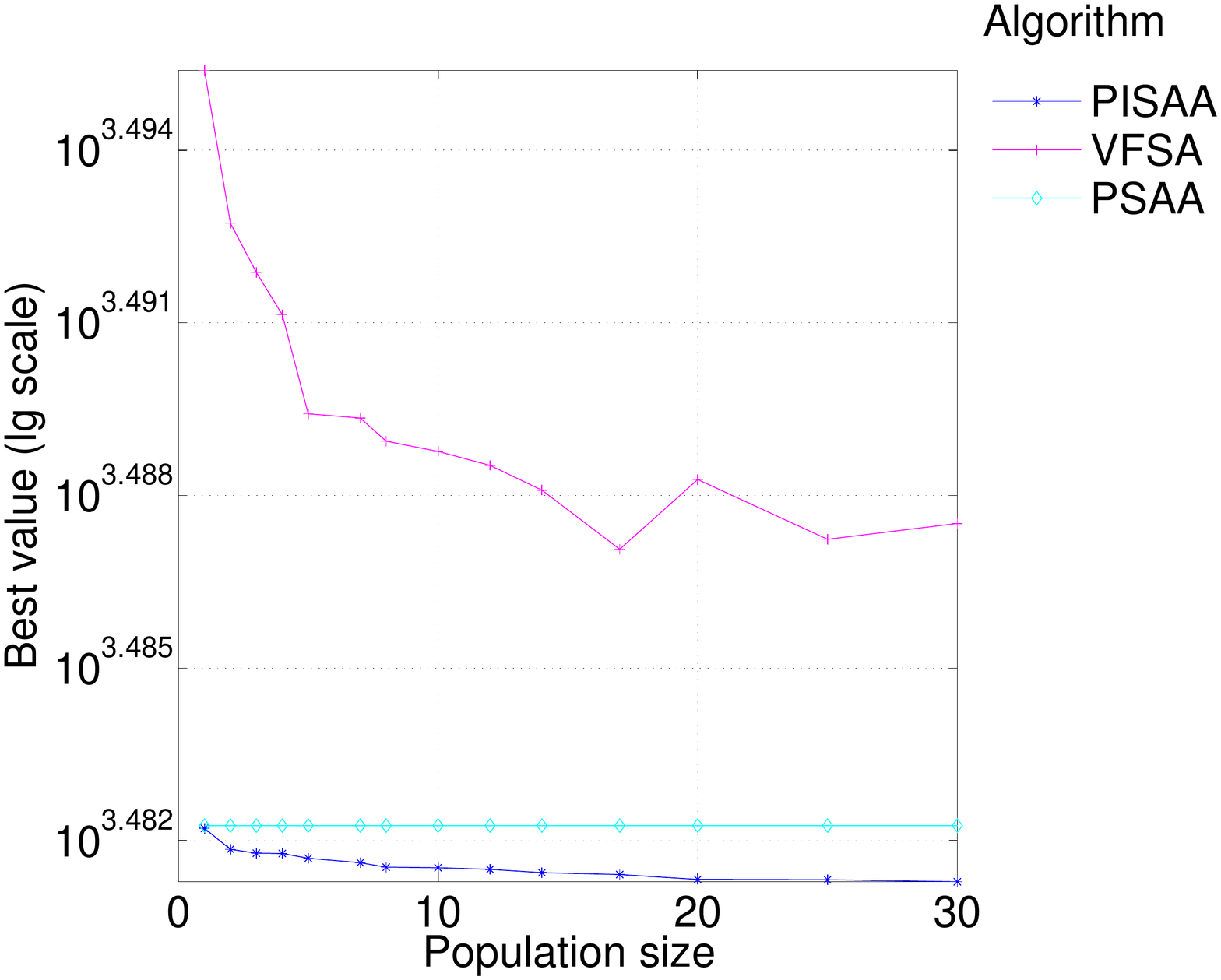}

}

\protect\caption{(Section \ref{sub:Bayesian-network-learning}) Performance and comparison
plots of PISAA, PSAA, and VFSA. \label{fig:BNLDD_Performance-and-comparison} }
\end{figure}

Figure \ref{fig:BNLDD_Average-progression_PISAA} presents the average
progression curves of the best values discovered by PISAA at different
population sizes. We observe that increasing the population size accelerates
the convergence of the algorithm towards smaller best values. Figure
\ref{fig:BNLDD_Average-progression_ALGS} shows the best function
values discovered by PISAA, PSAA, and VFSA using $30$ chains each.
We observe that PISAA tends to discover smaller best values quicker
than PSAA and VFSA. In Figure \ref{fig:BNLDD_Best-function-values_POP},
we present the best values discovered by the algorithms under comparison
after $2\cdot10^{8}$ iterations as functions of the population size.
We observe that PISAA has discovered smaller best values than PSAA
and VFSA. A reader, non-familiar to the Bayesian network modelling,
might argue that the observed improvement in performance of PISAA
due the population size increase is not that eye-catching in Figure
\ref{fig:BNLDD_Best-function-values_POP} because of the decisively
small slope of the curve. In Bayesian networks \citep{LiangZhang2009},
even slightly different negative log-posterior probabilities can correspond
to very different network structures leading to different statistical
inferences. 

\begin{SCfigure} 
\caption{\label{fig:MAP_SPECT} \small (Section \ref{sub:Bayesian-network-learning}) MAP estimate $\mathcal{G}_{\text{MAP}}$ of the putative network, as computed by running PISAA with population size 25 for $2\cdot 10^{8}$ iterations.
($U_{5}(\mathcal{G}_{\text{MAP}})=3026.935103$)\\ \\
The data set considers $267$ cardiac Single Proton Emission Computed Tomography (SPECT) images and particularly variables that corespond to features :\\ \\ 
The overal diagnosis, coded as `Overal diagnosis', that is a class attribute with values `normal' and `abnormal', \\ \\ 
The $j$-th partial diagnosis, coded as `F$j$', that takes values `normal' and `abnormal', where $j=1,...,22$. } 
\includegraphics[scale=0.2,angle =-90]{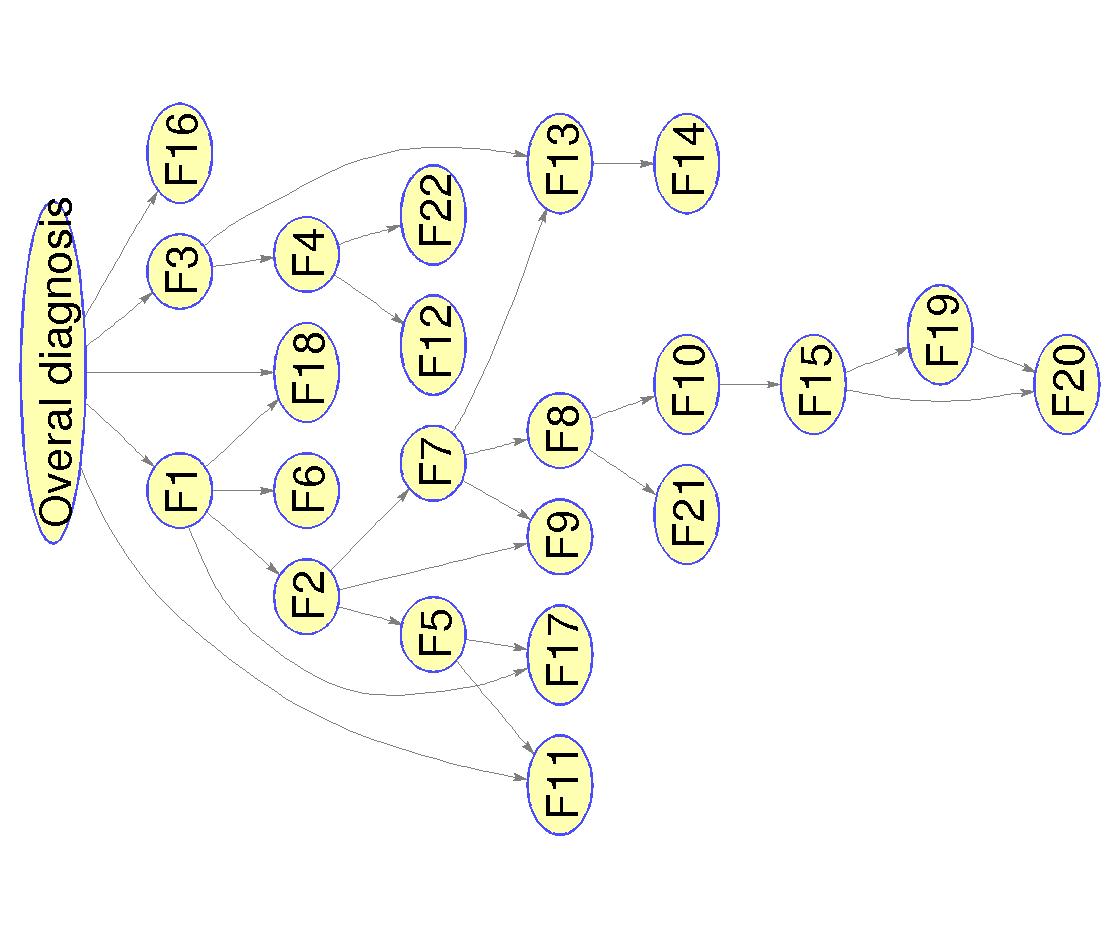}
\end{SCfigure}

The MAP putative network computed by running PISAA with population
size $30$ and $2\cdot10^{8}$ iterations is shown in Figure \ref{fig:MAP_SPECT}

\section{Summary and conclusions\label{sec:Conclusion}}

We developed the parallel and interacting stochastic approximation
annealing (PISAA) algorithm, a stochastic simulation procedure for
global optimisation, that builds upon the ideas of the stochastic
approximation annealing and population Monte Carlo samplers. PISAA
inherits from SAA a remarkable self-adjusting mechanism that operates
based on past samples and facilitates the system to escape from local
traps. Furthermore, the self-adjusting mechanism of PISAA is more
accurate and stable because it uses information from all the population
of chains. Yet, the sampling mechanism of PISAA is more effective
because it allows the use of advanced MCMC transitions such as the
crossover operations. Furthermore, it breaks sampling into multiple
parallel procedures able to search for minima at different sampling
space regions simultaneously. This allows PISAA to demonstrate a remarkable
performance, and be able to address challenging optimisation problems
with high dimensional and rugged cost functions that it would be quite
difficult for SAA to tackle acceptably. The computational overhead
due to the generation of multiple chains can be reduced dramatically
if parallel computing environment is available. 

We examined empirically the performance of PISAA against several challenging
optimisation problems. We observed that PISAA significantly outperforms
SAA in terms of convergence to the global minimum as it effectively
mitigates the problematic behaviour of SAA. Our results suggested
that, as the population size increases, the performance of PISAA improves
significantly in terms of discovering the global minimum and adjusting
the target density. Precisely, when the population size increases,
PISAA discovers the global minimum quicker, and the adjustment of
the target density is more stable. More importantly, we observed that
instead of running several SAA procedures completely independently,
it is preferable to run one PISAA procedure with the same number of
chains (or equiv. population size). In our examples, PISAA significantly
outperformed other competitors, such as SA and ASAMC, and their population
analogues, such as VFSA and AESAMC. In fact, it was observed that
as the population size increases, the performance of PISAA improves
significantly quicker than that of VFSA and AESAMC.

Under the framework of PISAA, we showed that theoretical results of
\citet{SongWuLiang2014} for pop-SAMC regarding the asymptotic efficiency
of the estimates of the unknown bias weights hold for PISAA as well,
and presented theoretical results of \citet{LiangChengLin2013} for
SAA regarding the convergence of the algorithm that hold for PISAA
as well. The empirical results confirmed that PISAA produces correct
estimates for the unknown bias weights $w_{*}$ as $\tau_{t}\rightarrow\tau_{*}$,
and that the efficiency of these estimates significantly improves
as the population size increases. Moreover, the theoretical limiting
ratio between the rates of convergence of their estimates generated
by PISAA and SAA was also confirmed by our empirical results.

Another important use of PISAA could be that of sampling from multi-modal
distributions and then performing inference via importance sampling
methods. PISAA can be extended to use an adaptive binning strategy
for automatically determining the partition of the sampling space
similar to \citep{BornnJacobMoralDoucet2013}, or a smoothing method
to estimate the frequency of visiting each subregion similar to \citep{Liang2009}.
Of particular interest would be to extend PISAA so that it can allow
different partition schemes and desired probabilities for each population
individual while ensuring the stability of the self-adjusted mechanism.

\section*{Supplementary material}

Supplementary material for the article is available online. 
\begin{lyxlist}{00.00.0000}
\item [{Appendix}] The appendix contains:

\begin{itemize}
\item Theoretical analysis of PISAA.
\item The pseudo-algorithms of the MCMC kernel mutation and MCMC kernel
crossover operations considered in the examples (Section \ref{sec:Applications})
\end{itemize}
\end{lyxlist}

\section*{Acknowledgements}

This work was supported by National Science Foundation Grant DMS-1115887,
and the United States Department of Energy, Office of Science, Office
of Advanced Scientific Computing Research, Applied Mathematics program
as part of the Collaboratory on Mathematics for Mesoscopic Modeling
of Materials, and Multifaceted Mathematics for Complex Energy Systems.
The research was performed by using the National Energy Research Scientific
Computing Center at Lawrence Berkeley National Laboratory.

\bibliographystyle{chicago}
\bibliography{paper}

\clearpage

\clearpage{}

\appendix

\section*{Appendix}

\section{Theoretical analysis of PISAA \label{sec:Theoretical-justification}}

\input{appendix_theory.tex}

\clearpage{}

\section{MCMC kernel crossover operations used in Section \ref{sec:Applications}\noun{\label{sec:Appendix_MCMC_operations}}}

\input{appendix_operations.tex}

\end{document}

%% file: appendix_theory.tex
The PISAA algorithm falls into the general class of the stochastic
approximation MCMC (SAMCMC) algorithms. In order to study the convergence
of PISAA, we adopt the technique developed by \citet{ChenZhu1986}.
Traditionally, the convergence of such algorithms is studied by reformulating
the equation in Step 2 of Algorithm \ref{Alg:PISAA} as $\theta'=\theta_{t-1}+\gamma_{t}(h_{\tau_{t}}^{(\kappa)}(\theta_{t-1})+\xi_{t}^{(\kappa)})$,
where $h_{\tau_{t}}^{(\kappa)}(\theta_{t-1})=\int H_{\tau_{t}}^{(\kappa)}(\theta_{t-1},x^{(1:\kappa)})f_{\theta_{t-1},\tau_{t}}^{(\kappa)}(x^{(1:\kappa)})\drv x^{(1:\kappa)}$
is called the mean field function, and $\xi_{t}^{(\kappa)}=H_{\tau_{t}}^{(\kappa)}(\theta_{t-1},x_{t}^{(1:\kappa)})-h_{\tau_{t}}^{(\kappa)}(\theta_{t-1})$
is called the observational noise.

Similar to SAA, PISAA solves the integral equation $h_{\tau_{*}}^{(\kappa)}(\theta)=0$
in the context of stochastic approximation, by solving sequentially
the system of equations $\{h_{\tau_{t}}^{(\kappa)}(\theta)=0;\ t=1,2,...\}$
defined along the temperature sequence $\{\tau_{t}\}$. The idea is
that if $\{\tau_{t}\}$ does not decrease too fast, the solution of
$h_{\tau_{t}}^{(\kappa)}(\cdot)=0$ can be used as an initial guess
for $h_{\tau_{t+1}}^{(\kappa)}(\cdot)=0$. Thus, in the limit, the
convergence $\theta_{t}\rightarrow\theta_{*}$ can hold under appropriate
conditions, where $\theta_{*}$ is the solution of the equation of
interest. For mathematical simplicity, in what follows, we treat the
temperature $\tau\in\mathcal{T}$ as a continuous variable instead
of a sequence, and assume that $\mathcal{T}$ is compact, $\mathcal{T}=[\tau^{*},\tau_{1}]$.
For parameter $\theta\in\Theta$, we assume $\Theta=\mathbb{R}^{m}$
where $m$ is the number of subregions.

For PISAA, we have
\begin{align}
h_{\tau}^{(\kappa)}(\theta) & =\int_{\mathcal{X}^{\kappa}}H_{\tau}^{(\kappa)}(\theta,x^{(1:\kappa)})f_{\theta,\tau}^{(\kappa)}(x^{(1:\kappa)})\drv x^{(1:n)};\label{eq:Mean_field_funtion_equality}\\
 & =\int_{\mathcal{X}^{\kappa}}[\frac{1}{\kappa}\sum_{i=1}^{\kappa}H_{\tau}(\theta,x^{(i)})]\prod_{j=1}^{\kappa}f_{\theta,\tau}(x^{(j)})\drv x^{(1:\kappa)};\nonumber \\
 & =\frac{1}{\kappa}\sum_{i=1}^{\kappa}\int_{\mathcal{X}}H_{\tau}(\theta,x^{(i)})f_{\theta,\tau}(x^{(i)})\drv x^{(i)};\nonumber \\
 & =\frac{1}{\kappa}\sum_{i=1}^{\kappa}h_{\tau}(\theta);\nonumber \\
 & =h_{\tau}(\theta),\nonumber 
\end{align}
 where $h_{\tau}(\theta)$ is the mean field function of SAA \citep{LiangChengLin2013}.
Likewise, it is easy to show that $\Var_{f_{\theta_{t-1},\tau_{t}}^{(\kappa)}}(\xi_{t}^{(\kappa)})=\frac{1}{\kappa}\Var_{f_{\theta_{t-1},\tau_{t}}^{(1)}}(\xi_{t}^{(1)})$.
Thus, for $\kappa\in\mathbb{N}-\{0\}$, PISAA solves the same set
of integration equations as the single-chain SAA, while reducing the
variation in the mean field approximation. Note that, if $\kappa=1$,
PISAA reduces to the single-chain SAA.

\subsection{Conditions for PISAA\label{sub:Conditions-for-PISAA}}

The convergence of PISAA is studied under conditions ($A_{1}$ - $A_{4}$)
assumed for the mean field function, observation noise, gain factor,
and temperature sequence. We recall from (\ref{eq:Mean_field_funtion_equality})
that $h_{\tau}^{(\kappa)}(\theta)=h_{\tau}(\theta)$ for $\kappa\ge1$.
To easy the notation we suppress indexes $\cdot^{(\kappa)}$, and
$\cdot^{(1:k)}$, when no confusion is caused. 
\begin{description}
\item [{$(A_{1})$}] (Lyapunov condition) 

\begin{description}
\item [{(i)}] The function $h_{\tau}(\theta)$ is bounded and continuously
differentiable with respect to both $\theta$ and $\tau$, and there
exists a non-negative, upper bounded, and continuously differentiable
function $v_{\tau}(\theta)$ such that for any $\Delta>\delta>0$,
\begin{equation}
\sup_{\delta\leq d((\theta,\tau),\mL)\leq\Delta}\nabla_{\theta}^{T}v_{\tau}(\theta)h_{\tau}(\theta)<0,\label{eq2.2.1}
\end{equation}
where $\mL=\{(\theta,\tau):h_{\tau}(\theta)=0,\theta\in\Theta,\tau\in\mT\}$
is the zero set of $h_{\tau}(\theta)$, and $d(z,S)=\inf_{y}\{\|z-y\|:y\in S\}$.
Further, the set $v(\mL)=\{v_{\tau}(\theta):(\theta,\tau)\in\mL\}$
is nowhere dense.
\item [{(ii)}] Both $\nabla_{\theta}v_{\tau}(\theta)$ and $\nabla_{\tau}v_{\tau}(\theta)$
are bounded over $\Theta\times\mT$. In addition, for any compact
set $\mK\subset\Theta$, there exists a constant $0<c<\infty$ such
that 
\begin{equation}
\begin{split}\sup_{(\theta,\theta')\in\mK\times\mK,\tau\in\mT} & \|\nabla_{\theta}v_{\tau}(\theta)-\nabla_{\theta}v_{\tau}(\theta')\|\leq c\|\theta-\theta'\|,\\
\sup_{\theta\in\mK,(\tau,\tau')\in\mT\times\mT} & \|\nabla_{\theta}v_{\tau}(\theta)-\nabla_{\theta}v_{\tau'}(\theta)\|\leq c|\tau-\tau'|,\\
\sup_{\theta\in\mK,(\tau,\tau')\in\mT\times\mT} & \|h_{\tau}(\theta)-h_{\tau'}(\theta)\|\leq c|\tau-\tau'|.
\end{split}
\label{Lyapeq2}
\end{equation}

\end{description}
\item [{$(A_{2})$}] (Doeblin condition)

For any given $\theta\in\Theta$ and $\tau\in\mT$, the Markov transition
kernel $P_{\theta,\tau}$ is irreducible and aperiodic. In addition,
there exist an integer $l$, $0<\delta<1$, and a probability measure
$\nu$ such that for any compact subset $\mK\subset\Theta$, 
\[
\inf_{\theta\in\mK,\tau\in\mT}P_{\theta,\tau}^{l}(x,A)\geq\delta\nu(A),\quad\forall x\in\mX,\ \forall A\in\mB_{\mX},
\]
where $\mB_{\mX}$ denotes the Borel set of $\mX$; that is, the whole
support $\mX$ is a \textit{small} set for each kernel $P_{\theta,\tau}$,
$\theta\in\mK$ and $\tau\in\mT$. 

\item [{$(A_{3})$}] (Stability Condition on $h_{\tau}(\theta)$)

For any value $\tau\in\mT$, the mean field function $h_{\tau}(\theta)$
is measurable and locally bounded on $\Theta$. There exist a stable
matrix $F_{\tau}$ (i.e., all eigenvalues of $F_{\tau}$ are with
negative real parts), $\rho>0$, and a constant $c$ such that, for
any $(\theta_{*},\tau)\in\mL$ (defined in $A_{1}$), 
\[
\|h_{\tau}(\theta)-F_{\tau}(\theta-\theta_{*})\|\leq c\|\theta-\theta_{*}\|^{2},\quad\forall\ \theta\in\{\theta:\|\theta-\theta_{*}\|\leq\rho\}.
\]

\item [{$(A_{4})$}] (\textit{Conditions on $\{\gamma_{t}\}$ and $\{\tau_{t}\}$}) 

\begin{description}
\item [{(i)}] The sequence $\{\gamma_{t}\}$, which is defined to be $\gamma(t)$
as a function of $t$ and is exchangeable with $\gamma(t)$ in this
paper, is positive, non-increasing and satisfies the following conditions:
\begin{equation}
\sum_{t=1}^{\infty}\gamma_{t}=\infty,\quad\frac{\gamma_{t+1}-\gamma_{t}}{\gamma_{t}}=O(\gamma_{t+1}^{\iota}),\quad\sum_{t=1}^{\infty}\frac{\gamma_{t}^{(1+\iota')/2}}{\sqrt{t}}<\infty,\label{coneq1}
\end{equation}
for some $\iota\in[1,2)$ and $\iota'\in(0,1)$.
\item [{(ii)}] The sequence $\{\tau_{t}\}$ is positive and non-increasing
and satisfies the following conditions: 
\begin{equation}
\lim_{t\to\infty}\tau_{t}=\tau_{*},\quad\tau_{t}-\tau_{t+1}=o(\gamma_{t}),\quad\sum_{t=1}^{\infty}\gamma_{t}|\tau_{t}-\tau_{t-1}|^{\iota''}<\infty,\label{coneq202}
\end{equation}
for some $\iota''\in(0,1)$, and 
\begin{equation}
\sum_{t=1}^{\infty}\gamma_{t}|\tau_{t}-\tau_{*}|<\infty,\label{coneq3}
\end{equation}

\item [{(iii)}] The function $\zeta(t)=\gamma(t)^{-1}$ is differentiable
such that its derivative varies regularly with exponent $\tilde{\beta}-1\geq-1$
(i.e., for any $z>0$, $\zeta'(zt)/\zeta'(t)\to z^{\tilde{\beta}-1}$
as $t\to\infty$), and either of the following two cases holds: 

\begin{description}
\item [{(iii.1)}] $\gamma(t)$ varies regularly with exponent $(-\beta)$,
$\frac{1}{2}<\beta<1$;
\item [{(iii.2)}] For $t\geq1$, $\gamma(t)=t_{0}/t$ with $-2\lambda_{F_{\tau}}t_{0}>\max\{1,\tilde{\beta}\}$
for any $\tau\in\mT$, where $\lambda_{F}$ denotes the largest real
part of the eigenvalue of the matrix $F_{\tau}$ (defined in condition
$A_{3}$) with $\lambda_{F_{\tau}}<0$. 
\end{description}
\end{description}
\end{description}

The Lyapunov condition ($A_{1}$) is related to the mean field function
$h_{\tau}$. The mean field function of PISAA is equal to that of
SAA as shown in (\ref{eq:Mean_field_funtion_equality}), and hence
condition ($A_{1}$) can be verified as a consequence of \citet[p. 850]{LiangChengLin2013}.
Briefly given (\ref{eq:Mean_field_funtion_equality}), it is $h_{\tau}^{(k)}(\theta)=(\frac{S_{\tau}^{(j)}(\theta)}{S_{\tau}(\theta)}-\pi_{j};j=1,...,m)$
where $S_{\tau}^{(j)}(\theta)=\sum_{j=1}^{m}e^{-U(\theta)/\tau}\drv x/e^{\theta^{(j)}}$
and $S_{\tau}(\theta)=\sum_{j=1}^{m}S_{\tau}^{(j)}(\theta)$, which
is bounded and continuously differentiable with respect to both $\theta\in\Theta$
and $\tau\in\mT$. We defined the Lyapunov function $v_{\tau}(\theta)=\frac{1}{2}\sum_{j=1}^{m}(\frac{S_{\tau}^{(j)}(\theta)}{S_{\tau}(\theta)}-\pi_{j})^{2}$,
which is non-negative, upper bounded, and continuously differentiable.
The gradient $\nabla_{\theta}v_{\tau}(\theta)$ is bounded over $\Theta\times\mT$,
following \citet[p. 318]{LiangLiuCarroll2007}; while $\nabla_{\tau}v_{\tau}(\theta)$
is bounded over $\Theta\times\mT$, provided that $U(x)$ has a finite
mean with respect to $f_{\tau}(x)$. Yet, the second partial derivatives
of $v_{\tau}(\theta)$ with respect to $\theta$ and $\tau$ are bounded
provided that $U(x)$ has a finite variance with respect to $f_{\theta,\tau}(x)$.
Then, (\ref{eq2.2.1}) is verified as in \citep{LiangLiuCarroll2007},
on the condition that the partition of the sampling space includes
at least two non-empty subregions. 

The observation noise condition ($A_{2}$) is equivalent to assuming
that the resulting Markov chain has a unique stationary and is uniformly
ergodic \citep{Nummelin2004}. It is not too restrictive for a PISAA
whose function $H_{\tau_{t}}^{(\kappa)}(\theta_{t-1},x^{(1:\kappa)})$
is bounded, and thus the mean-field function and observation noise
are bounded. Condition ($A_{2}$) is satisfied if $\mathcal{X}$ is
compact, $U(x)$ is bounded, and the proposal distribution used to
simulate from $P_{\theta,\tau}$ satisfies the local positive condition
$(Q)$: ``There exists $\delta_{q}>0$ and $q>0$ such that, for
every $x\in X$, $|x-y|\le\delta_{q}\Rightarrow q(x,y)\ge q$'';
following \citep[Theorem 2.2 of ][]{Roberts1996}. Condition ($A_{2}$)
may also be verified in cases that $\mathcal{X}$ is not compact,
e.g. \citep{Rosenthal1995}. Multistep Metropolis-Hastings moves,
such as those mentioned in Section \ref{sec:Parallel-interacting-stochastic-approximation-annealing},
can be shown to satisfy ($A_{2}$); see \citep[Lemma 7 of ][]{Rosenthal1995}
and \citep{Liang2009}. If ($A_{2}$) holds for the single-chain kernel
$P_{\theta,\tau}$, it must hold for the multiple-chain one as well;
see \citep[Supplementary material of ][]{SongWuLiang2014}. 

Condition ($A_{3}$) constrains the behaviour of the mean field function
around the solution points.

We remark that $(A_{4})$-(iii) can be applied to the usual gains
$\gamma_{t}=t_{0}/t^{\beta}$, $1/2<\beta\leq1$. Following \citet{Pelletier1998},
we deduce that 
\begin{equation}
\left(\frac{\gamma_{t}}{\gamma_{t+1}}\right)^{1/2}=1+\frac{\beta}{2t}+o(\frac{1}{t}).\label{Peleq1}
\end{equation}
In terms of $\gamma_{t}$, (\ref{Peleq1}) can be rewritten as 
\begin{equation}
\left(\frac{\gamma_{t}}{\gamma_{t+1}}\right)^{1/2}=1+\zeta\gamma_{t}+o(\gamma_{t}),\label{Peleq2}
\end{equation}
where $\zeta=0$ for the case (iii.1) and $\zeta=\frac{1}{2t_{0}}$
for $\beta=1$ for the case (iii.2). Clearly, the matrix $F_{\tau}+\zeta I$
is still stable. Furthermore, condition $(A_{4})$-(ii) implies that
$\{\tau_{t}\}$ cannot decrease too fast, and should be set according
to the gain factor sequence $\{\gamma_{t}\}$. A choice of $\tau_{t}=\frac{t_{1}}{\sqrt{t}}+\tau_{*}$,
with $t_{1}>0$, satisfies ($A_{4}$)-(ii).

\subsection{Main theorems hold in PISAA framework \label{sub:Main-theorems-applied}}

Under the conditions ($A_{1}$ - $A_{4}$), the following theorems
for the convergence of PISAA hold. Since Theorems \ref{mainth1},
\ref{SAAth2} and \ref{normalitytheorem} are applicable to both the
PISAA and single-chain SAA algorithms, we let $X_{t}$ denote the
sample(s) drawn at iteration $t$ and let $\BX$ denote the sample
space of $X_{t}$. For the PISAA algorithm, we have $\BX=\mX^{\kappa}$
and $X_{t}=x_{t}^{(1:k)}$. For the single-chain SAA algorithm, we
have $\BX=\mX$ and $X_{t}=x_{t}$. For any measurable function $f$:
$\BX\rightarrow\mathbb{R}^{d}$, $\bP_{\theta}f(X)=\int_{\BX}\bP_{\theta}(X,y)f(y)\drv y$.
\begin{thm}
\label{mainth1} (Restatement of Theorems 3.1 and 3.2 of \citet{LiangChengLin2013})
Assume that $\mT$ is compact and the conditions $(A_{1})$, $(A_{2})$,
$(A_{4})$-(i) and $(A_{4})$-(ii) hold. If $\tilde{\theta}_{0}$
used in the PISAA algorithm is such that $\sup_{\tau\in\mT}v_{\tau}(\tilde{\theta}_{0})<\inf_{\|\theta\|=c_{0},\tau\in\mT}v_{\tau}(\theta)$
for some $c_{0}>0$ and $\|\tilde{\theta}_{0}\|<c_{0}$, then the
number of truncations in PISAA is almost surely finite; that is, $\{\theta_{t}\}$
remains in a compact subset of $\Theta$ almost surely. In addition,
as $t\to\infty$, 
\[
d(\theta_{t},\mL_{\tau_{*}})\to0,\ \ \ a.s.,
\]
where $\mL_{\tau_{*}}=\{\theta\in\Theta:h_{\tau_{*}}(\theta)=0\}$
and $d(z,S)=\inf_{y}\{\|z-y\|:y\in S\}$.
\end{thm}

\begin{thm}
\label{SAAth2} (Restatement of Theorem 3.3 of \citet{LiangChengLin2013})
Assume the conditions of Theorem \ref{mainth1} hold. Let $x_{1},\ldots,x_{n}$
denote a set of samples simulated by PISAA in $n$ iterations. Let
$g$: $\BX\to\bbR$ be a measurable function such that it is bounded
and integrable with respect to $f_{\theta,\tau}(x)$. Then 
\[
\frac{1}{n}\sum_{t=1}^{n}g(x_{t})\to\int_{\BX}g(x)f_{\theta_{*},\tau_{*}}(x)\drv x,\quad a.s.
\]

\end{thm}

Therefore, given conditions ($A_{1}$ - $A_{4}$) and following \citet[ Corollary 3.1]{LiangChengLin2013},
PISAA can achieve the following convergence result with any individual:
For any $\epsilon>0$, as $t\rightarrow\infty$, and $\tau_{*}\rightarrow0$
\[
\text{P}(U(X_{t})\le u_{j}^{*}+\epsilon|J(X_{t})=j)\rightarrow1,\quad a.s.,
\]
 where $J(x)=j$ if $x\in E_{j}$, and $u_{j}^{*}=\min_{x\in E_{j}}U(x)$,
for $j=1,...,m$. Namely, given a square-root cooling schedule, as
the number of iterations $t$ becomes large, PISAA is able to locate
the minima of each subregion in a single run if $\tau_{*}$ is small.

Lemma \ref{lem51} concerns the decomposition of the noise $\xi_{t+1}$
in the PISAA framework. The proof of Lemma \ref{lem51} is presented
separably in Appendix \ref{sub:Proof-of-theoretical}. The importance
of this lemma is that by using Lemma \ref{lem51}, the Theorems \ref{normalitytheorem}
and \ref{efftheorem} can be proved to hold in PISAA framework as
consequences of the results from \citep{SongWuLiang2014}. Theorem
\ref{normalitytheorem} concerns the asymptotic normality of $\theta_{t}$.
With Lemma \ref{lem51}, the proof of Theorem \ref{normalitytheorem}
can be referred to the proof of \citep[Theorem 3,][]{SongWuLiang2014}
except for some notational changes, replacing $h(\theta_{t})$ by
$h_{\tau_{t+1}}(\theta_{t})$. Theorem \ref{efftheorem} concerns
the asymptotic relative efficiency of the PISAA estimator of $\theta_{t}$
versus that of SAA. The proof of Theorem \ref{efftheorem} is the
same as that of \citep[Theorem 4,][]{SongWuLiang2014} using Theorem
\ref{normalitytheorem} and Lemma \ref{lem51}. 
\begin{lem}
\label{lem51} (Noise decomposition) Assume the conditions of Theorem
\ref{mainth1} hold. Then there exist $\mR^{d_{\theta}}$-valued random
processes $\{e_{t}\}$, $\{\nu_{t}\}$, and $\{\varsigma_{t}\}$ defined
on a probability space $(\Omega,\mF,\mP)$ such that: 

(i) $\xi_{t+1}=e_{t+1}+\nu_{t+1}+\varsigma_{t+1}$, where $\xi_{t+1}=H_{\tau_{t+1}}(\theta_{t},X_{t+1})-h_{\tau_{t+1}}(\theta_{t})$
is the observation noise. 

(ii) For any constant $\rho>0$ (defined in condition $A_{2}$), 
\begin{align*}
E(e_{t+1}|\mF_{t})1_{\{\|\theta_{t}-\theta_{*}\|\leq\rho\}} & =0\\
\sup_{t\geq0}E(\|e_{t+1}\|^{\alpha}|\mF_{t})1_{\{\|\theta_{t}-\theta_{*}\|\leq\rho\}} & <\infty,
\end{align*}
 where $\mF_{t}$ is a family of $\sigma$-algebras satisfying $\sigma\{\theta_{0},X_{0};\theta_{1},X_{1};\ldots;\theta_{t},X_{t}\}=\mF_{t}\subseteq\mF_{t+1}$
for all $t\geq0$ and $\alpha\geq2$ is a constant. 

(iii) Almost surely on $\Lambda(\theta_{*})=\{\theta_{t}\to\theta_{*}\}$,
as $n\to\infty$, 
\begin{equation}
\frac{1}{n}\sum_{t=1}^{n}E(e_{t+1}e_{t+1}'|\mF_{t})\to\Gamma,\quad\mbox{a.s.},\label{Gammours}
\end{equation}
where $\Gamma$ is a positive definite matrix. 

(iv)$E(\|\nu_{t}\|^{2}/\gamma_{t})1_{\{\|\theta_{t}-\theta_{*}\|\leq\rho\}}\to0$,
as $t\to\infty$. 

(v) $E\|\gamma_{t}\varsigma_{t}\|\rightarrow0$, as $t\rightarrow\infty$. 
\end{lem}

\begin{thm}
\label{normalitytheorem}(Consequence of \citep[Theorem 2,][]{SongWuLiang2014}
and Lemma \ref{lem51}) Assume that $\mT$ is compact and the conditions
$(A_{1})$, $(A_{2})$, $(A_{3})$ and $(A_{4})$ hold. If $\tilde{\theta}_{0}$
used in the PISAA algorithm is such that $\sup_{\tau\in\mT}v_{\tau}(\tilde{\theta}_{0})<\inf_{\|\theta\|=c_{0},\tau\in\mT}v_{\tau}(\theta)$
for some $c_{0}>0$ and $\|\tilde{\theta}_{0}\|<c_{0}$, then, Conditioned
on $\Lambda(\theta_{*})=\{\theta_{t}\to\theta_{*}\}$, 
\begin{equation}
\frac{\theta_{t}-\theta_{*}}{\sqrt{\gamma_{t}}}\Longrightarrow\mN(0,\Sigma),\label{weakeq}
\end{equation}
with $\Longrightarrow$ denoting the weak convergence, $\mN$ the
Gaussian distribution and 
\begin{equation}
\Sigma=\int_{0}^{\infty}e^{(F_{\tau_{*}}'+\zeta I)t}\Gamma e^{(F_{\tau_{*}}+\zeta I)t}\drv t,\label{Sigmaeq}
\end{equation}
where $F_{\tau_{*}}$ is defined in $(A_{2})$, $\zeta$ is defined
in (\ref{Peleq2}), and $\Gamma$ is defined in Lemma \ref{lem51}. 
\end{thm}

\begin{thm}
\label{efftheorem} (Consequence of \citep[Theorem 3,][]{SongWuLiang2014})
Suppose that both the population PISAA (with pop. size $\kappa$)
and single-chain SAA algorithms satisfy the conditions given in Theorem
\ref{normalitytheorem}. Let $\theta_{t}^{p}$ and $\theta_{t}^{s}$
denote the estimates produced at iteration $t$ by the multiple-chain
PISAA and single-chain SAA algorithms, respectively. Given the same
gain factor sequence $\{\gamma_{t}\}$, then $(\theta_{t}^{p}-\theta_{*})/\sqrt{\gamma_{t}}$
and $(\theta_{\kappa t}^{s}-\theta_{*})/\sqrt{\kappa\gamma_{\kappa t}}$
have the same asymptotic distribution with the convergence rate ratio
\begin{equation}
\frac{\gamma_{t}}{\kappa\gamma_{\kappa t}}=\kappa^{\beta-1},\label{effequation}
\end{equation}
where $\kappa$ denotes the population size, and $\beta$ is defined
in $(A_{4})$. {[}Note: $1/2<\beta<1$ for the case $A_{4}$-(iii.1)
and $\beta=1$ for the case $A_{4}$-(iii.2).{]} 
\end{thm}

\subsection{Proof of theoretical results \label{sub:Proof-of-theoretical}}

In order to prove Lemma \ref{lem51}, we introduce Lemma \ref{lem1}
which is a restatement of Lemma 1.1 of \citet[online supplement]{LiangChengLin2013}
and Proposition 6.1 of \citet{AndrieuMoulinesPriouret2005}.

\begin{lem}
\label{lem1} (Restatement of Lemma 1.1 of \citet[online supplement]{LiangChengLin2013}
and Proposition 6.1 of \citet{AndrieuMoulinesPriouret2005}) Assume
that $\mT$ is compact and the condition $(A_{2})$ holds. Then the
following results hold for the PISAA algorithm: 

$(B_{1})$ For any $\theta\in\Theta$ and $\tau\in\mT$, the Markov
kernel $P_{\theta,\tau}$ has a single stationary distribution $f_{\theta,\tau}$.
In addition, $H:\Theta\times\mX\to\Theta$ is measurable for all $\theta\in\Theta$
and $\tau\in\mT$, $\int_{\mX}\|H_{\tau}(\theta,x)\|f_{\theta,\tau}(x)\drv x<\infty$.

$(B_{2})$ For any $\theta\in\Theta$ and $\tau\in\mT$, the Poisson
equation $u_{\theta,\tau}(X)-P_{\theta,\tau}u_{\theta,\tau}(X)=H_{\tau}(\theta,X)-h_{\tau}(\theta)$
has a solution $u_{\theta,\tau}(X)$, where $P_{\theta,\tau}u_{\theta,\tau}(X)=\int_{\mX}u_{\theta,\tau}(y)P_{\theta,\tau}(X,y)dy$.
For any constant $\eta\in(0,1)$ and any compact subset $\mK\subset\Theta$,
the following results hold: 
\[
\begin{split}(i) & \quad\sup_{\theta\in\mK,\tau\in\mT}(\|u_{\theta,\tau}(\cdot)\|+\|P_{\theta,\tau}u_{\theta,\tau}(\cdot)\|)<\infty,\\
(ii) & \quad\sup_{(\theta,\theta')\in\mK\times\mK,\tau\in\mT}\|\theta-\theta'\|^{-\eta}\left\{ \|u_{\theta,\tau}(\cdot)-u_{\theta',\tau}(\cdot)\|+\|P_{\theta,\tau}u_{\theta,\tau}(\cdot)-P_{\theta',\tau}u_{\theta',\tau}(\cdot)\|\right\} <\infty.\\
(iii) & \quad\sup_{\theta\in\mK,(\tau,\tau')\in\mT\times\mT}\|\tau-\tau'\|^{-\eta}\|P_{\theta,\tau}u_{\theta,\tau}(\cdot)-P_{\theta,\tau'}u_{\theta,\tau'}(\cdot)\|<\infty.
\end{split}
\]

$(B_{3})$ For any $\eta\in(0,1)$, 
\[
\sup_{(\theta,\theta')\in\Theta\times\Theta}\|\theta-\theta'\|^{-\eta}\|h_{\tau}(\theta)-h_{\tau}(\theta')\|<\infty.
\]

\end{lem}

\pagebreak{}

\paragraph*{\uline{Proof of Lemma \mbox{\ref{lem51}}}}
\begin{proof}
(i) Define 
\begin{equation}
\begin{split}e_{t+1} & =u_{\theta_{t},\tau_{t+1}}(x_{t+1})-P_{\theta_{t},\tau_{t+1}}u_{\theta_{t},\tau_{t+1}}(x_{t}),\\
\nu_{t+1} & =\big[P_{\theta_{t+1},\tau_{t+1}}u_{\theta_{t+1},\tau_{t+1}}(x_{t+1})-P_{\theta_{t},\tau_{t+1}}u_{\theta_{t},\tau_{t+1}}(x_{t+1})\big]+\frac{\gamma_{t+2}-\gamma_{t+1}}{\gamma_{t+1}}P_{\theta_{t+1},\tau_{t+1}}u_{\theta_{t+1},\tau_{t+1}}(x_{t+1})\\
 & +\frac{\gamma_{t+2}}{\gamma_{t+1}}\big[P_{\theta_{t+1},\tau_{t+2}}u_{\theta_{t+1},\tau_{t+2}}(x_{t+1})-P_{\theta_{t+1},\tau_{t+1}}u_{\theta_{t+1},\tau_{t+1}}(x_{t+1})\big],\\
\tilde{\varsigma}_{t+1} & =\gamma_{t+1}P_{\theta_{t},\tau_{t+1}}u_{\theta_{t},\tau_{t+1}}(x_{t}),\\
\varsigma_{t+1} & =\frac{1}{\gamma_{t+1}}(\tilde{\varsigma}_{t+1}-\tilde{\varsigma}_{t+2}),
\end{split}
\label{noisedecomeq}
\end{equation}
where $u(\cdot)$ is the solution of the Poisson equation. It is easy
to verify that $\xi_{t+1}=e_{t+1}+\nu_{t+1}+\varsigma_{t+1}$ holds.

(ii) By (\ref{noisedecomeq}), we have 
\begin{equation}
E(e_{t+1}|\mF_{t})=E(u_{\theta_{t},\tau_{t+1}}(X_{t+1})|\mF_{t})-P_{\theta_{t},\tau_{t+1}}u_{\theta_{t},\tau_{t+1}}(X_{t})=0,\label{Gammaeq1}
\end{equation}
Hence, $\{e_{t}\}$ forms a martingale difference sequence. Following
from Lemma \ref{lem1}-($B_{2}$), we have 
\begin{equation}
\sup_{t\geq0}E(\|e_{t+1}\|^{\alpha}|\mF_{t})1_{\{\|\theta_{t}-\theta_{*}\|\leq\rho\}}<\infty.\label{Gammaeq2}
\end{equation}
This concludes part (ii).

(iii) By (\ref{noisedecomeq}), we have 
\begin{equation}
\begin{split}E(e_{t+1}e_{t+1}^{T}|\mF_{t}) & =E\left[u_{\theta_{t}}(X_{t+1})u_{\theta_{t}}(X_{t+1})^{T}|\mF_{t}\right]-P_{\theta_{t}}u_{\theta_{t}}(X_{t})P_{\theta_{t}}u_{\theta_{t}}(X_{t})^{T}\\
 & \stackrel{\triangle}{=}l(X_{t}).
\end{split}
\label{lem3proofeq1}
\end{equation}
It follows from Lemma \ref{lem1}-($B_{2}$) that $l(X_{k})$ is bounded,
and then it follows from Theorem \ref{SAAth2} that 
\begin{equation}
\frac{1}{n}\sum_{t=1}^{n}l(X_{t})\rightarrow\int_{\BX}l(x)f_{\theta_{*},\tau_{*}}(x)dx=\Gamma,\quad a.s.\label{Gamma}
\end{equation}
for some positive definite matrix $\Gamma$. This concludes part (iii).

(iv) By condition $(A_{3})$-(i), we have 
\[
\frac{\gamma_{t+2}-\gamma_{t+1}}{\gamma_{t+1}}=O(\gamma_{t+2}^{\tau}),
\]
for some value $\tau\in[1,2)$. By (\ref{noisedecomeq}) and ($B_{2}$)
of Lemma \ref{lem1}, there exist constants $c_{1}$, $c_{1}'$ and
$\eta\in(0.5,1)$ such that the following inequality holds, 
\[
\|\nu_{t+1}\|\leq c_{1}\|\theta_{t+1}-\theta_{t}\|+O(\gamma_{t+2}^{\tau})+c_{1}'|\tau_{t+1}-\tau_{t+2}|^{\eta}=c_{1}\|\gamma_{t+1}H_{\tau_{t+1}}(\theta_{t},X_{t+1})\|+O(\gamma_{t+2}^{\tau})+o(\gamma_{t+1}^{\eta}),
\]
which implies, by the boundedness of $H_{\tau}(\theta,\cdot)$, that
there exists a constant $c_{2}$ such that 
\begin{equation}
\|\nu_{t+1}\|\leq c_{2}\gamma_{t+1}+o(\gamma_{t+1}^{\eta}).\label{nunormeq}
\end{equation}
Therefore, 
\[
E(\|\nu_{t}\|^{2}/\gamma_{t})1_{\{\|\theta_{t}-\theta_{*}\|\leq\rho\}}\to0.
\]
This concludes part (iv).

(v) A straightforward calculation shows that 
\[
\gamma_{t+1}\varsigma_{t+1}=\tilde{\varsigma}_{t+1}-\tilde{\varsigma}_{t+2}=\gamma_{t+1}P_{\theta_{t},\tau_{t+1}}u_{\theta_{t},\tau_{t+1}}(X_{t})-\gamma_{t+2}P_{\theta_{t+1},\tau_{t+2}}u_{\theta_{t+1},\tau_{t+2}}(X_{t+1}),
\]
By $(B_{2})$, $E\left[\|P_{\theta_{t},\tau_{t+1}}u_{\theta_{t},\tau_{t+1}}(X_{t})\|\right]$
is uniformly bounded with respect to $t$. Therefore, (v) holds. 
\end{proof}

\paragraph*{\uline{Proof of Theorem \mbox{\ref{normalitytheorem}}}}
\begin{proof}
With Lemma \ref{lem51}, the proof of this theorem can be referred
to the proof of \citep[Theorem 2,][]{SongWuLiang2014} except for
some notational changes, replacing $h(\theta_{t})$ by $h_{\tau_{t+1}}(\theta_{t})$.
\end{proof}

\paragraph*{\uline{Proof of Theorem \mbox{\ref{efftheorem}}}}
\begin{proof}
The proof of this theorem is the same as that of \citep[Theorem 3,][]{SongWuLiang2014},
with using Theorem \ref{normalitytheorem} and Lemma \ref{lem51}. \end{proof}

%% file: appendix_operations.tex
Let $\kappa$ denote the population size of the population $x^{(1:\kappa)}$,
and $d$ denote the number of dimensions of each individual $x^{(i)}$
for $i=1,...,\kappa$.

The pseudo-codes of the MCMC kernel crossover operations, used in
Sections \ref{sub:Gaussian-mixture-model} - \ref{sub:Spatial-imaging},
are presented below. More details can be found in \citep{Liang2011,LiangWong2000,LiangWong2001}. 
\begin{itemize}
\item $k$-point crossover operation (continuous or discrete target distributions):

\begin{enumerate}
\item draw $i\sim\varpi_{1}^{\text{KC}}(i;x^{(1:\kappa)})$ and $j|i\sim\varpi_{2}^{\text{KC}}(j|i;x^{(1:\kappa)})$
\item draw crossover points vector $v\sim\left\{ 1,...,d-1\right\} $, without
replacement and sort them
\item design $x'^{(i)}$ and $x'^{(j)}$ from $x'^{(i)}$ and $x'^{(j)}$
by swapping their elements between each odd and the next even crossover
points
\item accept $x'^{(1:\kappa)}:=(x^{(1:i-1)},x'^{(i)},x^{(i+1:j-1)},x'^{(j)},x^{(j+1:\kappa)})$
with prob. $a_{\text{KC}}=\min(1,\frac{f_{\theta,\tau}(x'^{(i)}|\mathcal{E})}{f_{\theta,\tau}(x^{(i)}|\mathcal{E})}\frac{f_{\theta_{t},\tau_{t}}(x'^{(j)}|\mathcal{E})}{f_{\theta_{t},\tau_{t}}(x^{(j)}|\mathcal{E})}\times\frac{\varpi_{1}^{\text{KC}}(i;x'^{(1:\kappa)})\varpi_{2}^{\text{KC}}(j|i;x'^{(1:\kappa)})+\varpi_{1}^{\text{KC}}(j;x'^{(1:\kappa)})\varpi_{2}^{\text{KC}}(i|j;x'^{(1:\kappa)})}{\varpi_{1}^{\text{KC}}(i;x^{(1:\kappa)})\varpi_{2}^{\text{KC}}(j|i;x^{(1:\kappa)})+\varpi_{1}^{\text{KC}}(j;x^{(1:\kappa)})\varpi_{2}^{\text{KC}}(i|j;x^{(1:\kappa)})})$ 
\end{enumerate}
\item Snooker crossover operation (continuous target distributions):

\begin{enumerate}
\item draw $i\sim\varpi_{1}^{\text{SC}}(i;x^{(1:\kappa)})$ and $j|i\sim\varpi_{2}^{\text{SC}}(j|i;x^{(1:\kappa)})$
\item compute $x'^{(i)}=x^{(i)}+\sigma_{\text{SC}}^{2}r_{\text{SC}}\frac{x^{(j)}-x^{(i)}}{\left\Vert x^{(j)}-x^{(i)}\right\Vert _{2}}$,
where $r_{\text{SC}}\sim\mathrm{N}(0,1)$
\item accept $x'^{(1:\kappa)}:=(x^{(1:i-1)},x'^{(i)},x^{(i+1:\kappa)})$
with prob. $a_{\text{SC}}=\min(1,\frac{f_{\theta,\tau}(x'^{(i)}|\mathcal{E})}{f_{\theta,\tau}(x^{(i)}|\mathcal{E})})$
\end{enumerate}
\item Linear crossover operation (continuous target distributions):

\begin{enumerate}
\item draw $i\sim\varpi_{1}^{\text{LC}}(i;x^{(1:\kappa)})$ and $j|i\sim\varpi_{2}^{\text{LC}}(j|i;x^{(1:\kappa)})$
\item compute $x'^{(i)}=x^{(i)}+r_{\text{LC}}x^{(j)}$, where $r_{\text{LC}}\sim\mathrm{U}(-1,1)$
\item accept $x'^{(1:\kappa)}:=(x^{(1:i-1)},x'^{(i)},x^{(i+1:\kappa)})$
with prob. $a_{\text{LC}}=\min(1,\frac{f_{\theta,\tau}(x'^{(i)}|\mathcal{E})}{f_{\theta,\tau}(x^{(i)}|\mathcal{E})})$
\end{enumerate}
\end{itemize}
For the crossover operations, we considered probabilities:
\begin{align*}
\varpi_{1}^{\text{KC}}(i;x^{(1:\kappa)}) & =\frac{\exp(-U(x^{(i)})/\tau_{\text{KC}})}{\sum_{\forall\ell}\exp(-U(x^{(\ell)})/\tau_{\text{KC}})}, & i\in\{1,...,\kappa\};\\
\varpi_{2}^{\text{KC}}(j|i;x^{(1:\kappa)}) & =\frac{\exp(-U(x^{(i)})/\tau_{\text{KC}})}{\sum_{\forall\ell\ne i}\exp(-U(x^{(\ell)})/\tau_{\text{KC}})}, & j\in\{1,...,i-1,i+1,...,\kappa\};\\
\varpi_{1}^{\text{SC}}(i;x^{(1:\kappa)}) & =\frac{1}{\kappa}, & i\in\{1,...,\kappa\};\\
\varpi_{2}^{\text{SC}}(j|i;x^{(1:\kappa)}) & =\frac{\exp(-U(x^{(i)})/\tau_{\text{SC}})}{\sum_{\forall\ell\ne i}\exp(-U(x^{(\ell)})/\tau_{\text{SC}})}, & j\in\{1,...,i-1,i+1,...,\kappa\};\\
\varpi_{1}^{\text{LC}}(i;x^{(1:\kappa)}) & =\frac{1}{\kappa}, & i\in\{1,...,\kappa\};\\
\varpi_{2}^{\text{LC}}(j|i;x^{(1:\kappa)}) & =\frac{\exp(-U(x^{(i)})/\tau_{\text{LC}})}{\sum_{\forall\ell\ne i}\exp(-U(x^{(\ell)})/\tau_{\text{LC}})}, & j\in\{1,...,i-1,i+1,...,\kappa\},
\end{align*}
with quantities $\tau_{\text{KC}}$, $\tau_{\text{SC}}$, and $\tau_{\text{LC}}$
equal to 0.1, in Section \ref{sec:Applications}.

The pseudo-codes of the MCMC kernel mutation operations, used in Sections
\ref{sub:Gaussian-mixture-model} - \ref{sub:Protein-folding} are
given below. More details can be found in \citep{Smith1984,ChenSchmeiser1993a,Liang2011,MetropolisRosenbluthRosenbluthTellerTeller1953}. 
\begin{itemize}
\item Metropolis mutation operation:

For $i=1,...,\kappa$: 
\begin{enumerate}
\item compute $x'^{(i)}=x^{(i)}+\sigma_{\text{MRW}}^{2}r_{\text{MRW}}$
where $r_{\text{MRW}}\sim\mathrm{N}(0,\idmat_{d})$
\item accept $x'^{(1:\kappa)}:=(x^{(1:i-1)},x'^{(i)},x^{(i+1:\kappa)})$
with prob. $a_{\text{MRW}}=\min(1,\frac{f_{\theta,\tau}(x'^{(i)}|\mathcal{E})}{f_{\theta,\tau}(x^{(i)}|\mathcal{E})})$
\end{enumerate}
\item Hit-and-run mutation operation:

For $i=1,...,\kappa$:
\begin{enumerate}
\item compute $x'^{(i)}=x^{(i)}+\sigma_{\text{HR}}^{2}r_{\text{HR}}e_{\text{HR}}$,
where $r_{\text{HR}}\sim\mathrm{N}(0,1)$ and $e_{\text{HR}}$ is
drawn randomly from a unit $d$-dimensional space
\item accept $x'^{(1:\kappa)}:=(x^{(1:i-1)},x'^{(i)},x^{(i+1:\kappa)})$
with prob. $a_{\text{HR}}=\min(1,\frac{f_{\theta,\tau}(x'^{(i)}|\mathcal{E})}{f_{\theta,\tau}(x^{(i)}|\mathcal{E})})$
\end{enumerate}
\item $k$-point mutation operation:

For $i=1,...,\kappa$:
\begin{enumerate}
\item compute $x'^{(i)}=x^{(i)}+\sigma_{\text{KM}}^{2}r_{\text{KM}}e_{\text{KM}}$,
where $r_{\text{KM}}\sim\mathrm{N}(0,1)$ and $e_{\text{KM}}$ is
a $k<d$ aces $0$-$1$ $d$-dimensional vector randomly drawn
\item accept $x'^{(1:\kappa)}:=(x^{(1:i-1)},x'^{(i)},x^{(i+1:\kappa)})$
with prob. $a_{\text{KM}}=\min(1,\frac{f_{\theta,\tau}(x'^{(i)}|\mathcal{E})}{f_{\theta,\tau}(x^{(i)}|\mathcal{E})})$
\end{enumerate}
\end{itemize}

The Gibbs update (updating one pixel at a time) in the Spatial imaging
example in Section \ref{sub:Spatial-imaging} is given below. 
\begin{itemize}
\item Gibbs mutation operation in Section \ref{sub:Spatial-imaging}:

For $i=1,...,\kappa$:
\begin{enumerate}
\item draw $j$ randomly in $\{1,...,d\}$
\item draw $x_{j}^{(i)}\sim\text{Bernulli}(\varpi_{GI}(j;x^{(i)}))$, where
$\varpi_{GI}(j;x^{(i)})=(1+\frac{f_{\theta,\tau}((x_{1}^{(1)},...,x_{j-1}^{(i)},0,x_{j+1}^{(i)},...,x_{d}^{(i)})|\mathcal{E})}{f_{\theta,\tau}((x_{1}^{(1)},...,x_{j-1}^{(i)},1,x_{j+1}^{(i)},...,x_{d}^{(i)})|\mathcal{E})})^{-1}$.
\end{enumerate}
\end{itemize}

The pseudo-codes of the MCMC kernel mutation operations, used for
the Bayesian network example in Section \ref{sub:Bayesian-network-learning},
are given below. More details can be found in \citep{LiangZhang2009,WallaceKorb1999}.
\begin{itemize}
\item Temporal order operation:

For $i=1,...,\kappa$:
\begin{enumerate}
\item compute $\mathcal{G}'^{(i)}$ by swapping the order of two randomly
selected neighbouring nodes; if there is an edge between them, reverse
its direction.
\item accept $\mathcal{G}'^{(1:\kappa)}=(\mathcal{G}^{(1:i-1)},\mathcal{G}'^{(i)},\mathcal{G}^{(i+1:\kappa)})$
with prob. $a_{\text{TO}}=\min(1,\frac{f_{\theta,\tau}(\mathcal{G}'^{(i)}|\mathcal{E})}{f_{\theta,\tau}(\mathcal{G}{}^{(i)}|\mathcal{E})})$
\end{enumerate}
\item Skeletal change:

For $i=1,...,\kappa$:
\begin{enumerate}
\item compute $\mathcal{G}'^{(i)}$ by adding or deleting an edge between
a pair of randomly selected nodes.
\item accept $\mathcal{G}'^{(1:\kappa)}=(\mathcal{G}^{(1:i-1)},\mathcal{G}'^{(i)},\mathcal{G}^{(i+1:\kappa)})$
with prob. $a_{\text{SC}}=\min(1,\frac{f_{\theta,\tau}(\mathcal{G}'^{(i)}|\mathcal{E})}{f_{\theta,\tau}(\mathcal{G}{}^{(i)}|\mathcal{E})})$
\end{enumerate}
\item Double skeletal change:

For $i=1,...,\kappa$:
\begin{enumerate}
\item compute $\mathcal{G}'^{(i)}$ by randomly choosing two different pairs
of nodes, and adding or deleting edges between each pair of the nodes.
\item accept $\mathcal{G}'^{(1:\kappa)}=(\mathcal{G}^{(1:i-1)},\mathcal{G}'^{(i)},\mathcal{G}^{(i+1:\kappa)})$
with prob. $a_{\text{DS}}=\min(1,\frac{f_{\theta,\tau}(\mathcal{G}'^{(i)}|\mathcal{E})}{f_{\theta,\tau}(\mathcal{G}{}^{(i)}|\mathcal{E})})$
\end{enumerate}
\end{itemize}

\begin{rem}
The scale parameters of the proposals of the operations were tuned
during pilot runs using the adaptation scheme:\\
 $\log(\sigma_{\text{MRW}}^{2})\leftarrow\log(\sigma_{\text{MRW}}^{2})+[a_{\text{MRW}}-0.234]$;
this ensures that the associated expected acceptance probabilities
will be around $0.234$. In our applications, the performance of this
adaptation scheme was acceptable, however more sophisticated schemes
can be used. For more adaptive Metropolis-Hastings schemes see \citep{AndrieuThoms2008}.\end{rem}

%% file: paper.bbl
\begin{thebibliography}{}

\bibitem[\protect\citeauthoryear{Ali, Khompatraporn, and Zabinsky}{Ali
  et~al.}{2005}]{AliKhompatrapornZabinsky2005}
Ali, M.~M., C.~Khompatraporn, and Z.~B. Zabinsky (2005).
\newblock A numerical evaluation of several stochastic algorithms on selected
  continuous global optimization test problems.
\newblock {\em Journal of Global Optimization\/}~{\em 31\/}(4), 635--672.

\bibitem[\protect\citeauthoryear{Andrieu, Moulines, and Priouret}{Andrieu
  et~al.}{2005}]{AndrieuMoulinesPriouret2005}
Andrieu, C., {\'E}.~Moulines, and P.~Priouret (2005).
\newblock Stability of stochastic approximation under verifiable conditions.
\newblock {\em SIAM Journal on control and optimization\/}~{\em 44\/}(1),
  283--312.

\bibitem[\protect\citeauthoryear{Andrieu and Thoms}{Andrieu and
  Thoms}{2008}]{AndrieuThoms2008}
Andrieu, C. and J.~Thoms (2008).
\newblock A tutorial on adaptive {MCMC}.
\newblock {\em Statistics and Computing\/}~{\em 18\/}(4), 343--373.

\bibitem[\protect\citeauthoryear{Bachmann, Arkin, and Janke}{Bachmann
  et~al.}{2005}]{BachmannArkinJanke2005}
Bachmann, M., H.~Arkin, and W.~Janke (2005, Mar).
\newblock Multicanonical study of coarse-grained off-lattice models for folding
  heteropolymers.
\newblock {\em Phys. Rev. E\/}~{\em 71}, 031906.

\bibitem[\protect\citeauthoryear{Besag}{Besag}{1977}]{Besag1977}
Besag, J. (1977).
\newblock On spatial-temporal models and {Markov} fields.
\newblock In {\em Transactions of the Seventh Prague Conference on Information
  Theory, Statistical Decision Functions, Random Processes and of the 1974
  European Meeting of Statisticians}, pp.\  47--55. Springer.

\bibitem[\protect\citeauthoryear{Besag}{Besag}{1986}]{Besag1986}
Besag, J. (1986).
\newblock On the statistical analysis of dirty pictures.
\newblock {\em Journal of the Royal Statistical Society. Series B
  (Methodological)\/}~{\em 48\/}(3), 259--302.

\bibitem[\protect\citeauthoryear{Bornn, Jacob, Moral, and Doucet}{Bornn
  et~al.}{2013}]{BornnJacobMoralDoucet2013}
Bornn, L., P.~E. Jacob, P.~D. Moral, and A.~Doucet (2013).
\newblock An adaptive interacting {W}ang-{L}andau algorithm for automatic
  density exploration.
\newblock {\em Journal of Computational and Graphical Statistics\/}~{\em
  22\/}(3), 749--773.

\bibitem[\protect\citeauthoryear{Casella and Berger}{Casella and
  Berger}{1990}]{CasellaBerger1990}
Casella, G. and R.~L. Berger (1990).
\newblock {\em Statistical inference}, Volume~70.
\newblock Duxbury Press Belmont, CA.

\bibitem[\protect\citeauthoryear{{\v{C}}ern{\`y}}{{\v{C}}ern{\`y}}{1985}]{CernyVladimir1985}
{\v{C}}ern{\`y}, V. (1985).
\newblock Thermodynamical approach to the traveling salesman problem: An
  efficient simulation algorithm.
\newblock {\em Journal of optimization theory and applications\/}~{\em
  45\/}(1), 41--51.

\bibitem[\protect\citeauthoryear{Chen and Zhu}{Chen and
  Zhu}{1986}]{ChenZhu1986}
Chen, H. and Y.~Zhu (1986).
\newblock Stochastic approximation procedures with randomly varying
  truncations.
\newblock {\em Science China Mathematics\/}~{\em 29\/}(9), 914.

\bibitem[\protect\citeauthoryear{Chen and Schmeiser}{Chen and
  Schmeiser}{1993}]{ChenSchmeiser1993a}
Chen, M.-H. and B.~Schmeiser (1993).
\newblock Performance of the {Gibbs}, hit-and-run, and metropolis samplers.
\newblock {\em Journal of computational and graphical statistics\/}~{\em
  2\/}(3), 251--272.

\bibitem[\protect\citeauthoryear{Chen and Schmeiser}{Chen and
  Schmeiser}{1996}]{ChenSchmeiser1993}
Chen, M.-H. and B.~W. Schmeiser (1996).
\newblock General hit-and-run {Monte} {Carlo} sampling for evaluating
  multidimensional integrals.
\newblock {\em Operations Research Letters\/}~{\em 19\/}(4), 161--169.

\bibitem[\protect\citeauthoryear{Cios, Wedding, and Liu}{Cios
  et~al.}{1997}]{CiosWeddingLiu1997}
Cios, K.~J., D.~K. Wedding, and N.~Liu (1997).
\newblock {CLIP3}: {C}over learning using integer programming.
\newblock {\em Kybernetes\/}~{\em 26\/}(5), 513--536.

\bibitem[\protect\citeauthoryear{Dieterich and Hartke}{Dieterich and
  Hartke}{2012}]{DieterichHartke2012}
Dieterich, J.~M. and B.~Hartke (2012).
\newblock Empirical review of standard benchmark functions using evolutionary
  global optimization.
\newblock {\em Applied Mathematics\/}~{\em 3}, 1552.

\bibitem[\protect\citeauthoryear{Ellis and Wong}{Ellis and
  Wong}{2008}]{EllisWong2008}
Ellis, B. and W.~H. Wong (2008).
\newblock Learning causal {Bayesian} network structures from experimental data.
\newblock {\em Journal of the American Statistical Association\/}~{\em
  103\/}(482), 778--789.

\bibitem[\protect\citeauthoryear{Geman and Geman}{Geman and
  Geman}{1984}]{GemanGeman1984}
Geman, S. and D.~Geman (1984).
\newblock Stochastic relaxation, {G}ibbs distributions, and the {B}ayesian
  restoration of images.
\newblock {\em Pattern Analysis and Machine Intelligence, IEEE Transactions
  on\/}~{\em PAMI-6\/}(6), 721--741.

\bibitem[\protect\citeauthoryear{Gilks, Roberts, and George}{Gilks
  et~al.}{1994}]{GilksRobertsGeorge1994}
Gilks, W.~R., G.~O. Roberts, and E.~I. George (1994).
\newblock Adaptive direction sampling.
\newblock {\em Journal of the Royal Statistical Society. Series D (The
  Statistician)\/}~{\em 43\/}(1), pp. 179--189.

\bibitem[\protect\citeauthoryear{Gladshtein, Larionova, and Belyaev}{Gladshtein
  et~al.}{2012}]{GladshteinLarionovaBelyaev2012}
Gladshtein, L., N.~Larionova, and B.~Belyaev (2012).
\newblock Effect of ferrite-pearlite microstructure on structural steel
  properties.
\newblock {\em Metallurgist\/}~{\em 56\/}(7-8), 579--590.

\bibitem[\protect\citeauthoryear{Goldberg}{Goldberg}{1989}]{Goldberg1989}
Goldberg, D.~E. (1989).
\newblock {\em Genetic algorithms in search, optimization, and machine
  learning}, Volume 412.
\newblock Addison-Wesley Reading Menlo Park.

\bibitem[\protect\citeauthoryear{Haario and Saksman}{Haario and
  Saksman}{1991}]{HaarioSaksman1991}
Haario, H. and E.~Saksman (1991).
\newblock Simulated annealing process in general state space.
\newblock {\em Advances in Applied Probability\/}~{\em 23\/}(4), 866--893.

\bibitem[\protect\citeauthoryear{Hastings}{Hastings}{1970}]{Hastings1970}
Hastings, W.~K. (1970).
\newblock Monte carlo sampling methods using markov chains and their
  applications.
\newblock {\em Biometrika\/}~{\em 57\/}(1), 97--109.

\bibitem[\protect\citeauthoryear{Heckerman, Geiger, and Chickering}{Heckerman
  et~al.}{1995}]{HeckermanGeigerChickering1995}
Heckerman, D., D.~Geiger, and D.~M. Chickering (1995).
\newblock Learning {Bayesian} networks: The combination of knowledge and
  statistical data.
\newblock {\em Machine learning\/}~{\em 20\/}(3), 197--243.

\bibitem[\protect\citeauthoryear{Higdon}{Higdon}{1998}]{Higdon1998}
Higdon, D.~M. (1998).
\newblock Auxiliary variable methods for {Markov} chain {Monte} {Carlo} with
  applications.
\newblock {\em Journal of the American Statistical Association\/}~{\em
  93\/}(442), 585--595.

\bibitem[\protect\citeauthoryear{Holland}{Holland}{1975}]{Holland1975}
Holland, J.~H. (1975).
\newblock {\em Adaptation in natural and artificial systems: An introductory
  analysis with applications to biology, control, and artificial intelligence.}
\newblock U Michigan Press.

\bibitem[\protect\citeauthoryear{Hsu, Mehra, and Grassberger}{Hsu
  et~al.}{2003}]{HsuMehraGrassberger2003}
Hsu, H.-P., V.~Mehra, and P.~Grassberger (2003).
\newblock Structure optimization in an off-lattice protein model.
\newblock {\em Physical Review E\/}~{\em 68\/}(3), 037703.

\bibitem[\protect\citeauthoryear{Ingber}{Ingber}{1989}]{Ingber1989}
Ingber, L. (1989).
\newblock Very fast simulated re-annealing.
\newblock {\em Mathematical and computer modelling\/}~{\em 12\/}(8), 967--973.

\bibitem[\protect\citeauthoryear{Irb{\"a}ck, Peterson, Potthast, and
  Sommelius}{Irb{\"a}ck et~al.}{1997}]{IrbackPetersonPotthastSommelius1997}
Irb{\"a}ck, A., C.~Peterson, F.~Potthast, and O.~Sommelius (1997).
\newblock Local interactions and protein folding: A three-dimensional
  off-lattice approach.
\newblock {\em The Journal of chemical physics\/}~{\em 107\/}(1), 273--282.

\bibitem[\protect\citeauthoryear{Ising}{Ising}{1925}]{Ising1925}
Ising, E. (1925).
\newblock Beitrag zur theorie des ferromagnetismus.
\newblock {\em Zeitschrift f{\"u}r Physik A Hadrons and Nuclei\/}~{\em
  31\/}(1), 253--258.

\bibitem[\protect\citeauthoryear{Jackson, Sen, and Stoffa}{Jackson
  et~al.}{2004}]{JacksonSenStoffa2004}
Jackson, C., M.~K. Sen, and P.~L. Stoffa (2004).
\newblock An efficient stochastic {Bayesian} approach to optimal parameter and
  uncertainty estimation for climate model predictions.
\newblock {\em Journal of Climate\/}~{\em 17\/}(14), 2828--2841.

\bibitem[\protect\citeauthoryear{Kim, Lee, and Lee}{Kim
  et~al.}{2005}]{KimLeeLee2005}
Kim, S.-Y., S.~B. Lee, and J.~Lee (2005, Jul).
\newblock Structure optimization by conformational space annealing in an
  off-lattice protein model.
\newblock {\em Phys. Rev. E\/}~{\em 72}, 011916.

\bibitem[\protect\citeauthoryear{Kirkpatrick, Gelatt~Jr, Vecchi, and
  McCoy}{Kirkpatrick et~al.}{1983}]{KirkpatrickVecchi1983}
Kirkpatrick, S., C.~Gelatt~Jr, M.~Vecchi, and A.~McCoy (1983).
\newblock Optimization by simulated annealing.
\newblock {\em Science\/}~{\em 220\/}(4598), 671--679.

\bibitem[\protect\citeauthoryear{Kurgan, Cios, Tadeusiewicz, Ogiela, and
  Goodenday}{Kurgan et~al.}{2001}]{KurganCiosTadeusiewiczOgielaGoodenday2001}
Kurgan, L.~A., K.~J. Cios, R.~Tadeusiewicz, M.~Ogiela, and L.~S. Goodenday
  (2001).
\newblock Knowledge discovery approach to automated cardiac spect diagnosis.
\newblock {\em Artificial intelligence in medicine\/}~{\em 23\/}(2), 149--169.

\bibitem[\protect\citeauthoryear{Liang}{Liang}{2004}]{Liang2004}
Liang, F. (2004).
\newblock Annealing contour {Monte} {Carlo} algorithm for structure
  optimization in an off-lattice protein model.
\newblock {\em The Journal of chemical physics\/}~{\em 120\/}(14), 6756--6763.

\bibitem[\protect\citeauthoryear{Liang}{Liang}{2007}]{Liang2007}
Liang, F. (2007).
\newblock {A}nnealing stochastic approximation {M}onte {C}arlo algorithm for
  neural network training.
\newblock {\em Machine Learning\/}~{\em 68\/}(3), 201--233.

\bibitem[\protect\citeauthoryear{Liang}{Liang}{2009}]{Liang2009}
Liang, F. (2009).
\newblock Improving {SAMC} using smoothing methods: theory and applications to
  {Bayesian} model selection problems.
\newblock {\em The Annals of Statistics\/}~{\em 37\/}(5B), 2626--2654.

\bibitem[\protect\citeauthoryear{Liang}{Liang}{2011}]{Liang2011}
Liang, F. (2011).
\newblock {A}nnealing evolutionary stochastic approximation {M}onte {C}arlo for
  global optimization.
\newblock {\em Statistics and Computing\/}~{\em 21\/}(3), 375--393.

\bibitem[\protect\citeauthoryear{Liang}{Liang}{2014}]{Liang2014}
Liang, F. (2014).
\newblock An overview of stochastic approximation {Monte} {Carlo}.
\newblock {\em Wiley Interdisciplinary Reviews: Computational
  Statistics\/}~{\em 6\/}(4), 240--254.

\bibitem[\protect\citeauthoryear{Liang, Cheng, and Lin}{Liang
  et~al.}{2014}]{LiangChengLin2013}
Liang, F., Y.~Cheng, and G.~Lin (2014).
\newblock Simulated stochastic approximation annealing for global optimization
  with a square-root cooling schedule.
\newblock {\em Journal of the American Statistical Association\/}~{\em
  109\/}(506), 847--863.

\bibitem[\protect\citeauthoryear{Liang, Liu, and Carroll}{Liang
  et~al.}{2007}]{LiangLiuCarroll2007}
Liang, F., C.~Liu, and R.~J. Carroll (2007).
\newblock {S}tochastic approximation in {M}onte {C}arlo computation.
\newblock {\em Journal of the American Statistical Association\/}~{\em
  102\/}(477), 305--320.

\bibitem[\protect\citeauthoryear{Liang, Liu, and Carroll}{Liang
  et~al.}{2010}]{LiangLiuCarroll2010}
Liang, F., C.~Liu, and R.~J. Carroll (2010).
\newblock {S}tochastic approximation {M}onte {C}arlo.
\newblock {\em Advanced {Markov} Chain {Monte} {Carlo} Methods: Learning from
  Past Samples\/}, 199--303.

\bibitem[\protect\citeauthoryear{Liang and Wong}{Liang and
  Wong}{2000}]{LiangWong2000}
Liang, F. and W.~H. Wong (2000).
\newblock Evolutionary {Monte} {Carlo}: Applications to $c_p$ model sampling
  and change point problem.
\newblock {\em Statistica sinica\/}~{\em 10\/}(2), 317--342.

\bibitem[\protect\citeauthoryear{Liang and Wong}{Liang and
  Wong}{2001}]{LiangWong2001}
Liang, F. and W.~H. Wong (2001).
\newblock Real-parameter evolutionary {Monte} {Carlo} with applications to
  {B}ayesian mixture models.
\newblock {\em Journal of the American Statistical Association\/}~{\em
  96\/}(454), 653--666.

\bibitem[\protect\citeauthoryear{Liang and Zhang}{Liang and
  Zhang}{2009}]{LiangZhang2009}
Liang, F. and J.~Zhang (2009).
\newblock Learning {Bayesian} networks for discrete data.
\newblock {\em Computational Statistics \& Data Analysis\/}~{\em 53\/}(4),
  865--876.

\bibitem[\protect\citeauthoryear{Liang, Qin, Suganthan, and Baskar}{Liang
  et~al.}{2006}]{LiangQinSuganthanBaskar2006}
Liang, J.~J., A.~K. Qin, P.~N. Suganthan, and S.~Baskar (2006).
\newblock Comprehensive learning particle swarm optimizer for global
  optimization of multimodal functions.
\newblock {\em Evolutionary Computation, IEEE Transactions on\/}~{\em 10\/}(3),
  281--295.

\bibitem[\protect\citeauthoryear{Madigan and Raftery}{Madigan and
  Raftery}{1994}]{MadiganRaftery1994}
Madigan, D. and A.~E. Raftery (1994).
\newblock Model selection and accounting for model uncertainty in graphical
  models using {Occam}'s window.
\newblock {\em Journal of the American Statistical Association\/}~{\em
  89\/}(428), 1535--1546.

\bibitem[\protect\citeauthoryear{Metropolis, Rosenbluth, Rosenbluth, Teller,
  and Teller}{Metropolis
  et~al.}{1953}]{MetropolisRosenbluthRosenbluthTellerTeller1953}
Metropolis, N., A.~W. Rosenbluth, M.~N. Rosenbluth, A.~H. Teller, and E.~Teller
  (1953).
\newblock Equation of state calculations by fast computing machines.
\newblock {\em The journal of chemical physics\/}~{\em 21\/}(6), 1087--1092.

\bibitem[\protect\citeauthoryear{M{\"u}hlenbein, Schomisch, and
  Born}{M{\"u}hlenbein et~al.}{1991}]{MuhlenbeinSchomischBorn1991}
M{\"u}hlenbein, H., M.~Schomisch, and J.~Born (1991).
\newblock The parallel genetic algorithm as function optimizer.
\newblock {\em Parallel computing\/}~{\em 17\/}(6), 619--632.

\bibitem[\protect\citeauthoryear{M{\"u}ller}{M{\"u}ller}{1991}]{Muller1991}
M{\"u}ller, P. (1991).
\newblock A generic approach to posterior integration and {Gibbs} sampling.
\newblock Technical report, Purdue University, Department of Statistics,
  Indiana.

\bibitem[\protect\citeauthoryear{Neal}{Neal}{1996}]{Neal1996}
Neal, R. (1996).
\newblock Sampling from multimodal distributions using tempered transitions.
\newblock {\em Statistics and computing\/}~{\em 6\/}(4), 353--366.

\bibitem[\protect\citeauthoryear{Nummelin}{Nummelin}{2004}]{Nummelin2004}
Nummelin, E. (2004).
\newblock {\em General irreducible {M}arkov chains and non-negative operators},
  Volume~83.
\newblock Cambridge University Press.

\bibitem[\protect\citeauthoryear{Pelletier}{Pelletier}{1998}]{Pelletier1998}
Pelletier, M. (1998).
\newblock Weak convergence rates for stochastic approximation with application
  to multiple targets and simulated annealing.
\newblock {\em Annals of Applied Probability\/}~{\em 8\/}(1), 10--44.

\bibitem[\protect\citeauthoryear{Robbins and Monro}{Robbins and
  Monro}{1951}]{RobbinsMonro1951}
Robbins, H. and S.~Monro (1951).
\newblock A stochastic approximation method.
\newblock {\em The annals of mathematical statistics\/}~{\em 22\/}(3),
  400--407.

\bibitem[\protect\citeauthoryear{Robert}{Robert}{2007}]{Robert2007}
Robert, C.~P. (2007, May).
\newblock {\em The Bayesian Choice: From Decision-Theoretic Foundations to
  Computational Implementation\/} (2nd ed.).
\newblock Springer.

\bibitem[\protect\citeauthoryear{Robert and Casella}{Robert and
  Casella}{2004}]{RobertCasella2004}
Robert, C.~P. and G.~Casella (2004).
\newblock {\em {Monte} {Carlo} statistical methods}, Volume 319.
\newblock Springer-Verlag, New York.

\bibitem[\protect\citeauthoryear{Roberts and Tweedie}{Roberts and
  Tweedie}{1996}]{Roberts1996}
Roberts, G.~O. and R.~L. Tweedie (1996).
\newblock Geometric convergence and central limit theorems for multidimensional
  hastings and metropolis algorithms.
\newblock {\em Biometrika\/}~{\em 83\/}(1), 95--110.

\bibitem[\protect\citeauthoryear{Rosenthal}{Rosenthal}{1995}]{Rosenthal1995}
Rosenthal, J.~S. (1995).
\newblock Minorization conditions and convergence rates for markov chain monte
  carlo.
\newblock {\em Journal of the American Statistical Association\/}~{\em
  90\/}(430), 558--566.

\bibitem[\protect\citeauthoryear{Salomon}{Salomon}{1996}]{Salomon1996}
Salomon, R. (1996).
\newblock Re-evaluating genetic algorithm performance under coordinate rotation
  of benchmark functions. a survey of some theoretical and practical aspects of
  genetic algorithms.
\newblock {\em BioSystems\/}~{\em 39\/}(3), 263--278.

\bibitem[\protect\citeauthoryear{Sen and Stoffa}{Sen and
  Stoffa}{1996}]{SenStoffa1996}
Sen, M.~K. and P.~L. Stoffa (1996).
\newblock {Bayesian} inference, {Gibbs} sampler and uncertainty estimation in
  geophysical inversion.
\newblock {\em Geophysical Prospecting\/}~{\em 44\/}(2), 313--350.

\bibitem[\protect\citeauthoryear{Smith}{Smith}{1984}]{Smith1984}
Smith, R.~L. (1984).
\newblock Efficient {Monte} {Carlo} procedures for generating points uniformly
  distributed over bounded regions.
\newblock {\em Operations Research\/}~{\em 32\/}(6), 1296--1308.

\bibitem[\protect\citeauthoryear{Song, Wu, and Liang}{Song
  et~al.}{2014}]{SongWuLiang2014}
Song, Q., M.~Wu, and F.~Liang (2014, 12).
\newblock Weak convergence rates of population versus single-chain stochastic
  approximation mcmc algorithms.
\newblock {\em Advances in Applied Probability\/}~{\em 46\/}(4), 1059--1083.

\bibitem[\protect\citeauthoryear{Stillinger and Head-Gordon}{Stillinger and
  Head-Gordon}{1995}]{StillingerHeadGordon1995}
Stillinger, F.~H. and T.~Head-Gordon (1995).
\newblock Collective aspects of protein folding illustrated by a toy model.
\newblock {\em Physical Review E\/}~{\em 52\/}(3), 2872.

\bibitem[\protect\citeauthoryear{Stillinger, Head-Gordon, and
  Hirshfeld}{Stillinger et~al.}{1993}]{StillingerHeadGordonHirshfeld1993}
Stillinger, F.~H., T.~Head-Gordon, and C.~L. Hirshfeld (1993).
\newblock Toy model for protein folding.
\newblock {\em Physical review E\/}~{\em 48\/}(2), 1469.

\bibitem[\protect\citeauthoryear{T{\"o}rn and Zilinskas}{T{\"o}rn and
  Zilinskas}{1989}]{TornZilinskas1989}
T{\"o}rn, A. and A.~Zilinskas (1989).
\newblock {\em Global Optimization (or Lecture Notes in Computer Science; Vol.
  350)}.
\newblock Springer-Verlag, Berlin.

\bibitem[\protect\citeauthoryear{Wallace and Korb}{Wallace and
  Korb}{1999}]{WallaceKorb1999}
Wallace, C.~S. and K.~B. Korb (1999).
\newblock Learning linear causal models by mml sampling.
\newblock In {\em Causal models and intelligent data management}, pp.\
  89--111. Springer.

\bibitem[\protect\citeauthoryear{Wang and Landau}{Wang and
  Landau}{2001}]{WangLandau2001}
Wang, F. and D.~P. Landau (2001).
\newblock Efficient, multiple-range random walk algorithm to calculate the
  density of states.
\newblock {\em Physical Review Letters\/}~{\em 86\/}(10), 2050.

\bibitem[\protect\citeauthoryear{Wermuth and Lauritzen}{Wermuth and
  Lauritzen}{1982}]{WermuthLauritzen1982}
Wermuth, N. and S.~L. Lauritzen (1982).
\newblock {\em Graphical and Recursive Models for Contigency Tables}.
\newblock Biometrika Trust.

\bibitem[\protect\citeauthoryear{Wu and Liang}{Wu and
  Liang}{2011}]{WuLiang2011}
Wu, M. and F.~Liang (2011).
\newblock Population {SAMC} vs {SAMC}: Convergence and applications to gene
  selection problems.
\newblock {\em J Biomet Biostat S\/}~{\em 1}, 2.

\end{thebibliography}
